\documentclass[onecolumn]{article}

\usepackage[numbers,sort&compress]{natbib}
\usepackage{amsmath,bm}	
\usepackage{authblk}							
\usepackage{float}
\usepackage{graphicx}
\usepackage{verbatim}
\usepackage{subfigure}
\usepackage{amssymb}
\usepackage{color}
\usepackage[top=25mm,bottom=25mm,left=20mm,right=20mm,headsep=10mm]{geometry}
\usepackage[colorlinks,linkcolor=blue,anchorcolor=blue,citecolor=blue]{hyperref}
\usepackage{indentfirst}
\usepackage{enumitem}
\usepackage{lineno}
\usepackage{setspace}
\usepackage{algorithm}
\usepackage{algorithmicx}
\usepackage{algpseudocode}
\usepackage{soul}

\newcommand\keywords[1]{\textbf{Keywords}: #1}

\title{Continuum model for disconnection structures of hetero-interfaces}

\author[1]{Luchan Zhang} 
\author[2,3]{Xiaoxue Qin}
\author[4] {Dongsong Tao}
\author[4] {Jian Han}
\author[5,6] {David J. Srolovitz }
\author[7,8]{Yang Xiang \thanks{Corresponding author: Yang Xiang (maxiang@ust.hk)}}

\affil[1]{\small School of Mathematical Sciences, Shenzhen University, Shenzhen, China}
\affil[2]{\small Department of Mathematics, Shanghai University, Shanghai, China}
\affil[3]{\small Newtouch Center for Mathematics of Shanghai University, Shanghai, China}
\affil[4]{\small Department of Materials Science and Engineering, City University of Hong Kong, Hong Kong SAR, China.}
\affil[5]{\small Department of Mechanical Engineering, The University of Hong Kong, Hong Kong SAR, China.}
\affil[6]{\small Materials Innovation Institute for Life Sciences and Energy (MILES), The University of Hong Kong, Shenzhen, China.}
\affil[7]{\small Department of Mathematics, The Hong Kong University of Science and Technology, Clear Water Bay, Hong Kong SAR, China}
\affil[8]{\small HKUST Shenzhen-Hong Kong Collaborative Innovation Research Institute, Futian, Shenzhen, China}

\date{}

\begin{document}
\maketitle

\begin{abstract}
Hetero-interfaces, as planar defects between dissimilar materials, accommodate the lattice misfit through networks of line defects of disconnections (a combination of structural ledges/step and dislocation characters).
We develop a continuum model that predicts the hetero-interface structures, including disconnection networks, the interface inclination and misorientation angles. The model is based on the interfacial energy minimization with respect to all possible disconnection modes, subject to the interface structure constraint of the generalized Frank-Bilby equation that describes the lattice correspondence across the interface including both dislocation and step characters of disconnections and the inclination and misorientation angles.
Numerical simulations are performed to determine the equilibrium configurations of three different types of hetero-interfaces: the $\alpha$/$\beta$  (orthorhombic/monoclinic) interface in ZrO$_2$,
$\alpha$/$\beta$ hetero-interfaces in Ti and  Au/Pd (001) twist-misfit interfaces.
Comparisons with experiments, atomistic simulations and theoretical predictions show that our continuum model provides excellent predictions of the hetero-interface structures including disconnection networks and inclination, misorientation angles.
Furthermore, systematic analysis reveals the functional dependence of hetero-interface structural properties on inclination and misorientation angles.
\end{abstract}

\keywords{hetero-interface structure, disconnection, generalized Frank-Bilby equation.}


\section{Introduction}

{
Hetero-interfaces are planar defects across which the composition, crystal lattice structure (lattice parameters and/or crystal symmetry), and/or crystallographic orientation change. The change in crystal structure/lattice parameters across the interface (misfit) may be accommodated by a network of line defects~\cite{Merwe1950,Matthews1966,Sutton1995,Bollmann1970,Hirth19964749,Han2018}.
These line defects are, in general, disconnections which may be accommodated by a combination of structural ledges, characterized by step height $h$, and dislocations, characterized by a Burgers vector $\bm{b}$, i.e.,  $(\bm{b}, h)$.
The hetero-interface structure strongly influences the mechanical, electronic and plastic deformation properties of multiphase alloys and composite materials  \cite{Merwe1950,Matthews1966,Sutton1995,Bollmann1970,Hirth19964749,Han2018,Hwang1980,Chiao19901163,Aaronson19912873,Aaronson19912857,
Aaronson19942133,Mills1998295,Hirth20035385,Bhattacharyya20076765,Pond20075289,Hirth2010-1,Wang20106685,Quek2011,Shi20124172,Banerjee2013844,
Hirth2013,Demkowicz2013,Wang201440,Demkowicz2015234,quantizedFB2018,Shao2018,Zheng201825,Ackerman2020013602,Zhang2021110277,Gordon-Sills2024,Tao2023}.}

Banerjee and Williams \cite{Banerjee2013844}  reviewed the experiments, theory, and simulations of phase transformations in titanium and its alloys, and demonstrated the important role played by the $\alpha$/$\beta$ interface in determining the  mechanical behaviour of these materials.
Hwang {\it et al.} \cite{Hwang1980} investigated the interfacial microstructure of twist interphase boundaries in Au-Ag and Au-Pd bicrystal thin films. Chiao and Chen \cite{Chiao19901163} investigated the single-interface transformation in the martensitic growth in ZrO$_2$ by experiments.
Furuhara {\it et al.} \cite{Aaronson19912873} studied the body-centered cubic (BCC)/hexagonal close-packed (HCP) interface structure of Ti-Cr alloy, which is characterized by broad faces and structural/growth ledges.
Their analysis of these interfaces led to a prediction of the most probable interfacial structures, which they characterized in terms of  misfit dislocations with structural ledges \cite{Aaronson19912857} based on the Bollmann O-lattice theory \cite{Aaronson19942133}.
Mills {\it et al.} \cite{Mills1998295} studied the structure of $\alpha$/$\beta$ interfaces in Ti alloys by experiments.
Zheng {\it et al.} \cite{Zheng201825}, Ackerman {\it et al.} \cite{Ackerman2020013602}, and Bhattacharyya {\it et al.} \cite{Bhattacharyya20076765} characterized  $\alpha$/$\beta$ interfaces in Ti in terms of a disconnection network (spacings and step heights),  Burgers vectors and habit plane  inclination.
Hirth and Pond  \cite{Hirth19964749} and Pond {\it et al.}  \cite{Hirth20035385,Pond20075289} developed a topological model to predict  disconnection interfacial structures, habit planes and orientation relationships.
Shi {\it et al.} \cite{Shi20124172} and Zhou {\it et al.} \cite{Wang20106685} performed phase field simulations which incorporate  interfacial misfit dislocations and structural ledges to predict  equilibrium precipitate shapes and stress fields.

The misfit, associated with an equilibrium hetero-interface, may be accommodated by dislocation arrays that are constrained  by the Frank-Bilby equation~\cite{Frank1950,Bilby1955,Sutton1995}, i.e., predicting the net Burgers vector content $\bm{B}(\bm{p})$ crossing a (any) probe vector $\bm{p}$ that lies in  the interface, { and the long-range elastic
fields vanish when the Frank-Bilby equation is satisfied.
Vattre and Demkowicz \cite{Demkowicz2013} focused on the determination of the reference state for the interfacial misfitting dislocation arrays, using the far field strain vanishing condition and solving an anisotropic elasticity problem for a given interface. They did not consider the disconnection structure, and only considered dislocation arrays with up to two possible Burgers vectors for which the Frank-Bilby equation gives a unique dislocation structure.
  Sangghaleh and  Demkowicz developed a software for  enumeration of the dislocation structure solutions  to the Frank-Bilby equation up to three Burgers vectors on a hetero-interface \cite{quantizedFB2018}.
Zhang {\it et al.} developed optimization methods for finding dislocation structures of low angle grain boundaries \cite{Zhang2017,Qin2022} and hetero-interfaces \cite{Zhang2021110277} for dislocations with all possible Burgers vectors. These studies do not consider the effects of disconnections.
}

Since interfaces are generally described by disconnection networks rather than networks of pure dislocations (that only accommodate misfit), description of the disconnection step character is essential. A distribution of step content (step height and spacing) provides a description of the interface inclination (i.e., the product of the disconnection step height and density). {
   Pond and Hirth~\cite{Hirth2010-1} derived a nonlinear equation that determines the interface inclination under the condition of cancellation of the long-range strain fields for a single mode disconnections based on the misfit and rotation across the interface, and solved the nonlinear equation numerically by an iterative method.  Gordon and Sills~\cite{Gordon-Sills2024} proposed a root finding method that solves such a single-disconnection-mode based nonlinear equation self-consistently, under the condition of additional two sets of dislocations for which the three-Burgers-vector Frank-Bilby equation has a unique solution~\cite{Knowles1982}, and developed a software that enumerates all the possible solutions to the Frank-Bilby equation up to three Burgers vectors combined with the constraint of the nonlinear equation for the single mode disconnections. In the phase field simulations of equilibrium precipitate shapes by Shi {\it et al.}~\cite{Shi20124172} and Zhou {\it et al.}~\cite{Wang20106685},  they also prescribed the single mode disconnections and two sets of Burgers vectors for dislocations as well as the inter-distance of each array of disconnections or dislocations.

In realistic hetero-interfaces, the interfacial defect network may contain multiple modes of disconnections and dislocations that have different Burgers vectors and step heights,  whose competition and interactions determine the interfacial energy, interface mobility, defect accommodation, and morphological evolution~\cite{Hwang1980,Han2018,Wei2020}.  We have proposed a generalized Frank-Bilby equation based on the constraint on disconnection step character together with the original Frank-Bilby equation~\cite{Tao2023}, unlike the available single-disconnection based theories~\cite{Hirth2010-1,Gordon-Sills2024}. This  setting of multiple modes of disconnections allows the existence of disconnection arrays to accommodate the constraint on disconnection step character on any interface, and the original Frank-Bilby equation can be further satisfied by dislocation arrays with multiple possible Burgers vectors. Thus, the generalized Frank-Bilby equation enables full description of the structure of most interfaces of practical interest.

While the generalized Frank-Bilby equation \cite{Tao2023} provides a quantitative model for the general setting of multiple modes of disconnections for the structure of hetero-interfaces, it does not provide a unique prediction of the interface disconnection network,  interface inclination  and misorientation across the interface when multiple modes of disconnections are considered; see the details given at the end of Sec.~\ref{sec:setting}.  
This is, in large part,  associated with the fact that there are many possible disconnections allowed by the bicrystallography of the two phases that meet at the heterophase interface.
A special case of one-dimensional solution curve in the five-dimensional interface
inclination-misorientation space for the interface in the $\alpha$ nucleus in a $\beta$ Ti alloy with two possible modes of disconnections was examined using the generalized Frank-Bilby equation \cite{Tao2023}. The dimension of the solution space of the generalized Frank-Bilby equation may be even higher when more possible modes of disconnections (including more possible Burgers vectors of dislocation arrays) are present. For practical application of the theory, a systematic approach is desired to predict the physically relevant, energetically favorable interface configurations and disconnection structures out of the solution space of the generalized Frank-Bilby equation with infinitely many solutions, instead of exploring these infinitely many solutions directly.



In this paper, we present a continuum model to predict the  hetero-interface disconnection network, as well as the interface inclination  and the misorientation across the interface.  The equilibrium interface configuration and disconnection structure are obtained by minimizing the interfacial energy associated with the disconnection network. Specifically, we treat the generalized Frank-Bilby equation~\cite{Tao2023} as a constraint  and then minimize the interfacial energy with respect to all possible disconnection modes, interface inclination and the  misorientation across the interface, to obtain the interface configuration and its disconnection network structure. The interface misorientation and inclination angles are naturally incorporated in  the interfacial energy formulation and through the generalized Frank-Bilby equation constraint.
The interfacial energy of the hetero-interface consists of only the local dislocation energy (core energy and short-range elastic fields) and the step energy of the disconnection network; { recall that \cite{Hirth2001,Sutton1995}} the long-range elastic fields vanish when the Frank-Bilby equation is satisfied.
We  employ the local energy  of a dislocation framework that was derived in \cite{Zhu2014175,Zhang2017}, validated by atomistic simulations \cite{Zhang2017,Wan2023}, and applied to dislocation structures of low angle grain boundaries \cite{Zhang2017,Qin2022} and hetero-interfaces without disconnections \cite{Zhang2021110277}.
}

We numerically solve this continuum constrained minimization problem by a penalty method, and obtain  evolution equations for the disconnection densities with orientations, misorientation and interface inclination angles.
A numerical scheme is designed to find the equilibrium interface state  corresponding to the global  minimum of the energy among all possible disconnection modes of the interface allowed by the bicrystallography.
We perform numerical simulations of the  $\alpha$/$\beta$  (orthorhombic/monoclinic) interface in zirconia ZrO$_2$,
$\alpha$/$\beta$ (bcc/hcp) hetero-interfaces in Ti and  Au/Pd (001) twist-misfit interfaces.
Comparisons with experiments and atomistic simulations show that our continuum model  provides excellent predictions of the structure of hetero-interfaces.
Based on our continuum model, we provide a systematic analysis of the properties of the hetero-interfaces as functions of interface inclination and misorientations. 

This paper is organized as follows. In Sec.~\ref{sec:setting}, we review the structure of hetero-interfaces and the generalized Frank-Bilby equation~\cite{Tao2023}.
In Sec.~\ref{sec:model}, we present our continuum model for the disconnection networks, interface inclination and misorienation that describe the interface, and a numerical scheme for finding the global energy minimum interface structure.
In Sec.~\ref{sec:simulation}, we apply our continuum model to several example  hetero-interfaces, and compare our predictions with experimental data, theoretical predictions and atomistic simulations \cite{Ackerman2020013602,Hirth20035385,Hirth19964749,Dahmen198263,Mills1998295,Zheng201825,Aaronson19912857,Aaronson19912873}.
Conclusions and discussion are presented in Sec.~\ref{sec:conclusion}.

\section{Hetero-interfaces and the generalized Frank-Bilby equation}\label{sec:setting}

\begin{figure}[tb]
\centering
\includegraphics[width=0.98\linewidth]{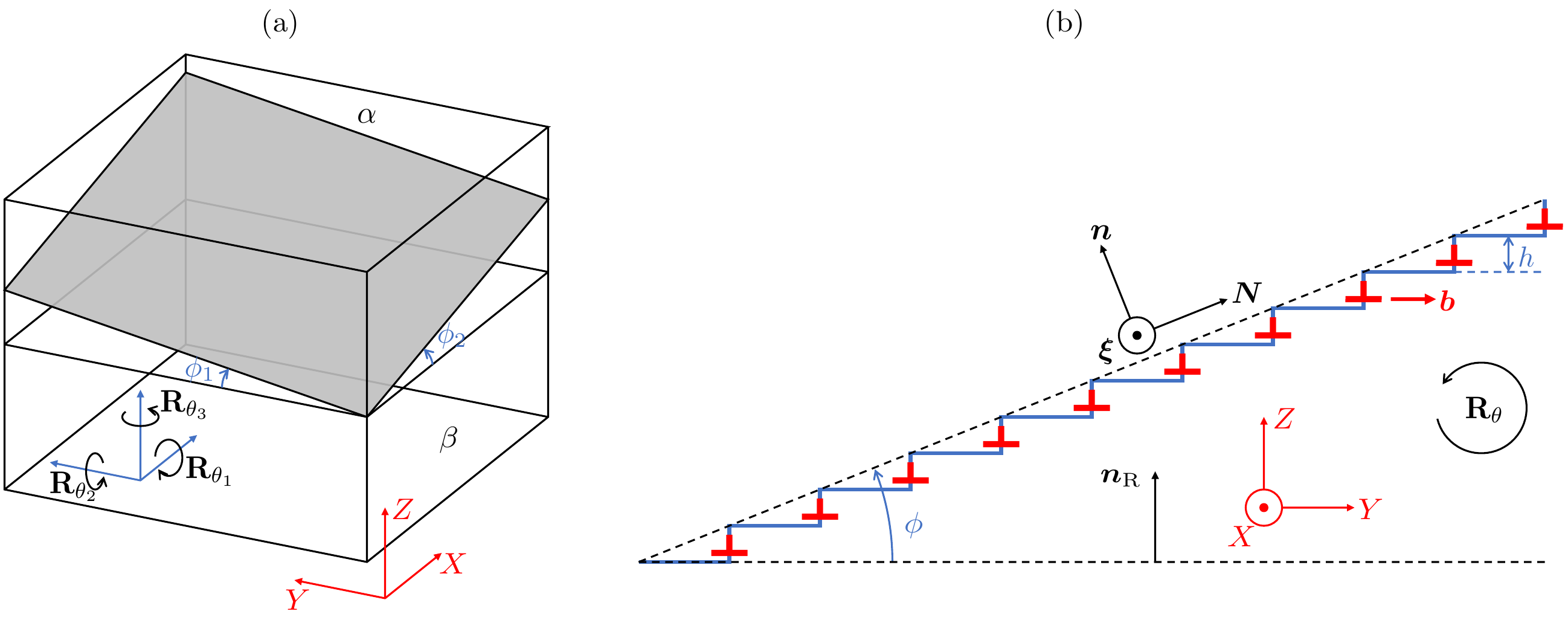}
\caption{(a) Illustration of the bicrystal geometry with a hetero-interface. The inclination is described by $\bm{\phi} = (\phi_1, \phi_2)$, where $\phi_1$ and $\phi_2$ are tilt angles of the reference interface plane (i.e., the $XY$-plane) about the $X$ and $Y$ axes, respectively.
(b) Discrete disconnection representation of the interface. {The interface plane is obtained by rotating the $XY$-plane successively about the $X$-axis by an angle $\phi_1$ first and then about the $Y$-axis by $\phi_2$.
The lattices adjacent to the interface are allowed to rotate relative to each other described by the misorientation $\bm{\theta} = (\theta_1,\theta_2,\theta_3)$, where $\theta_1$, $\theta_2$, and $\theta_3$ are rotation angles around the $X$, $Y$, and $Z$ axes, respectively.}
}
\label{fig:geometry_demo}
\end{figure}

Figure~\ref{fig:geometry_demo}(a) illustrates a bicrystal with a hetero-interface (grey plane) and two adjacent crystals, labeled $\alpha$ and $\beta$ (with different crystal structures and/or lattice parameters).
When the misfit in lattice spacings of the adjacent crystals at the interface is not so large that disconnection cores overlap, the interface may be described as alternating  terraces and steps described as a network of disconnections with misfit dislocation character and steps~\cite{Aaronson19912873,Hirth19964749,Ackerman2020013602,Shi20124172,Hirth20035385,Hirth2013}; see Fig.~\ref{fig:geometry_demo}(b).
{
We define the reference plane as that parallel to the terraces, and establish a coordinate system $XYZ$ such that the $XY$ plane aligns with the reference plane. The actual (natural-state) interface plane is obtained by rotating the $XY$-plane successively about the $X$-axis by an inclination angle $\phi_1$ first and then about the $Y$-axis by an inclination angle $\phi_2$; thus, the interface inclination is described by the vector $\bm{\phi} = (\phi_1, \phi_2)$.
The crystals adjacent to the interface are allowed to rotate relative to each other described by the misorientation across the interface $\bm{\theta} = (\theta_1,\theta_2,\theta_3)$, where $\theta_1$, $\theta_2$, and $\theta_3$ are rotation angles around the $X$, $Y$, and $Z$ axes, respectively. See Figs.~\ref{fig:geometry_demo}(a) and (b) for an illustration of the geometry of the hetero-interface and its disconnection structure.}

{
On an interface with normal direction $\bm{n}$, there are multiple possible modes of disconnection and dislocations.
Each mode of disconnection has a Burgers vector $\bm{b}_j$ and a step height $h_j$, and each array of dislocations have Burgers vector $\bm{b}_i$. We denote all the possible modes of disconnections and dislocations by the Burgers vector-step height pair $(\bm{b}_j, h_j)$, $j=1, \cdots, J$, where $h_j=0$ for dislocations. The possible Burgers vectors and step heights of disconnections and possible Burgers vectors of dislocations are given by experiments or atomistic calculations; see the application examples in Sec.~\ref{sec:simulation}.
}

The Frank-Bilby equation provides a geometric constraint on the Burgers vector in the disconnection array at the equilibrium interface~\cite{Frank1950,Bilby1955}; more specifically, it describes the net Burgers vector $\bm{B}(\bm{p})$ crossing a probe vector $\bm{p}$ lying in the interface.
On one hand, consideration of the misfit across the interface suggests that  the net Burgers vector is $\bm{B}(\bm{p})=(\mathbf{S}_{\alpha}^{-1} - \mathbf{S}_{\beta}^{-1})\bm{p}$, where $\mathbf{S}_{\alpha/\beta}$ is the transformation matrix that maps the unstrained reference lattice to that of $\alpha$/$\beta$ phase ($\mathbf{S}_{\alpha/\beta}^{-1}$ is the inverse of $\mathbf{S}_{\alpha/\beta}$).
The net Burgers vector $\bm{B}(\bm{p})$ can also be calculated directly from the Burgers vectors of the disconnections in the interface.
Given the possible disconnection Burgers vectors $\bm{b}_j$ ($j=1, \cdots, J$), the structures of the dislocation arrays can be determined from the corresponding reciprocal vectors $\bm{N}_j$ ($j=1, \cdots, J$), where the magnitude $N_j$ is the disconnection density and its direction is perpendicular to the disconnection line $\bm{\xi}_j$ and the interface unit normal $\bm{n}$, i.e., $\bm{\xi}_j = (\bm{N}_j / N_j) \times \bm{n}$.
In this case, the net Burgers vector is
$\bm{B}(\bm{p}) = \sum_{j=1}^J (\bm{N}_j \cdot \bm{p})\bm{b}_j$.
From these two approaches, the Frank-Bilby equation implies that the net Burger density may be described as
\begin{equation}\label{eqn:GFB1}
\bm{B}(\bm{p})
=
\sum_{j=1}^J
\left(\bm{N}_j \cdot \bm{p}\right) \bm{b}_{j}
=
\left(\mathbf{S}_{\alpha}^{-1} - \mathbf{S}_{\beta}^{-1}\right) \bm{p}
\end{equation}
for any probe vector $\bm{p}$ lying in the interface.

While the dislocation character of the disconnection network at the hetero-interface must satisfy the Frank-Bilby equation, the step character must also satisfy a  constraint that the component of the probe vector normal to the reference plane equals the product of the disconnection step heights and the disconnection density
\begin{equation}\label{eqn:GFB2}
H(\bm{p})
=
\sum_{j=1}^J
\left(\bm{N}_j \cdot \bm{p}\right) h_{j}
=
\bm{n}_{\rm R} \cdot \bm{p}
\end{equation}
for any interfacial probe vector $\bm{p}$.
Here $h_{j}$ is the step height of disconnection of the $j$th mode and $\bm{n}_{\rm R}$ is the reference plane normal. { Note that the step height $h_{j}$ of the $j$th mode disconnections is defined as positive if the height is increasing in the direction of the reciprocal vector $\bm{N}_j$, otherwise $h_{j}$ is negative; see Fig.~\ref{fig:geometry_demo}(b). For the special case of  only single mode disconnections being present, i.e., $N_j=1$, the interface morphology constraint in Eq.~\eqref{eqn:GFB2} leads to a  relationship between disconnection density and inclination angle
$\phi=\tan^{-1}(h/d_1)$, where $d_1$ is the inter-disconnection distance in the terrace plane, which is consistent with the nonlinear equation in \cite{Gordon-Sills2024} for the interface with single-mode disconnections.  }

The combination of Eqs.~\eqref{eqn:GFB1} and \eqref{eqn:GFB2} constitutes the \emph{generalized Frank-Bilby equation}~\cite{Tao2023} (GFBE), which describes the constraint on the disconnection network of a hetero-interface.
{
Equations~\eqref{eqn:GFB1} and \eqref{eqn:GFB2} are coupled: they must be solved simultaneously for the disconnection densities $\bm{N}_j$, $j = 1, \cdots, J$.
In general, the GFBE does not uniquely determine the disconnection network, interface inclination, or interfacial misorientation. Each of the $J$ disconnection modes contributes two degrees of freedom (DOFs), corresponding to the two components of the reciprocal vector $\bm{N}_j$. In addition, the interface inclination $\bm{\phi}$ and misorientation $\bm{\theta}$ may contribute two and three DOFs, respectively. The GFBE provides eight scalar constraints: two independent probe vectors $\bm{p}$ can be chosen, and for each probe vector Eqs.~\eqref{eqn:GFB1} and \eqref{eqn:GFB2} yield four scalar equations.

Two classes of problems can be distinguished.
First, when the interface inclination and misorientation $(\bm{\phi},\bm{\theta})$ are prescribed, the number of unknowns is $2J$. Since the GFBE provides eight scalar constraints, at least four disconnection modes are generally required to accommodate the interfacial misfit, except in special cases where some constraints are linearly dependent. Second, when $(\bm{\phi},\bm{\theta})$ are allowed to vary, the number of unknowns becomes $2J+5$. In this case the GFBE is typically underdetermined; for $J\ge 2$, the number of unknowns already exceeds the number of constraints, leading to infinitely many solutions.
For the first class of problems, a unique solution can in principle be obtained by including a sufficient number of candidate disconnection modes. For the second class, additional criteria are required to select a physically relevant solution from the admissible solution space.
Here, we treat the GFBE as a constraint and determine the equilibrium interface structure by minimizing the interfacial energy subject to this constraint. For fixed $(\bm{\phi},\bm{\theta})$, a sufficiently large set of candidate disconnection modes ($J\ge 4$) can be supplied, and the energy minimization automatically selects the optimal modes. When $(\bm{\phi},\bm{\theta})$ are allowed to vary, the minimization simultaneously searches for the energetically optimal interface inclination, misorientation, and disconnection network. This framework enables systematic solution of the GFBE without skillful, case-by-case analysis.
}


\section{Continuum model}\label{sec:model}

We present a continuum model to obtain the hetero-interface structure of the disconnection networks, interface inclination and the misorientation across the GB, with the provided possible disconnection modes with Burgers vectors $\bm{b}_j$ and step heights $h_j$, $(j=1, \cdots, J)$.
In our continuum model, we minimize the interfacial energy subject to the constraint defined by the GFBE~\cite{Tao2023}~{ to predict the unique, physically relevant hetero-interface structure}.
The constrained minimization problem is to minimize the interfacial energy:
\begin{equation}\label{eqn:gb_density}
\gamma
= \sum_{j=1}^J
\left[
\frac{\mu b_{j}^2}{4\pi(1-\nu)} \!\left(1 - \nu\frac{(\bm{N}_j \times \bm{n} \cdot \bm{b}_{j})^2}{b_{j}^2 N_j^2}\right)\!N_j \log\! \frac{1}{r_g\sqrt{N_j^2+\epsilon}}
 + \gamma_{\rm s} |h_j| N_j{+\gamma_0}
 \right],
\end{equation}
subject to the GFBE:
\begin{align}
&\bm{c}_1
= \sum_{j=1}^J \left(\bm{N}_j \cdot \bm{p}\right)\bm{b}_{j}
- \left(\mathbf{S}_{\alpha}^{-1} - \mathbf{S}_{\beta}^{-1}\right) \bm{p}
= \bm{0}, \label{eqn:frank0}
\\
&c_2
= \sum_{j=1}^J \left(\bm{N}_j \cdot \bm{p}\right)h_{j} - \bm{n}_{\rm R} \cdot \bm{p} = 0. \label{eqn:frank1}
\end{align}
{
The minimization is over all possible interface disconnection network
structure described by the reciprocal
vectors of disconnections arrays $\bm{N}_j$ ($j = 1, \cdots, J$), the interface inclination $\bm{\phi}$, and the misorientation $\bm{\theta}$.
Recall that here dislocations are considered as special disconnections with step height $h=0$.
Note that $N_j=0$ in a solution of this minimization problem means that the spacing of the disconnections of the $j$th mode goes to infinity and thus this mode is not present in the interface.
}

For the interfacial energy estimate (Eq.~\eqref{eqn:gb_density}), we assume that the material is elastically isotropic, homogeneous, and linear elastic; $\mu$ is the shear modulus, $\nu$ is the Poisson ratio, $\gamma$ is the interfacial energy, { $\gamma_{\rm s}$ is the energy density per unit step height length of each disconnection, $\gamma_0$ is the energy density of the terrace plane without disconnections and dislocations,} $b_{j}$ and $h_{j}$ are the magnitudes of the Burgers vector and step height of the $j$th disconnection mode, $r_g$ is a parameter associated with the dislocation core size, and $\epsilon$ is a small positive regularization parameter introduced to smooth the numerical singularity at $N_j=0$.
The simplifications of elastic isotropy and homogeneity may be relaxed, but unduly complicate the presentation.
The first term in the interfacial energy expression (Eq.~\eqref{eqn:gb_density}) is the local energy associated with the disconnection Burgers vector~\cite{Zhu2014175}, and the second term $\gamma_{\rm s} |h_j| N_j$ is the disconnection step energy (height $h_j$) density associated.
Both terms are written in terms of the disconnection densities~\cite{Zhu2014175}.
The constraints $\bm{c}_1=\bm{0}$ and $c_2= 0$ in Eqs.~\eqref{eqn:frank0} and \eqref{eqn:frank1} are given by the GFBE~\cite{Tao2023}.
{ The energy density of the terrace plane $\gamma_0$ is due to the chemical bonding, which is small and does not affect the disconnection structure on the interface.}

{
For the interfacial energy associated with dislocations, i.e., the first term in Eq.~\eqref{eqn:gb_density},
we employ the interfacial energy estimate within a dislocation framework that was derived in Ref.~\cite{Zhu2014175,Zhang2017}; the predictions of dislocation structure and energy of low angle grain boundaries based on this framework agree with  the atomistic simulations~\cite{Zhang2017,Wan2023} and the theoretical predictions of special grain boundaries \cite{Hirth2001,Sutton1995}.
This method has been applied to dislocation structures of low angle grain boundaries~\cite{Zhang2017,Qin2022} and hetero-interfaces without disconnections~\cite{Zhang2021110277}.
Note that since the long-range elastic energy of the interface vanishes at equilibrium (this is guaranteed by by the Frank-Bilby equation~\cite{Frank1950,Bilby1955,Hirth2001,Sutton1995}), the total interfacial energy at equilibrium is a local energy (as  for  grain boundaries in homogeneous materials~\cite{Hirth2001,Sutton1995}).}

Given the reference state of the interface, the transformation matrices, $\mathbf{S}_\alpha$ and $\mathbf{S}_\beta$, can be obtained.
For example, when we consider the $\alpha$/$\beta$ interface and choose $\beta$  as the reference lattice, the transformation matrices are $\mathbf{S}_{\alpha} = \mathbf{R} \mathbf{U}_{\alpha}$ and $\mathbf{S}_{\beta} = \mathbf{I}$, where $\mathbf{I}$ is the identity matrix, $\mathbf{R}$ is the rotation matrix, and $\mathbf{U}_{\alpha}$ is the stretch matrix.
The rotation matrix can be expressed as $\mathbf{R} = \mathbf{R}_{\theta_1} \mathbf{R}_{\theta_2} \mathbf{R}_{\theta_3}$, where $\mathbf{R}_{\theta_1}$, $\mathbf{R}_{\theta_2}$, and $\mathbf{R}_{\theta_3}$ are the rotation matrices around { the $X$, $Y$, and $Z$ axes,} respectively, with rotation angles $\theta_1$, $\theta_2$, and $\theta_3$, i.e.,
\begin{equation}
\mathbf R_{\theta_1} = \left(
\begin{array}{lll}
1 & 0 & 0\\
0 & \cos\theta_1 & -\sin\theta_1 \\
0 & \sin\theta_1 & \cos\theta_1
\end{array}
\right), \;
\mathbf R_{\theta_2} = \left(
\begin{array}{lll}
\cos\theta_2 & 0 & -\sin\theta_2\\
0 & 1 & 0 \\
\sin\theta_2 & 0 & \cos\theta_2
\end{array}
\right), \;
\mathbf R_{\theta_3} = \left(
\begin{array}{lll}
\cos\theta_3 & -\sin\theta_3 & 0\\
\sin\theta_3 & \cos\theta_3 & 0\\
0 & 0 & 1
\end{array}
\right).
\end{equation}
The stretch matrix $\mathbf{U}_{\alpha}$ describes the misfit between the two lattices meeting at the interface.
If the misfit can be eliminated by extension or compression along the $X$, $Y$, and $Z$ axes alone, $\mathbf U_{\alpha}$ is diagonal:
\begin{flalign}
\mathbf U_{\alpha} = \left(
\begin{array}{lll}
\Omega_1 & 0 & 0\\
0 & \Omega_2 & 0 \\
0 & 0 & \Omega_3
\end{array}
\right),
\end{flalign}
where $\Omega_1$, $\Omega_2$, and $\Omega_3$ are the lattice correspondences { in the $X$, $Y$, and $Z$ directions.}

The constrained minimization problem in Eqs.~\eqref{eqn:gb_density}-\eqref{eqn:frank1} can be solved numerically with the penalty method~\cite{Chong2013}, in which it is approximated by the following unconstrained minimization problem:
\begin{equation}\label{penalty}
\min_{\{\bm{N}_j\}, \bm{\theta}, (\bm{\phi})}
\left[
\gamma(\{\bm{N}_j\}, \bm{\phi})
+
\frac{\alpha_p}{2} \left\Vert \bm{c}(\{\bm{N}_j\}, \bm{\phi}, \bm{\theta}) \right\Vert^2
\right],
\end{equation}
where { the vector $\bm{c} \equiv (\bm{c}_1, c_2)$} and $\alpha_p >0$ is the penalty parameter.
{The optimization can be performed over $\{\bm{N}_j\}$, $\bm{\phi}$, and $\bm{\theta}$ simultaneously, or over $\{\bm{N}_j\}$ and $\bm{\theta}$ with $\bm{\phi}$ held fixed.}
Note that as the penalty parameter $\alpha_p\rightarrow+\infty$, the solution of this unconstrained minimization problem converges to the solution of the constrained minimization problem~\cite{Chong2013}.
(Other methods, such as the augmented Lagrangian method, may also be used  to solve this constrained minimization problem~\cite{Chong2013,Qin2022}.)


{
Recall that the minimization is over all possible interface disconnection network
structure described by the reciprocal
vectors of disconnections/dislocation arrays $\{\bm{N}_j\}$ ($j = 1, \cdots, J$), the interface inclination $\bm{\phi}$, and the misorientation $\{\theta\}$.
The interface normal direction is related to the inclination as
$\bm{n} = ( -\cos\phi_1\sin\phi_2, -\sin\phi_1, \cos\phi_1\cos\phi_2 )$.
When the interface inclination $\bm{\phi}$ and accordingly the interface normal $\bm{n}$ are known, there are only two DOFs for each reciprocal vector $\bm{N}_j$, which lies in the interface.
Following the definition of the inclination angles $\phi_1$ and $\phi_2$ by rotation of the $XY$ plane to the interface plane given at the beginning of the previous section, the $X$ axis is rotated to the vector $\hat{\bm{x}}^p=(\cos\phi_2,0,\sin\phi_2)$ in the interface, and the $Y$ axis to the vector $\hat{\bm{y}}^p=(-\sin\phi_1\sin\phi_2,\cos\phi_1,\sin\phi_1\cos\phi_2)$ in the interface.
Since $\bm{N}_j$ must lie in the interface plane (i.e., the $\hat{\bm{x}}^p$-$\hat{\bm{y}}^p$ plane), it can be written as
\begin{equation}
\bm{N}_j
=
N_{j1}^p\hat{\bm{x}}_p+N_{j2}^p\hat{\bm{y}}_p
=( N_{j1}^p\cos\phi_2-N_{j2}^p\sin\phi_1\sin\phi_2, N_{j2}^p\cos\phi_1, N_{j1}^p\sin\phi_2 + N_{j2}^p\sin\phi_1\cos\phi_2 ), \label{eq:N_global}
\end{equation}
where $\bm{N}_j^p = (N_{j1}^p, N_{j2}^p)$ is defined as the representation of $\bm{N}_j$ under the interfacial frame; hence, $N_{j1}^p$ and $N_{j2}^p$ represent two independent variables of the reciprocal vector $\bm{N}_j$.
}

The unconstrained problem \eqref{penalty} can be solved by gradient minimization
with respect to variables $\{\bm{N}_j^p\}$,  $\bm{\phi}$, and $\bm{\theta}$; this leads to the following evolution equations with an artificial time $t$:
\begin{align}
&\frac{\partial \bm{N}_j^p}{\partial t}
=
-\left( \frac{\partial \gamma}{\partial \bm{N}_j^p} + \alpha_p\frac{\partial g}{\partial \bm{N}_j^p}\right),
\quad (j = 1, \cdots, J)
\label{eqn:evolution01}\\
&\frac{\partial \bm{\phi}}{\partial t}
=
-\left( \frac{\partial \gamma}{\partial \bm{\phi}} + \alpha_p\frac{\partial g}{\partial \bm{\phi}}\right),
\label{eqn:evolution03}\\
&\frac{\partial \bm{\theta}}{\partial t}
=
-\alpha_p\frac{\partial g}{\partial \bm{\theta}},
\label{eqn:evolution02}
\end{align}
where $g \equiv \Vert \bm{c} \Vert^2/2$.
The continuum model in Eqs.~\eqref{eqn:evolution01}-\eqref{eqn:evolution03} includes the evolution equations for three types of variables: disconnection densities, interface inclination, and misorientation.

In the numerical implementation, we find that when the numerical simulation is applied to all  three variable types in Eqs.~\eqref{eqn:evolution01}-\eqref{eqn:evolution03} at the same time in one evolution loop, it does not converge to the global minimum.
Hence, we designed the numerical scheme shown in Algorithm~\ref{alg:cap}.
The input parameters include all possible disconnection modes $(\bm{b}_j, h_j)$ and the lattice correspondence matrix $\mathbf{U}_{\alpha}$.
Finally, the simulation determines the interface inclination $\bm{\phi}$, disconnection network $\{\bm{N}_{j}^p\}$ ($j = 1, \cdots, J$), and misorientation $\bm{\theta}$ at equilibrium.
There are two evolution loops in our numerical scheme: the inner evolution loop evolves  the disconnection densities $\{\bm{N}_{j}^p\}$ and the misorientation $\bm{\theta}$, by using Eqs.~\eqref{eqn:evolution01} and \eqref{eqn:evolution02} to obtain the interfacial structure for a particular inclination $\bm{\phi}$;
the outer  loop evolves the inclination $\bm{\phi}$ through the interfacial energy gradient with respect to $\bm{\phi}$.

\begin{algorithm}
\caption{Numerical scheme}\label{alg:cap}
\begin{algorithmic}
\Require disconnection modes $\{\bm{b}_j, h_j\}$ ($j = 1, \cdots, J$), and lattice correspondence matrix $\mathbf{U}_{\alpha}$
\Ensure disconnection network $\{\bm{N}_j^p\}$ ($j = 1, \cdots, J$), inclination $\bm{\phi}$, and misorientation $\bm{\theta}$\\
\State \textbf{Initialization:}
$\bm{\phi} = \bm{\phi}_0$
\\
\State
\textbf{Evolve inclination:}
\\
\hspace{4em}
$\bm{\phi}_{t+1} = \bm{\phi}_{t} - \lambda_1 \left(\dfrac{\partial \gamma'}{\partial \bm{\phi}} + \alpha_p \dfrac{\partial g'}{\partial \bm{\phi}}\right)_t$,
\\
where
\\
\hspace{4em}
$\gamma'(\bm{\phi}) \equiv \min\limits_{\{\bm{N}_j^p\}} \gamma(\{\bm{N}_j^p\}, \bm{\phi})$ and
$g'(\bm{\phi}) \equiv \min\limits_{\{\bm{N}_j^p\},\bm{\theta}} g(\{\bm{N}_j^p\}, \bm{\phi}, \bm{\theta})$,
\\
\State
\textbf{through evolving disconnection network and misorientation until convergence:}
\\
\hspace{4em}
$(\bm{N}_{j}^p)_{t+1} = (\bm{N}_{j}^p)_{t} - \lambda_2 \left( \dfrac{\partial \gamma}{\partial \bm{N}_{j}^p} + \alpha_p \dfrac{\partial g}{\partial \bm{N}_{j}^p}\right)_t$;
\\
\hspace{4em}
$\bm{\theta}_{t+1} = \bm{\theta}_{t} - \lambda_2 \alpha_p \left(\dfrac{\partial g}{\partial \bm{\theta}}\right)_t$,
\\
where $\lambda_1$ and $\lambda_2$ are two different step lengths.
\end{algorithmic}
\end{algorithm}

\section{Example applications}
\label{sec:simulation}

We apply the continuum model described above to determine the interface structure (disconnection network, interface inclination, and misorientation) of two different example interfaces:
(i) an $\alpha$/$\beta$ (orthorhombic/monoclinic) interface in zirconia ZrO$_2$;
and (ii) a Au-Pd (001) twist-misfit interface. { Multiple modes of disconnections are present on these interfaces and the existing methods for single-mode disconnections do not apply~\cite{Hirth2010-1,Gordon-Sills2024}.}
{An additional example on the $\alpha$/$\beta$ (HCP/BCC) interfaces in titanium alloys with the Burgers orientation relationship (OR) is given in Appendix.}

To apply the continuum model described above, we first employ the numerical scheme in Algorithm~\ref{alg:cap}, allowing all the interface structure variables, including the disconnection densities, interface inclination, and misorientation, to evolve to achieve the equilibrium interface configuration corresponding to the global minimum of the interface energy.
{This situation may arise when one phase forms a small inclusion within the matrix of the other phase, such that both the inclusion orientation and the inclusion/matrix interface morphology can adjust to reduce the interfacial energy~\cite{Ackerman2020013602,peng2015two}.}
The variables are initialized as $\bm{N}_{j}^p = \bm{0}$ ($j = 1, \cdots, J$), $\bm{\phi} = \bm{0}$, and $\bm{\theta} = \bm{0}$.
The penalty parameter $\alpha_p$ in Eq.~\eqref{penalty} is gradually increased until the disconnection network no longer changes.

We then systematically investigate the dependence of the disconnection structure and interfacial energy on misorientation for a prescribed interface inclination using the energy-gradient flows in Eqs.~\eqref{eqn:evolution01} and \eqref{eqn:evolution02}.
{This scenario may correspond to a crystallite of one phase supported on the surface of the other phase, where the crystallite is free to rotate relative to the substrate while maintaining a fixed substrate surface in order to reduce the interfacial energy~\cite{chan1985study,kuang2022orientation}.}
For each prescribed inclination $\bm{\phi}$, the variables are initialized as $\bm{N}_{j}^p = \bm{0}$ ($j = 1, \cdots, J$) and $\bm{\theta} = \bm{0}$.
The evolution equations are then solved to obtain the equilibrium disconnection densities and misorientation, from which the interfacial energy is evaluated as a function of misorientation and interface inclination.

\subsection{Interface associated with the martensite transformation of zirconia}

We apply our continuum model to predict the structure of $\alpha$/$\beta$ orthorhombic/monoclinic $(100)_{\rm o/m}$ interfaces in zirconia ZrO$_2$, which are formed during the orthorhombic-to-monoclinic martensite transformation.
We set $[010]_{\rm o/m}$ as the $X$-axis with unit vector $\bm{e}_1$ and $[001]_{\rm o/m}$ as the $Y$-axis with unit vector $\bm{e}_2$.
Consider three disconnection modes characterized by the Burgers vectors and step heights:
\begin{align}
&\bm{b}_1 = -\frac{c^{\rm o}}{c^{\rm m}}2a^{\rm m}\cos{\beta^{\rm m}}\bm{e}_2
\quad\text{and}\quad
h_1 = a^{\rm o};\label{eqn:BV_step4}
\\
&\bm{b}_2 = -\frac{c^{\rm o}}{c^{\rm m}}2a^{\rm m}\cos{\beta^{\rm m}}\bm{e}_2
\quad\text{and}\quad
h_2 = -3a^{\rm o};
\\
&\bm{b}_3 = -\frac{c^{\rm o}}{c^{\rm m}}\sqrt{2}a^{\rm m}\cos{\beta^{\rm m}} \bm{e}_1
\quad\text{and}\quad
h_3 = 0.\label{eqn:BV_step6}
\end{align}
The lattice constants of orthorhombic ZrO$_2$ are $a^{\rm o}=1.0140$~nm, $b^{\rm o} = 0.5210$~nm, and $c^{\rm o}=0.5110$~nm; the lattice constants of monoclinic ZrO$_2$ are $a^{\rm m}=0.5145$~nm, $b^{\rm m} = 0.5208$~nm, $c^{\rm m}=0.5311$~nm, and  $\beta^{\rm m}=99.25^{\circ}$~\cite{Chiao19901163}.
{Since the martensitic transformation involves the nucleation of the monoclinic phase from the orthorhombic matrix, it is natural to choose the orthorhombic phase as the reference lattice.}
The lattice correspondence parameters are $\Omega_1 = 1,$ $\Omega_2 = c^{\rm m}/c^{\rm o}$, and $\Omega_3 = 2(a^{\rm m}/a^{\rm o}) \sin{\beta}^{\rm m}$.

\subsubsection{Equilibrium interface structure}
\label{sec:equ1}

\begin{figure}[htbp]
\centering
\subfigure[]{ \includegraphics[width=0.4\linewidth]{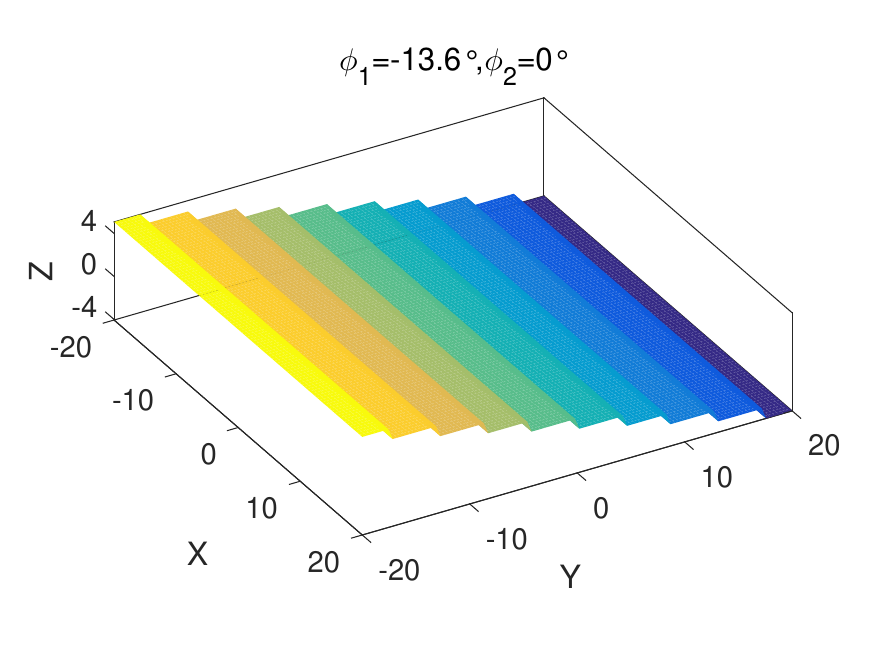}}
\subfigure[]{ \includegraphics[width=0.4\linewidth]{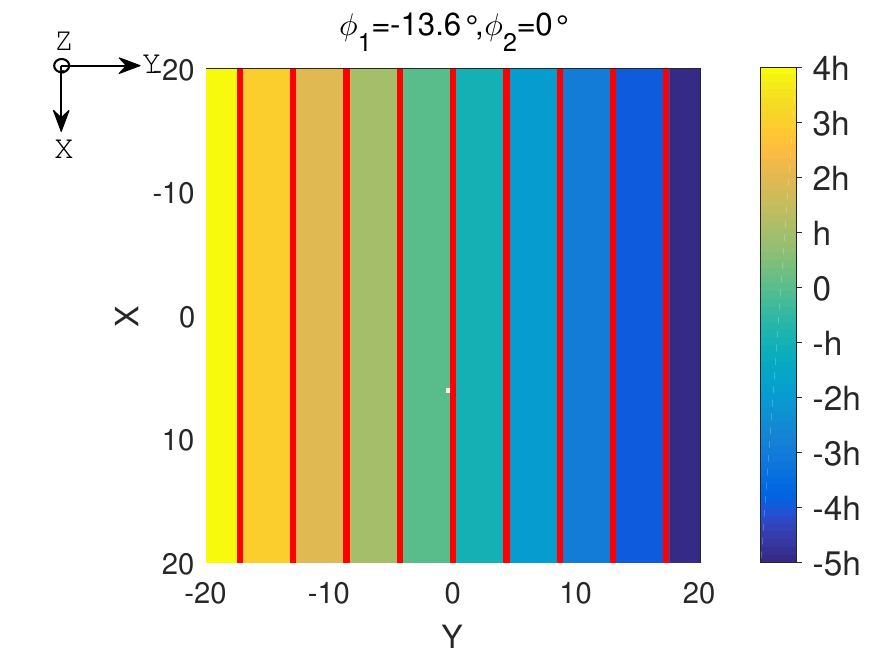}}
\caption{The equilibrium structure of the $(100)$ orthorhomic/monoclinic interface in zirconia ZrO$_2$ from (a) oblique view and
(b) plan view along the $Z$-direction (normal to the reference interface, where the red lines denote disconnection type $(\bm{b}_1,h_1)$ and the shading between the lines indicates the height profile.}
\label{fig:simu10}
\end{figure}

At equilibrium, the numerical simulation converges to an interface structure characterized by inclination angles $\phi_1=-13.5884^{\circ}$ and $\phi_2=0$ and  misorientation angles $\theta_1=-0.0221^{\circ}$, $\theta_2=0$, and $\theta_3=0$, with a disconnection structure  shown in Fig.~\ref{fig:simu10}.
The disconnection network consists only one array of disconnections $(\bm{b}_1,h_1)$, represented by  reciprocal vector $\bm{N}_1=(0, -0.2317)$~nm$^{-1}$.
The dislocation characters of this disconnection array are edge dislocations with line directions parallel to the  $X$ direction with a spacing of  $D_1=1/{N_1} = 4.3159$~nm.
The disconnections with $(\bm{b}_2,h_2)$ and  $(\bm{b}_3,h_3)$ (represented by reciprocal vectors $\bm{N}_2= \bm{0}$ and $\bm{N}_3=\bm{0}$) do not appear.

The predicted equilibrium interface structure (disconnection network, inclination, and misorientation) are in excellent agreement with  experimental observations~\cite{Chiao19901163} and consistent with the phenomenological theory of martensite crystallography (PTMC) model~\cite{Hirth20035385} and the topological model (TM)~\cite{Pond20075289}.
There are no disconnections with  Burgers vectors other than $\bm{b}_1$ or step height other than $h_1$, i.e., there is only one disconnection mode  $(\bm{b}_1,h_1)$ in Eq.~\eqref{eqn:BV_step4} on this interface.
The misorientation angle $\theta_1$ agrees with the PTMC prediction $-0.0221^{\circ}$~\cite{Hirth20035385}.
The predicted disconnection spacing $4.3159$~nm agrees with the experimental measurement $4.2$~nm~\cite{Chiao19901163} and the TM prediction $4.3$~nm~\cite{Pond20075289}.
The inclination angle $\phi_1$ is very close to the experimental observation $13.5^{\circ}$~\cite{Chiao19901163}, the PTMC prediction $13.5884^{\circ}$~\cite{Hirth20035385}, and the TM prediction $13.28^{\circ}$~\cite{Pond20075289}.

\subsubsection{Interfaces with fixed inclinations}

\begin{figure}[htbp]
\centering
\subfigure[]{  \includegraphics[width=0.33\linewidth, trim=3cm 9.5cm 3.5cm 8cm, clip]{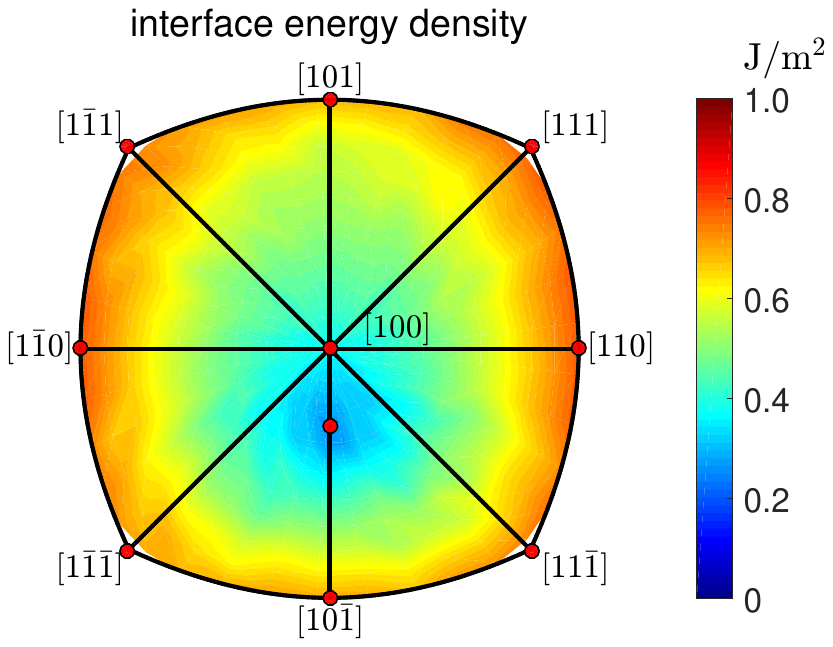}}
\subfigure[]{  \includegraphics[width=0.33\linewidth, trim=3cm 9.5cm 3.5cm 8cm, clip]{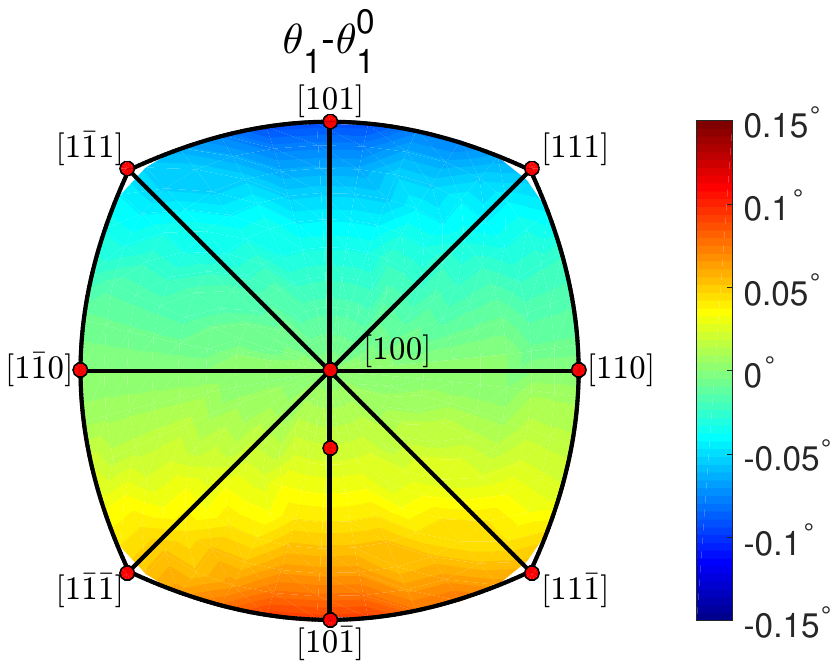}}
\subfigure[]{  \includegraphics[width=0.33\linewidth, trim=3cm 9.5cm 3.5cm 8cm, clip]{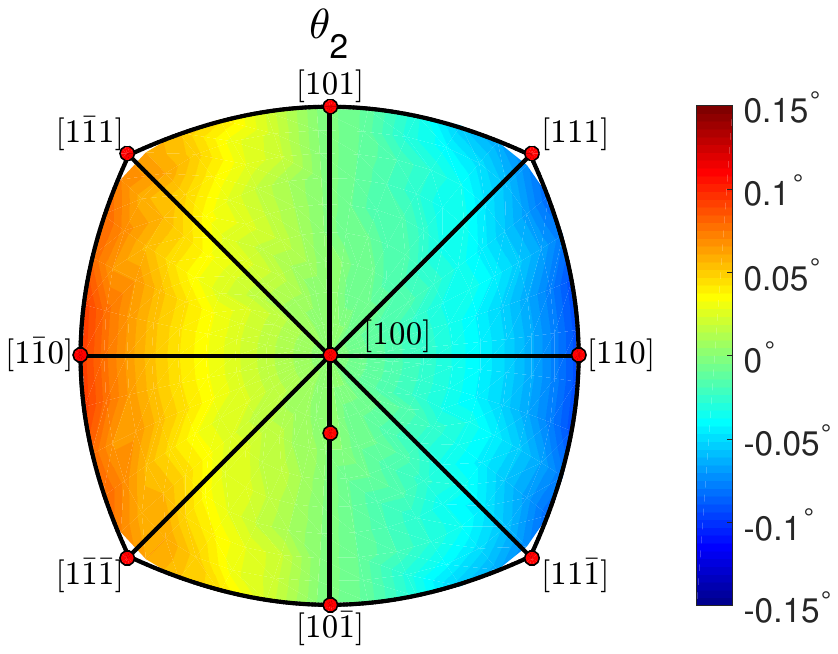}}
\subfigure[]{  \includegraphics[width=0.33\linewidth, trim=3cm 9.5cm 3.5cm 8cm, clip]{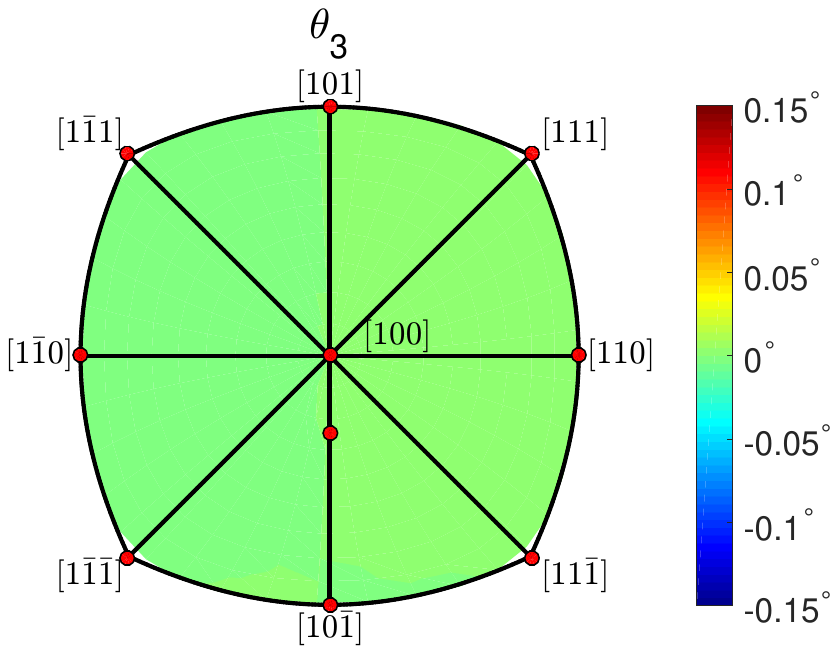}}
\caption{(a) Interfacial energy and (b-d) misorientation angles of the $(100)$ orthorhombic/monoclinic interface of zirconia as functions of interface inclination angles $(\phi_1,\phi_2)$ in stereographic projections.
The inclination angles, $\phi_1$ and $\phi_2$, are the angles relative to the $X$- and $Y$-axes of the tangent plane of the stereographic projection sphere. The value of $\phi_1$ increases from bottom to top, while the value of $\phi_2$ increases from left to right.
Here, $\theta_1=0.0221^{\circ}$ is the equilibrium misorientation (see Sec.~\ref{sec:equ1}).
}
\label{fig:simu11}
\end{figure}

We next examine how  the interface disconnection network and the misorientations across the interface vary at fixed interface inclination $\bm{\phi}$ by minimizing the energy via  energy gradient flows, Eqs.~\eqref{eqn:evolution01} and \eqref{eqn:evolution02}.
The interfacial energy and  misorientation  are functions of the inclination angles ($\phi_1,\phi_2$), as shown in Fig.~\ref{fig:simu11} in a stereographic projection.
Figure~\ref{fig:simu11}(a) shows that the  interfacial energy is, indeed, minimum at  inclination angles  $\phi_1=-13.6^{\circ}$ and $\phi_2=0$.
Figures~\ref{fig:simu11}(b-d) demonstrate how the misorientation angles $\theta_1$, $\theta_2$, and $\theta_3$ vary with interface inclination.
We observe that the tilt angle $\theta_1$ increases linearly and $\theta_2$ changes little with increasing inclination angle $\phi_1$, while with increasing inclination angle $\phi_2$, the  tilt angle $\theta_1$ changes little and $\theta_2$ increases linearly.
The twist angle $\theta_3$ is insensitive to change in the inclination.
{We may connect this analysis to the case of a relatively unconstrained monoclinic crystallite growing on the surface of a rigid orthorhombic substrate, as in the nucleation of a monoclinic phase on the surface of an orthorhombic bulk. In this scenario, the monoclinic crystallite can select its crystallographic orientation to minimize the interfacial energy. According to the results presented above, the energetically preferred interface plane deviates from the $(100)$ plane. When the exposed orthorhombic surface is tilted away from $(100)$ about the $X$-axis, the preferred orientation of the monoclinic crystallite correspondingly rotates about the same $X$-axis. Likewise, tilting the interface about the $Y$-axis induces a rotation of the crystallite about the same $Y$-axis. In contrast, changes in the interface inclination do not generate any twist of the crystallite about the $Z$-axis.}

\begin{figure}[htbp]
\centering
\subfigure[]{ \includegraphics[width=0.32\linewidth]{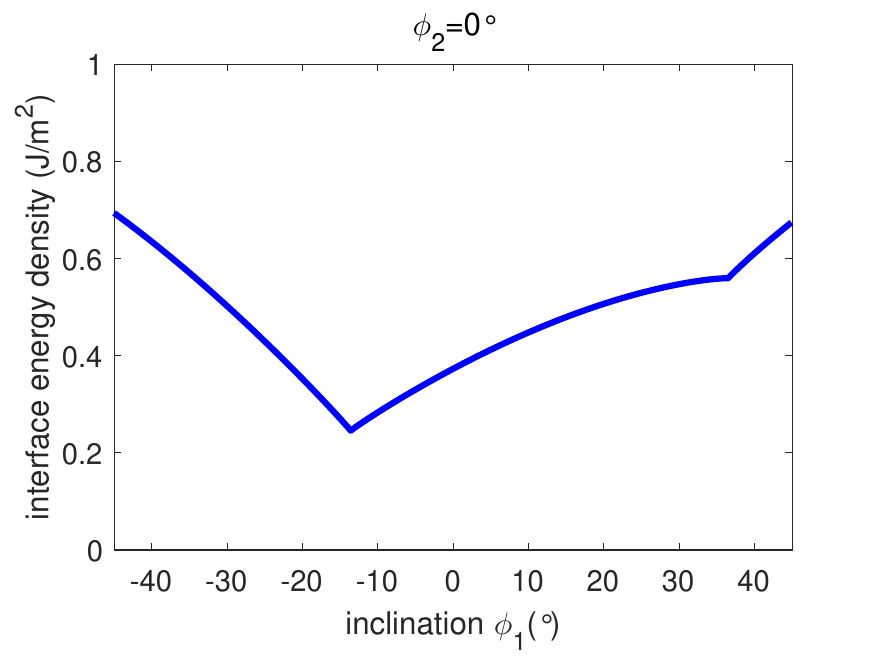}}
\subfigure[]{ \includegraphics[width=0.32\linewidth]{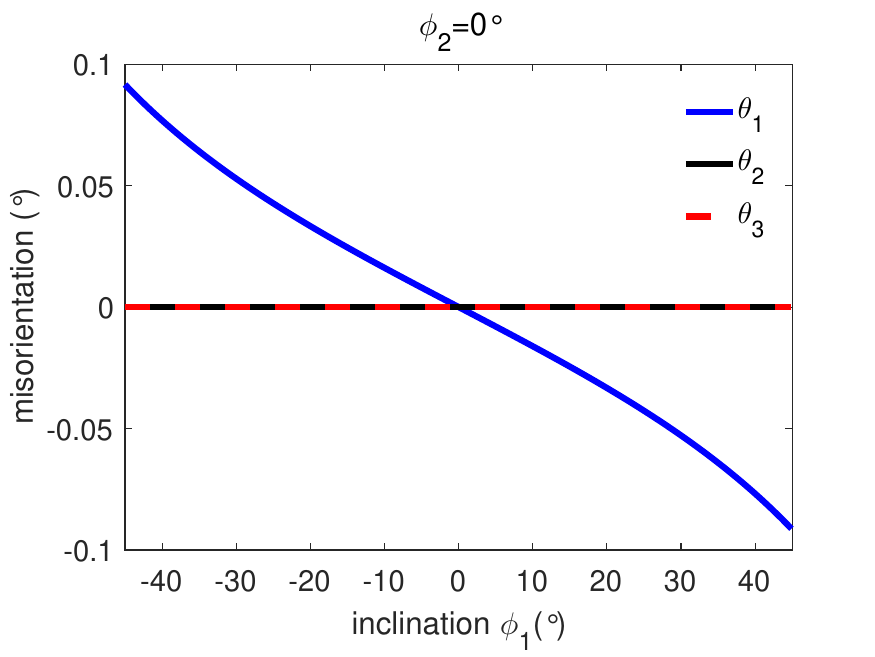}}
\subfigure[]{ \includegraphics[width=0.32\linewidth]{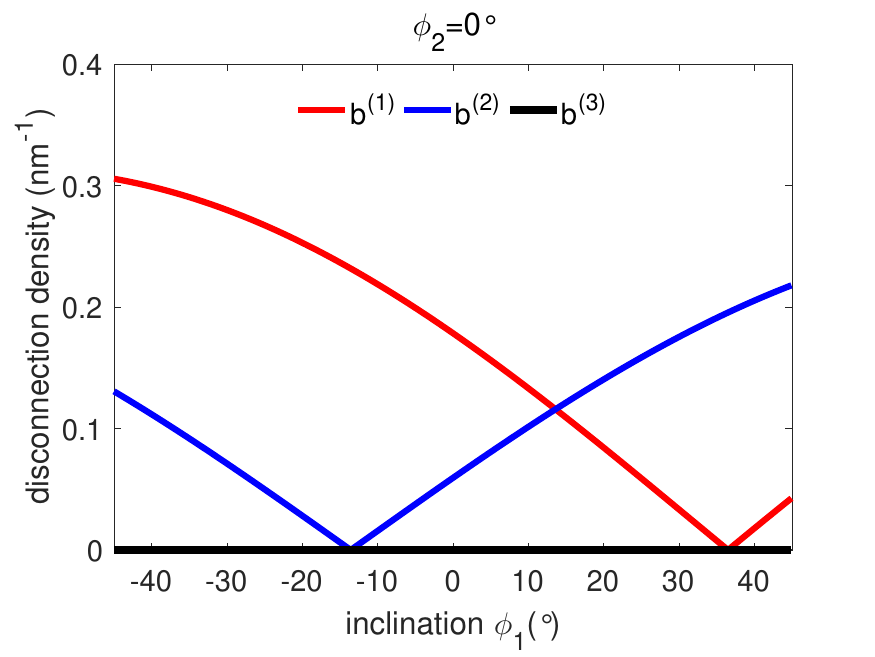}}
\subfigure[]{ \includegraphics[width=0.32\linewidth]{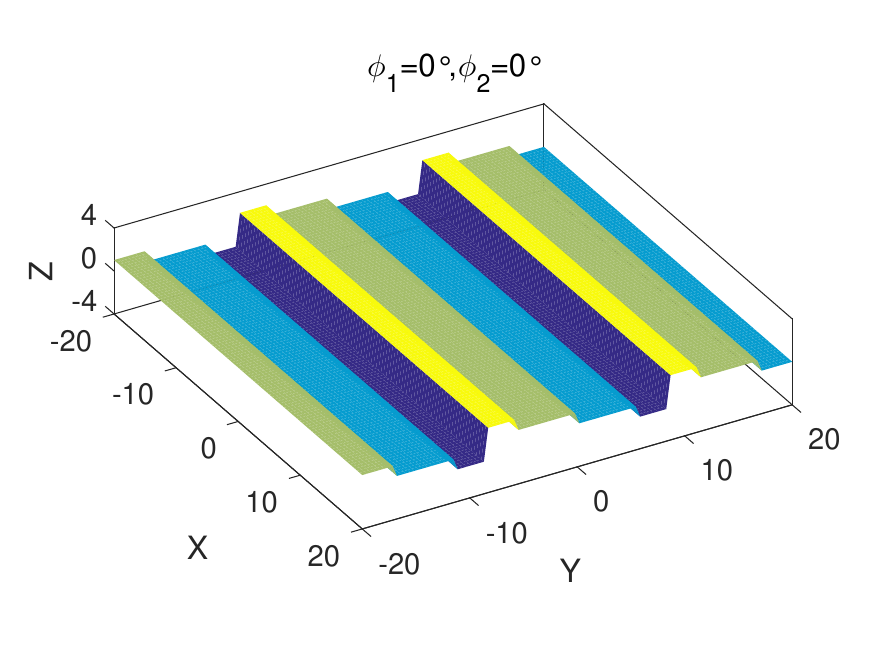}}
\subfigure[]{ \includegraphics[width=0.32\linewidth]{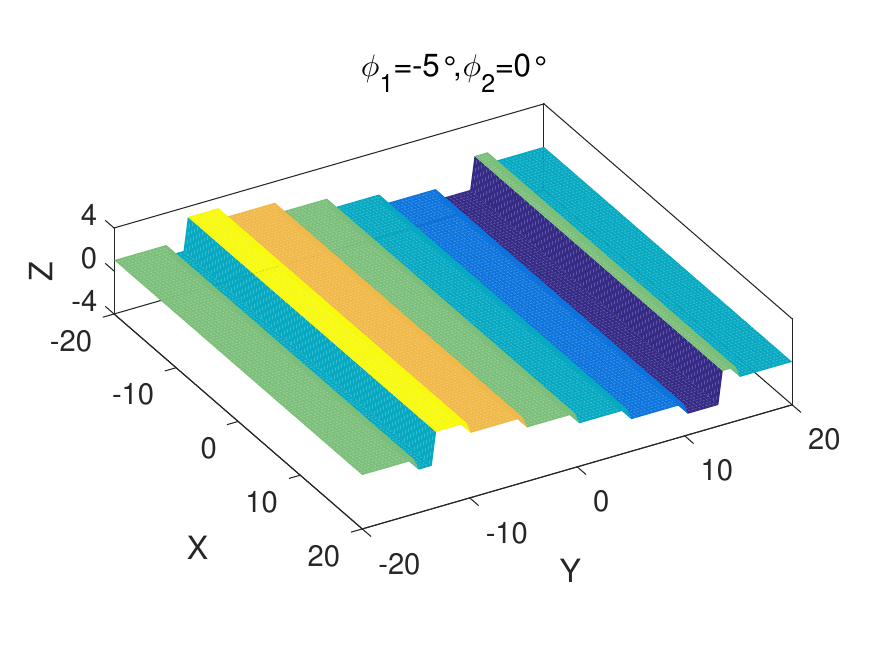}}
\subfigure[]{ \includegraphics[width=0.32\linewidth]{structure_inclination_phi00_ZrO2_3d.pdf}}
\subfigure[]{ \includegraphics[width=0.32\linewidth]{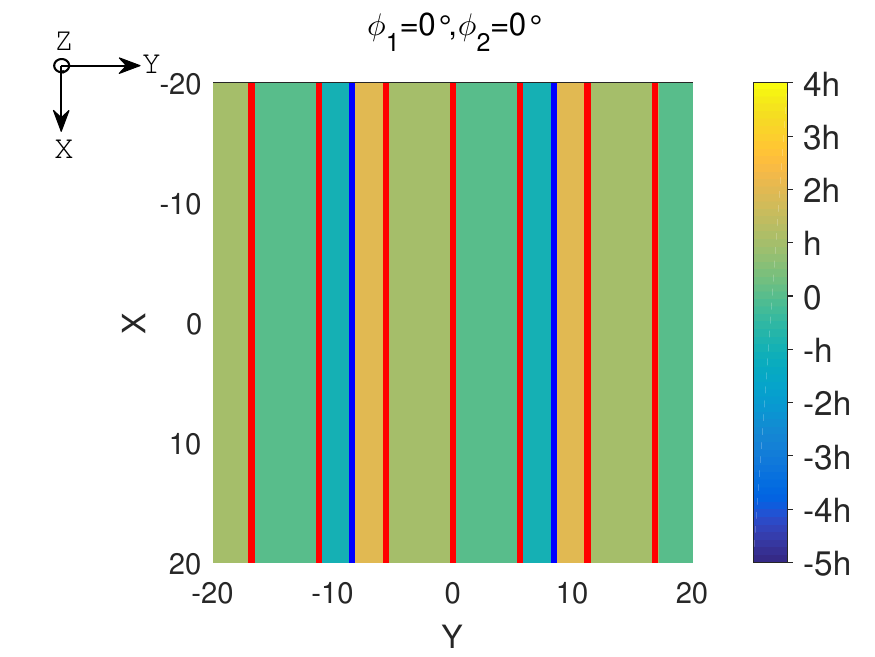}}
\subfigure[]{ \includegraphics[width=0.32\linewidth]{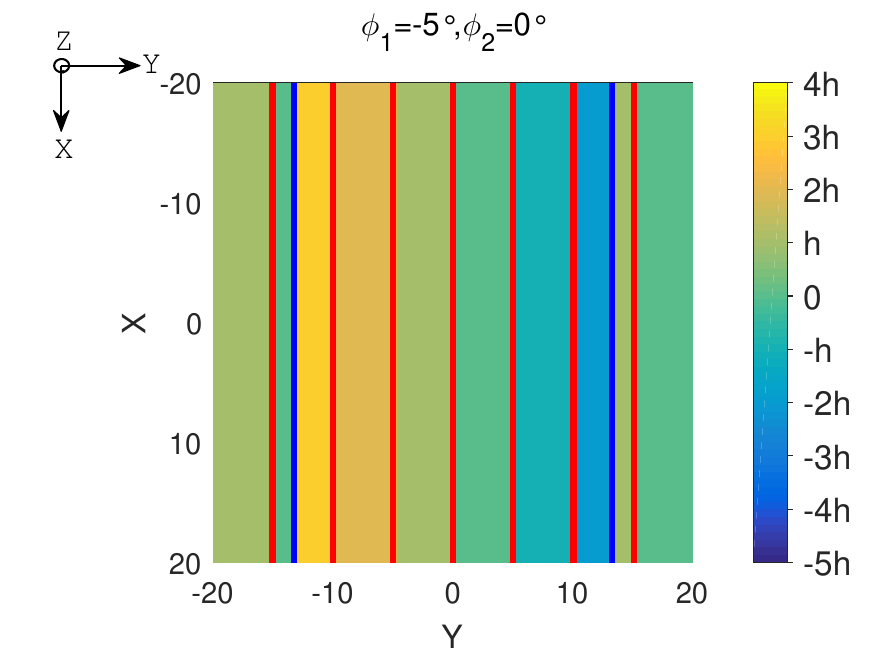}}
\subfigure[]{ \includegraphics[width=0.32\linewidth]{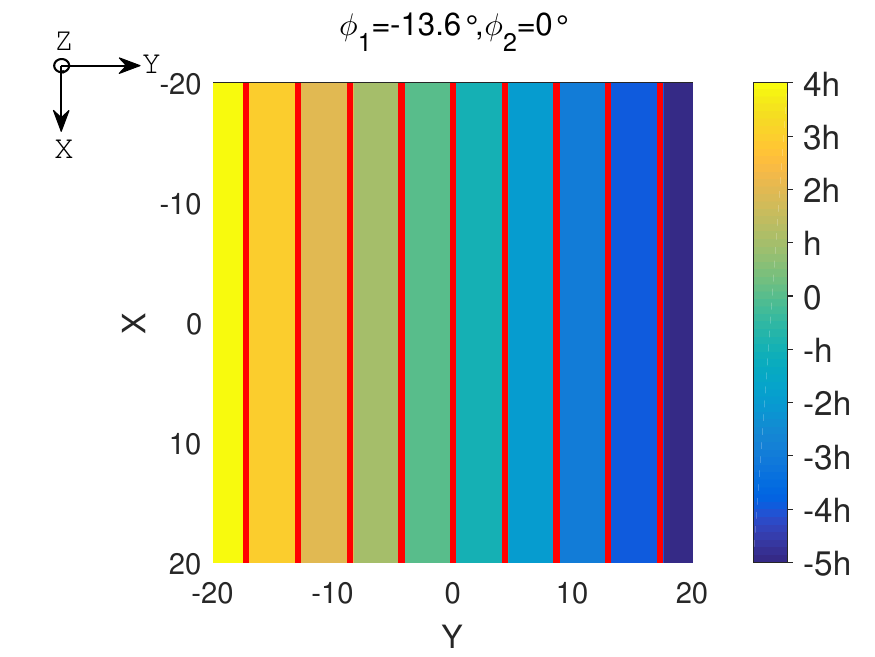}}
\caption{The interface structure of $(100)$ orthorhombic/monoclinic interface in zirconia for $\phi_2=0$, as a function of inclination $\phi_1$:  (a) interfacial energy, (b) misorientations across the interface,  and (c) disconnection densities;
(d-f) oblique view of the interface disconnection structure and height/step profile;
 (g-i) plan view of the interfaces along the $Z$-direction (normal to the reference interface), where the red and blue lines respectively denote  disconnection types $(\bm{b}_1,h_1)$ and $(\bm{b}_2,h_2)$, and the shading between the lines indicates the height profile.
}
    \label{fig:simu12}
\end{figure}

We now examine the interface structure for several particular inclinations.
Figure~\ref{fig:simu12} shows the interface structures as functions of the inclination angle $\phi_1$ for fixed inclination angle $\phi_2=0$.
As $\phi_1$ increases, the interfacial energy  decreases to the global minimum at $\phi_1=-13.6^{\circ}$, and then increases to another cusp at $\phi_1=36.5^{\circ}$; see Fig.~\ref{fig:simu12}(a).
Changing the inclination angle $\phi_1$ leads to a nearly linear change in the misorientation angle  $\theta_1$, while the other misorientation angles are unaffected (remain  zero); see Fig.~\ref{fig:simu12}(b).
Figure~\ref{fig:simu12}(c) shows the disconnection networks  for several inclinations, $\phi_1$.
When the interfacial energy reaches the minimum, $\phi_1=-13.6^{\circ}$ and the densities of the disconnections $(\bm{b}_1,h_1)$ and $(\bm{b}_2,h_2)$ are $0.2317$~nm$^{-1}$ and $0$, respectively.
On the other hand, at the subsidiary interfacial energy cusp ($\phi_1=36.5^{\circ}$, see Fig.~\ref{fig:simu12}(a)) the density of disconnection $(\bm{b}_1,h_1)$ is zero, that of disconnection $(\bm{b}_3,h_3)$ is always zero and only the density of disconnection $(\bm{b}_2,h_2)$ is non-zero.
Figure~\ref{fig:simu12}(d)-(i) and (g)-(i) show oblique and plan views  of the disconnection networks for inclination angles $\phi_1=0$, $-5^{\circ}$, and $-13.6^{\circ}$.
As noted above, there are only two disconnection modes, $(\bm{b}_1,h_1)$ and $(\bm{b}_2,h_2)$, and these  disconnection lines are all parallel to the $X$-axis.
As the inclination angle $\phi_1$ decreases from $0^{\circ}$ to $-13.6^{\circ}$, the spacing between disconnections $(\bm{b}_1,h_1)$ decreases and between disconnections $(\bm{b}_2,h_2)$ increases.
At $\phi_1=-13.6^{\circ}$,  disconnection $(\bm{b}_2,h_2)$ vanishes, and the disconnection network has only disconnection $(\bm{b}_1,h_1)$.
The contribution of disconnections $(\bm{b}_1,h_1)$ and $(\bm{b}_2,h_2)$ to the interface inclination are opposite to each other, because the disconnections $(\bm{b}_1,h_1)$ with a positive step height and the reciprocal vector in the $-Y$-direction decreases the inclination, while the disconnections $(\bm{b}_2,h_2)$ with a negative step height and the reciprocal vector in $-Y$-direction increases the inclination.

\begin{figure}[htbp]
\centering
\subfigure[]{ \includegraphics[width=0.32\linewidth]{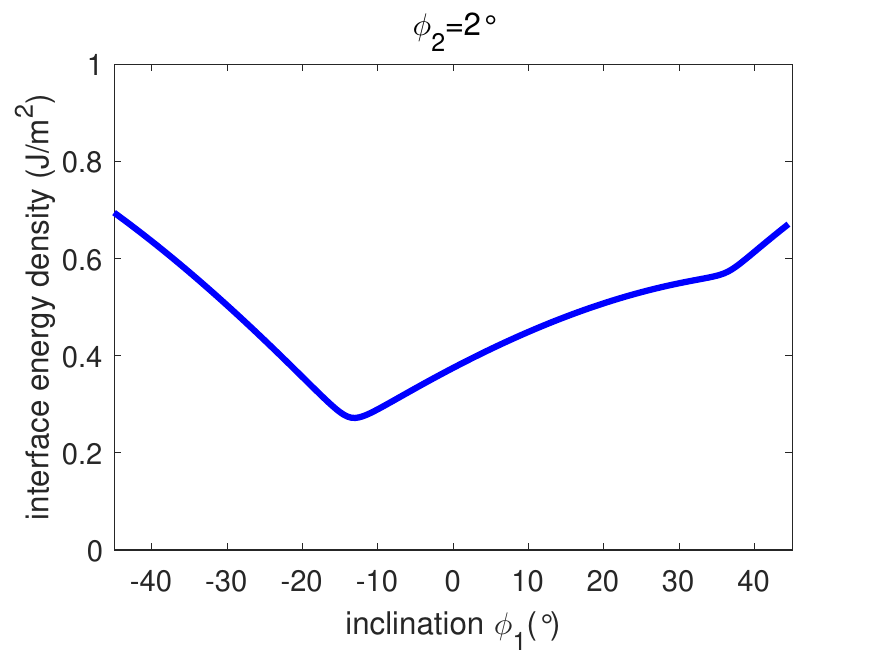}}
\subfigure[]{ \includegraphics[width=0.32\linewidth]{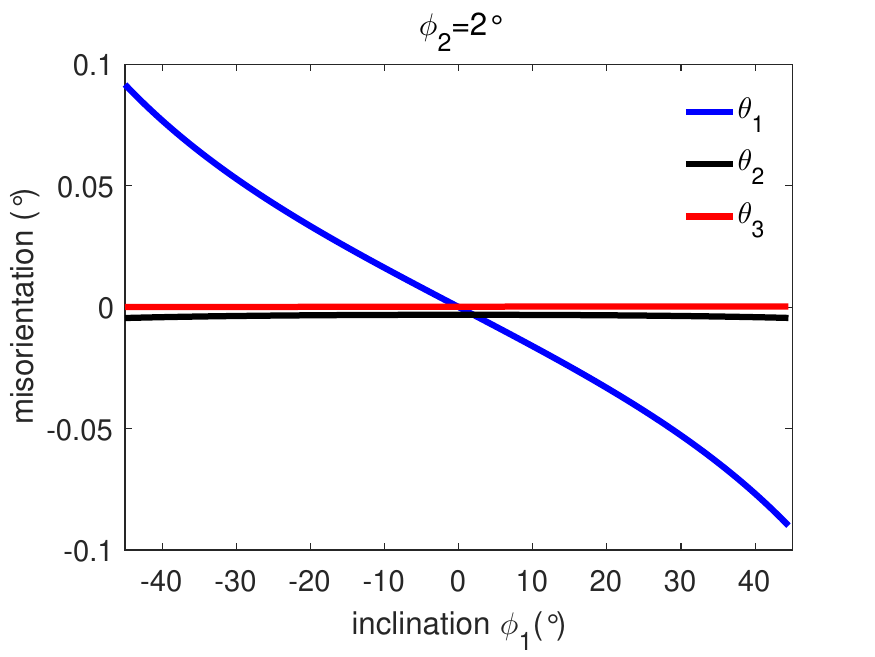}}
\subfigure[]{ \includegraphics[width=0.32\linewidth]{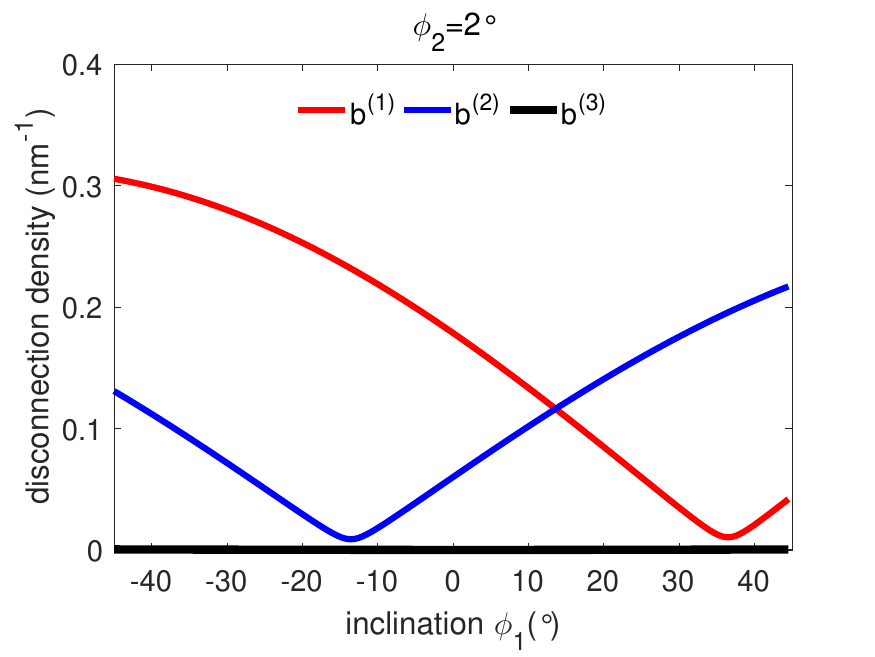}}
\subfigure[]{ \includegraphics[width=0.32\linewidth]{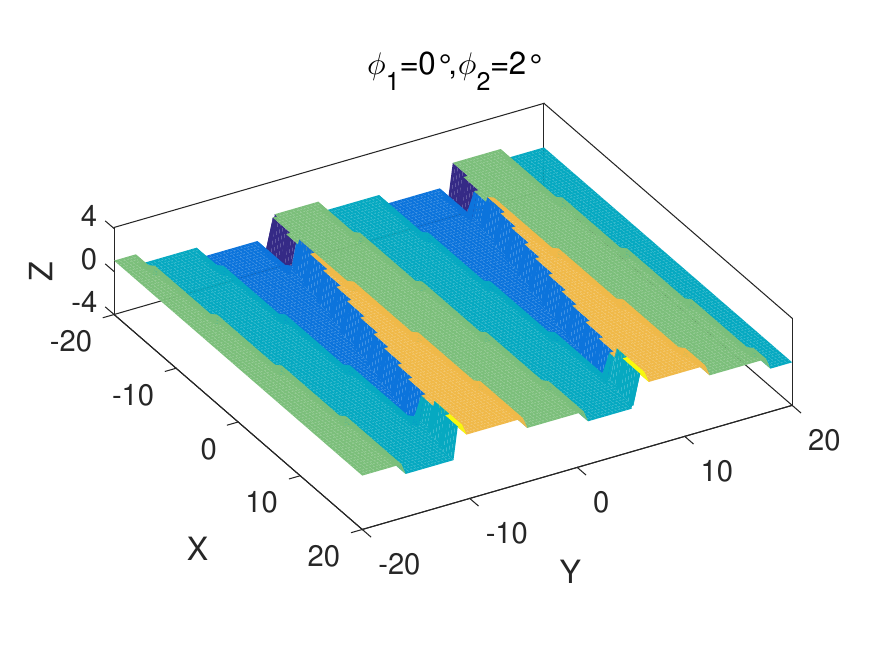}}
\subfigure[]{ \includegraphics[width=0.32\linewidth]{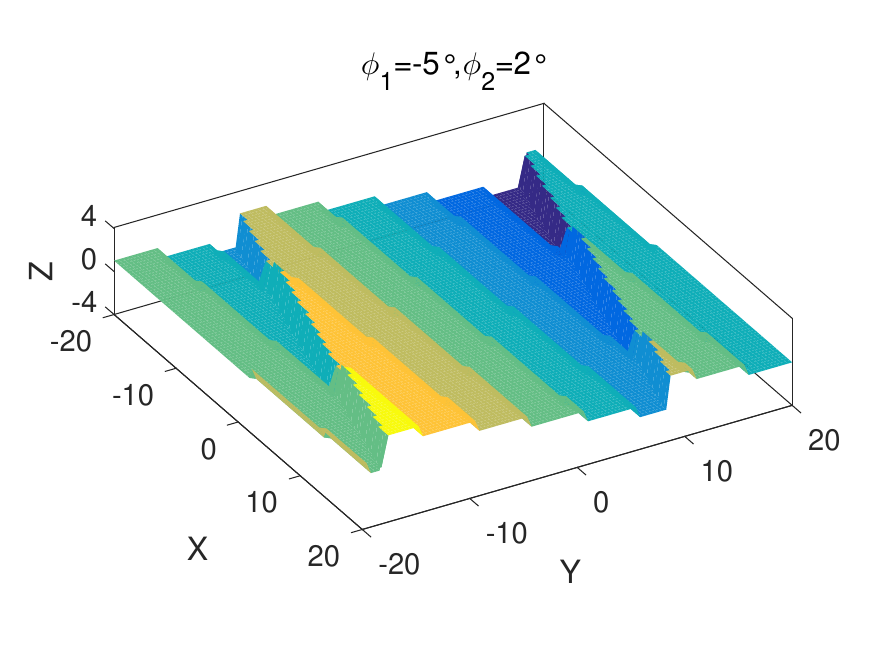}}
\subfigure[]{ \includegraphics[width=0.32\linewidth]{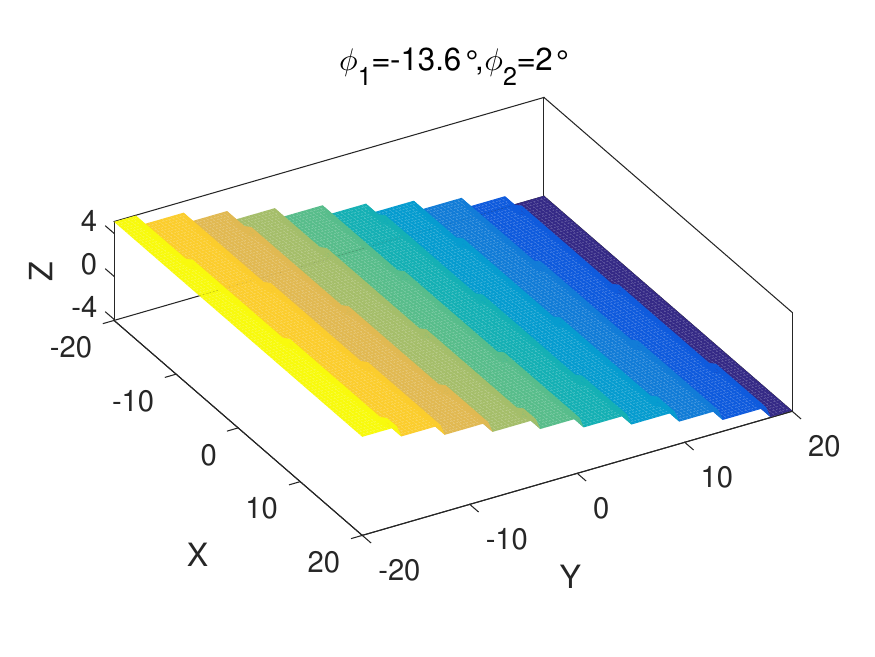}}
\subfigure[]{ \includegraphics[width=0.32\linewidth]{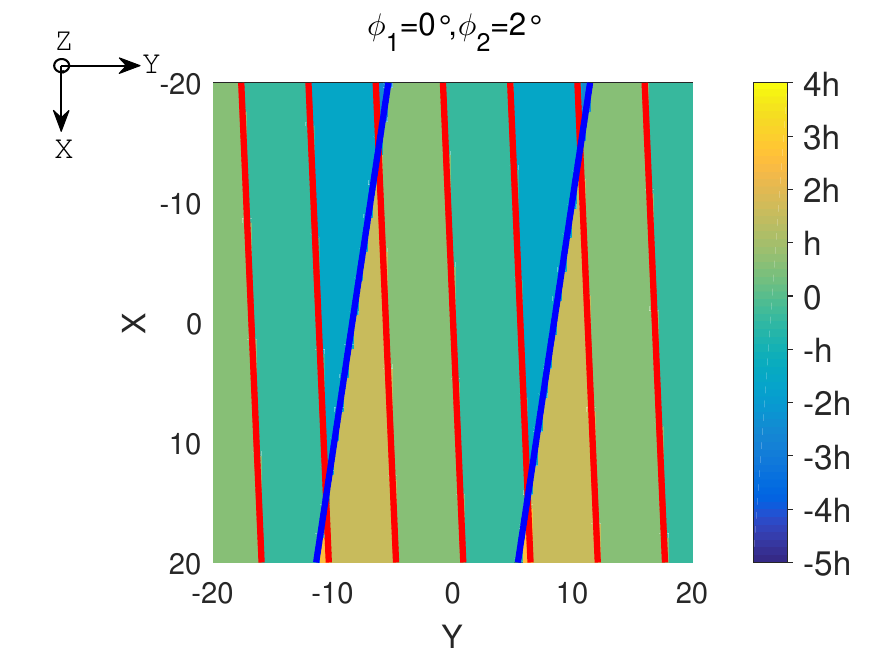}}
\subfigure[]{ \includegraphics[width=0.32\linewidth]{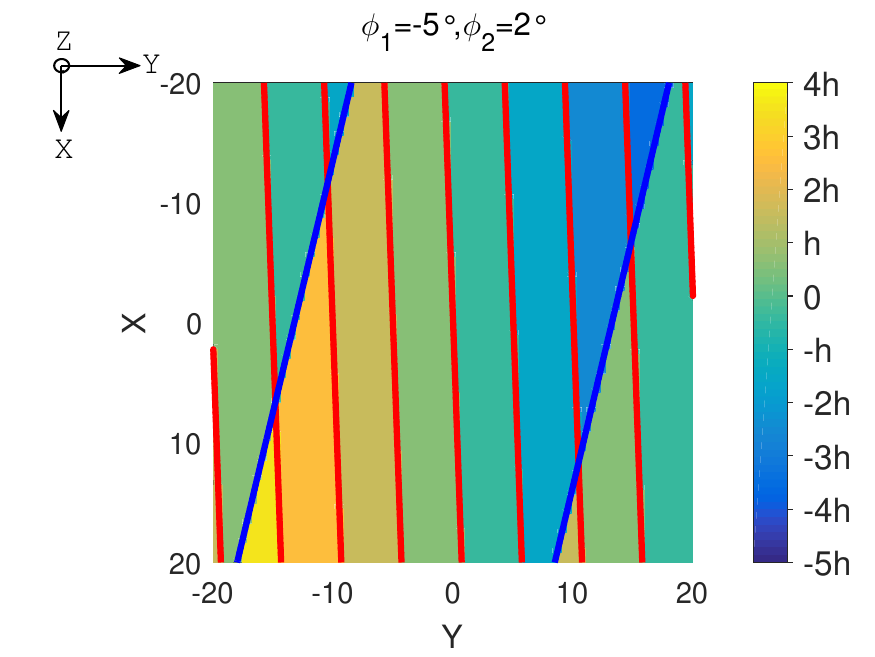}}
\subfigure[]{ \includegraphics[width=0.32\linewidth]{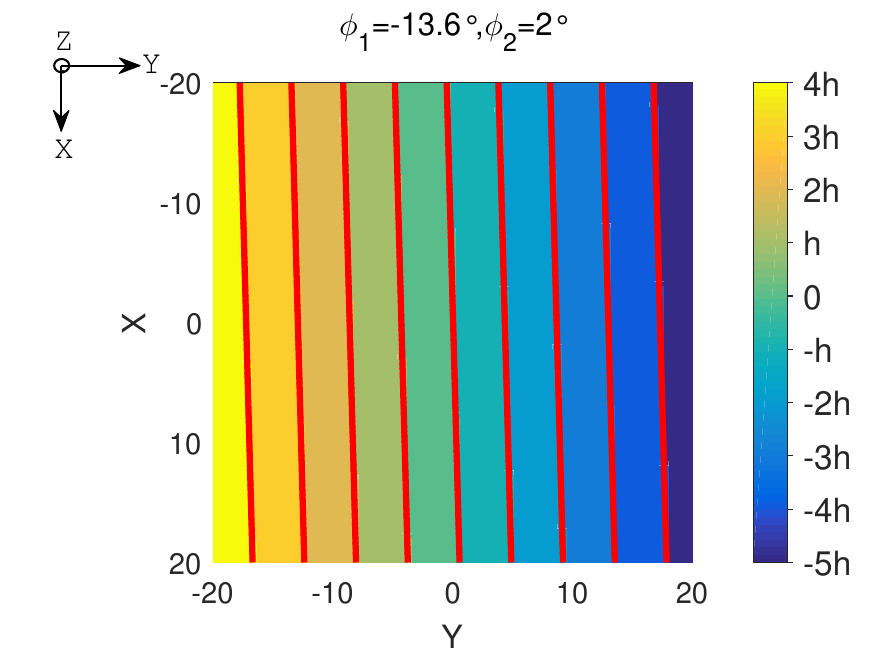}}
\caption{
The interface structure of $(100)$ orthorhombic/monoclinic interface in zirconia for $\phi_2=2^{\circ}$, as a function of inclination $\phi_1$: (a) interfacial energy, (b) misorientations across the interface,  and (c) disconnection densities;
(d-f) oblique view of the interface disconnection structure and height/step profile;
(g-i) plan view of the interfaces along the $Z$-direction (normal to the reference interface, where the red and blue lines respectively denote  disconnection types $(\bm{b}_1,h_1)$ and $(\bm{b}_2,h_2)$ and the shading between the lines indicates the height profile).
}
\label{fig:simu13}
\end{figure}

Figure~\ref{fig:simu13} shows the interface structures as functions of the inclination angle $\phi_1$ when  the inclination angle $\phi_2$ is fixed at $\phi_2=2^{\circ}$.
Comparing this case with that of $\phi_2=0$ (Fig.~\ref{fig:simu12}), we see that the misorientation angle $\theta_2$ is non-zero.
The densities of disconnections $(\bm{b}_1,h_1)$ and $(\bm{b}_2,h_2)$ are finite for all inclinations $\phi_1$ and   $\phi_2=2^{\circ}$.
The profiles of the interfacial energy, misorientation angles ($\theta_1$ and $\theta_3$), and the densities of disconnections versus $\phi_1$ are similar to the case of  $\phi_2=0$ (although the cusps in the interfacial energy are rounded at finite $\phi_2$ and sharp at $\phi_2=0$).
Figure~\ref{fig:simu13}(d)-(f)  and (g)-(i) show the oblique and plan view of the  interface disconnection networks with inclination angles $\phi=0$, $-5^{\circ}$, and $-13.6^{\circ}$.
Since the density of disconnection $(\bm{b}_3,h_3)$ is always zero,  the networks are only composed of two types of disconnections, $(\bm{b}_1,h_1)$ and $(\bm{b}_2,h_2)$.
For finite $\phi_2$, the disconnection lines of the two modes are neither parallel with each other nor with the $X$-axis.
One mode corresponds to inclinations about the $X$-axis and the other about the $Y$-axis.
The disconnection $(\bm{b}_2,h_2)$ is rotated such that it has both an edge component that contributes to the inclination $\phi_1$ and a screw component that  contributes to the inclination $\phi_2$.
As the magnitude of the inclination angle  changes from $\phi_1=0$ to $-5^{\circ}$ and to$-13.6^{\circ}$, the angle between the the two sets of disconnections increases.
At the same time, the spacing between disconnections $(\bm{b}_1,h_1)$ decreases and the spacing between disconnections $(\bm{b}_2,h_2)$ increases.
When $\phi_1=-13.6^{\circ}$, the disconnection lines of these two modes are nearly perpendicular to each other (the density of the second mode is too small to be seen in the figure).
Since  Burgers vectors $\bm{b}_1$ and $\bm{b}_2$ are both in the $Y$-direction, the dislocation character of disconnections $(\bm{b}_1,h_1)$ is near an edge orientation  and the character of $(\bm{b}_2,h_2)$ is near a screw orientation.

\subsubsection{Interfaces with fixed misorientations}

{We now examine the interfacial energy, interface inclination, and disconnection densities as functions of the interfacial misorientation. For each misorientation, the interfacial energy is minimized with respect to both the interface structure and the interface inclination. This scenario may correspond to a monoclinic particle embedded within an orthorhombic matrix. Under a fixed crystallographic orientation relationship between the particle and the matrix, the particle/matrix interface is free to adjust its morphology to attain the lowest-energy configuration. In certain circumstances, the crystallographic orientation of the inclusion may also undergo limited relaxation, subject to the constraints imposed by the surrounding matrix.}

\begin{figure}[htbp]
\centering
\subfigure[]{  \includegraphics[width=0.32\linewidth]{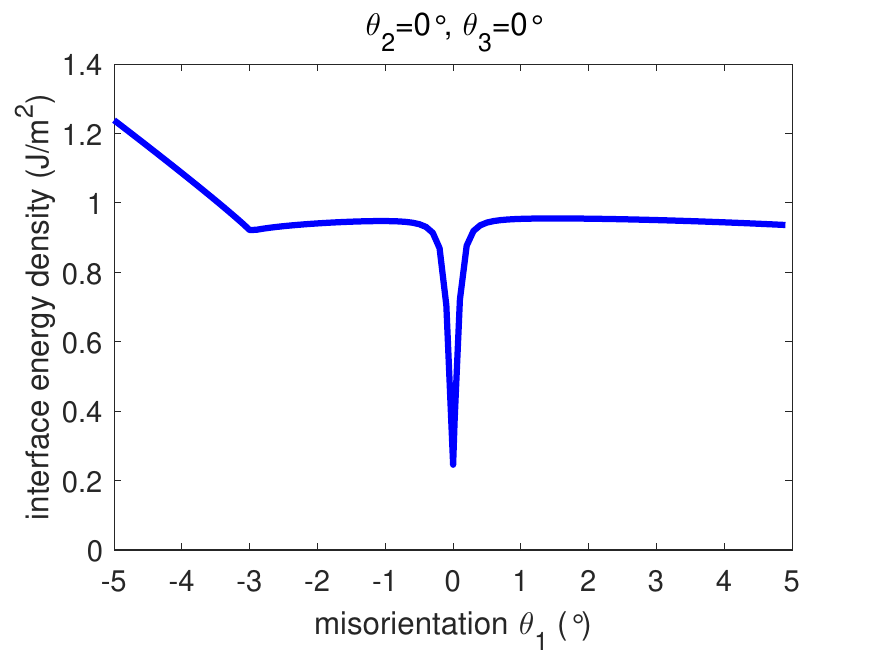}}
\subfigure[]{  \includegraphics[width=0.32\linewidth]{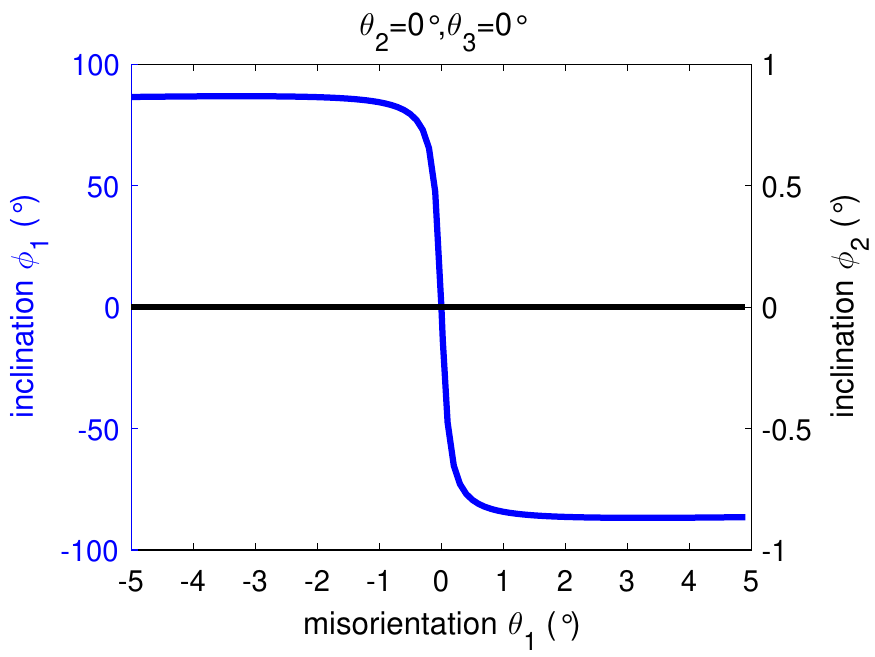}}
\subfigure[]{ \includegraphics[width=0.32\linewidth]{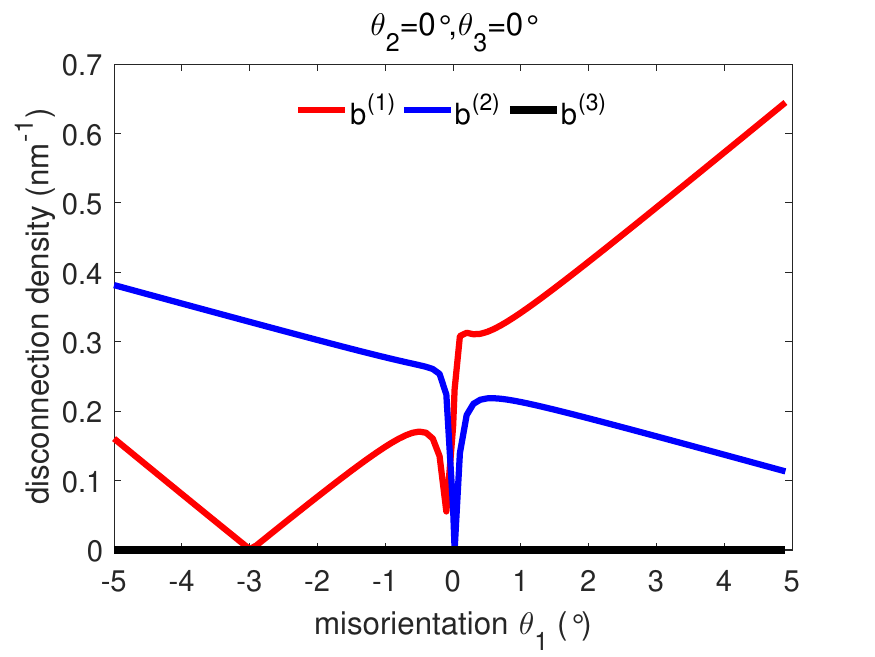}}
\subfigure[]{ \includegraphics[width=0.32\linewidth]{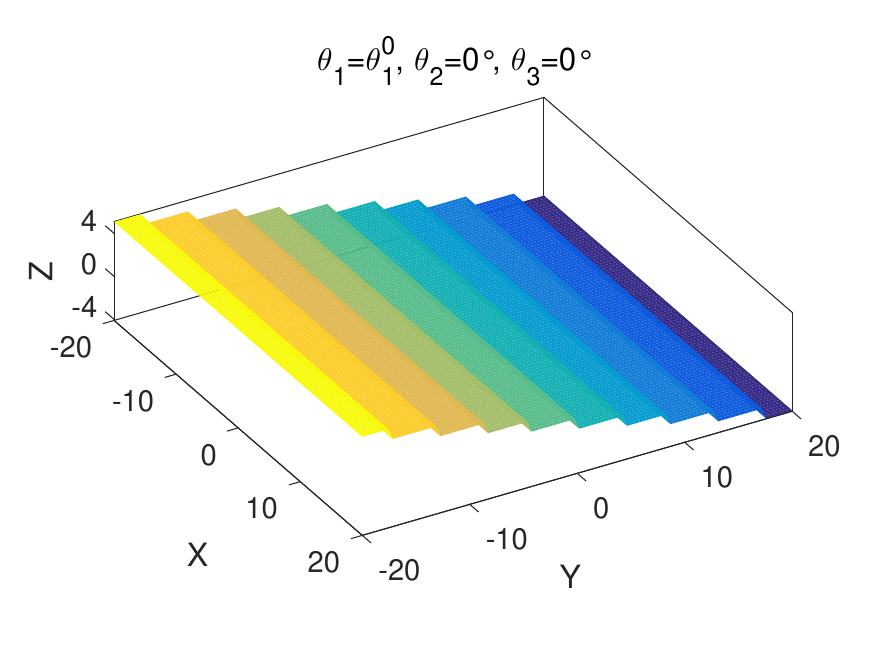}}
\subfigure[]{ \includegraphics[width=0.32\linewidth]{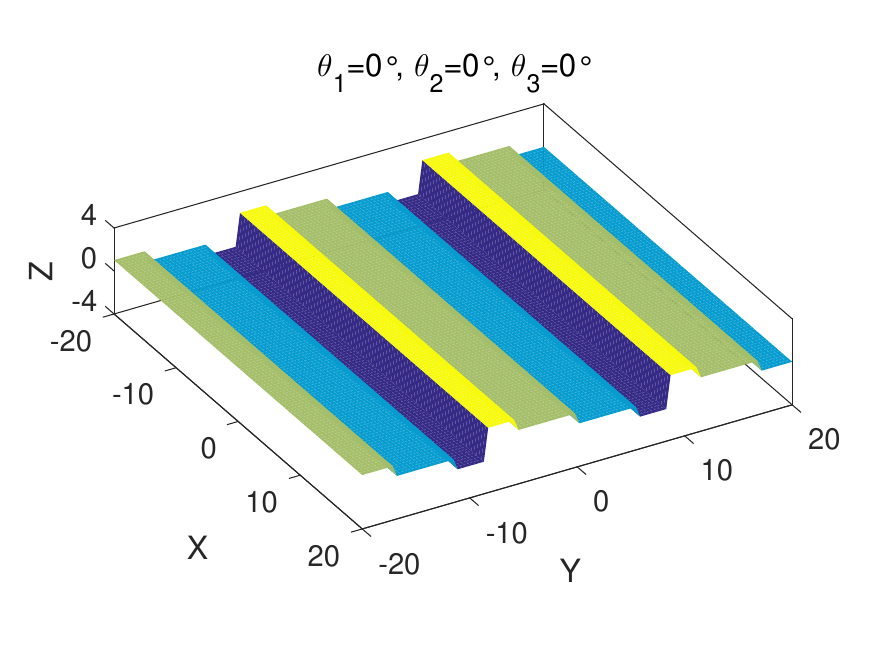}}
\subfigure[]{ \includegraphics[width=0.32\linewidth]{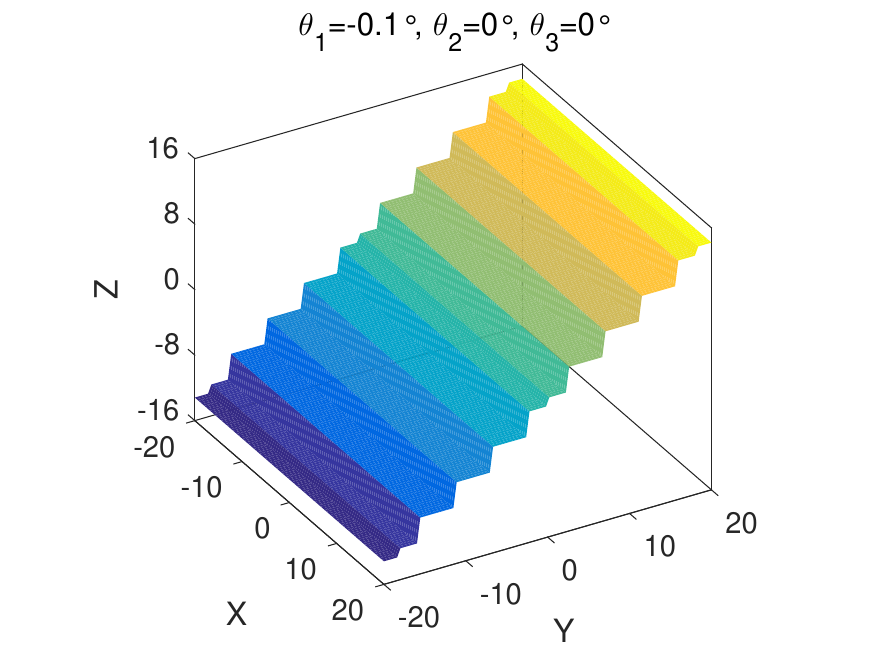}}
\subfigure[]{ \includegraphics[width=0.32\linewidth]{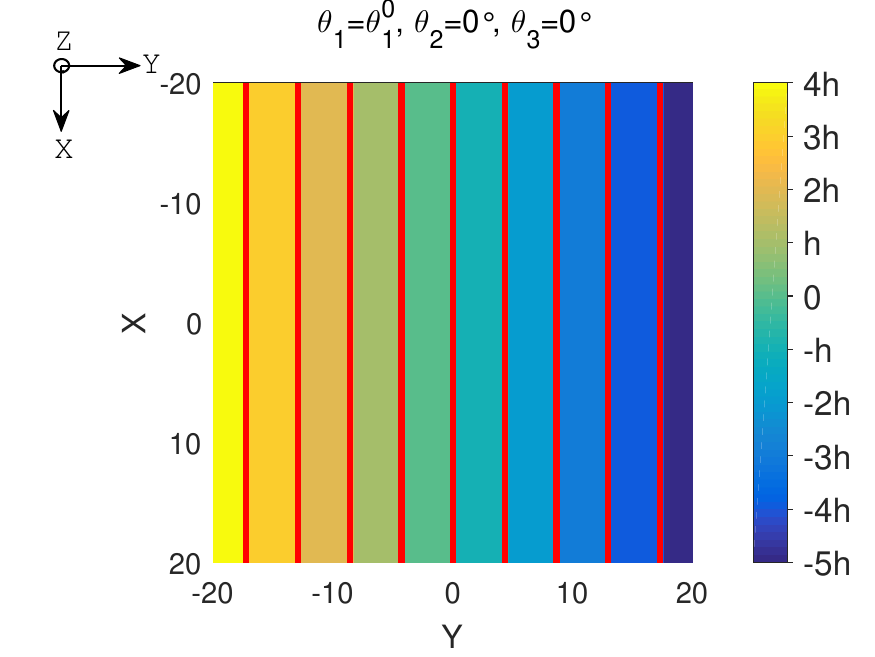}}
\subfigure[]{ \includegraphics[width=0.32\linewidth]{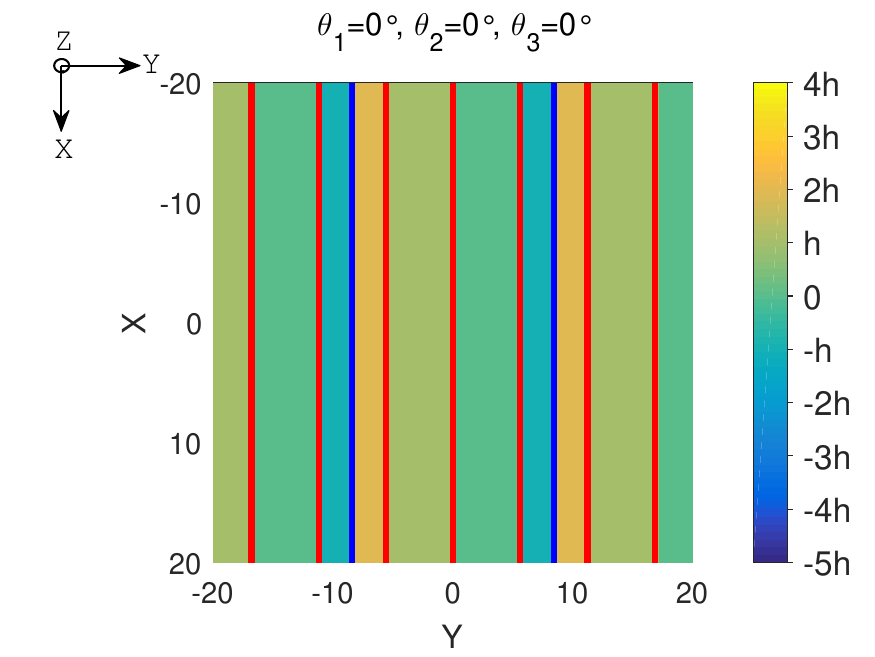}}
\subfigure[]{ \includegraphics[width=0.32\linewidth]{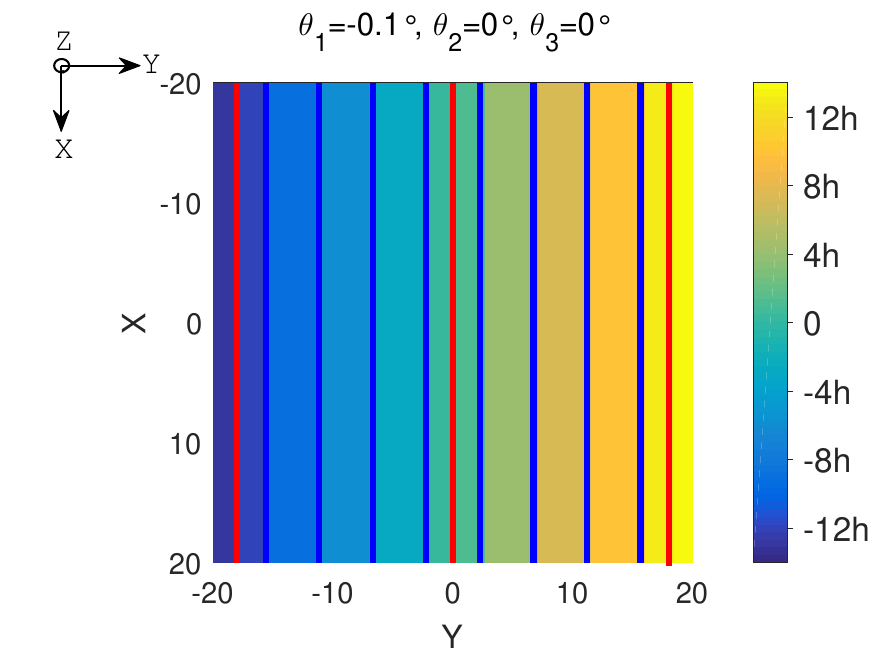}}
\caption{The interface structure of $(100)$ orthorhombic/monoclinic interface in zirconia, as a function of  misorientation $\theta_1$: (a)  interfacial energy, (b)  inclinations, (c)  disconnection densities; (d-f) oblique view of the interface disconnection structure and height/step profile;
(g-i) plan views of the interface structures along the $Z$-direction (normal to the reference interface, where the red and blue lines respectively denote disconnection types $(\bm{b}_1,h_1)$ and $(\bm{b}_2,h_2)$ and the shading between the lines indicates the height profile), for misorientations $\theta_1=\theta_1^0=0.0221^{\circ}$, $0^{\circ}$, and $-0.1^{\circ}$, $\theta_2=0$, and $\theta_3=0$.
}
\label{fig:simu14}
\end{figure}

Figure~\ref{fig:simu14} shows the interface structures as functions of  the misorientation tilt angles $\theta_1$, while the other two misorientation angles are fixed at zero.
In Fig.~\ref{fig:simu14}(a), we  see that the global energy minimum corresponds to $\theta_1=0.0221^{\circ}$, where the disconnection network only includes one disconnection mode $(\bm{b}_1,h_1)$; see Fig.~\ref{fig:simu14}(d) and (g).
Figure~\ref{fig:simu14}(b) shows that as the misorientation angle $\theta_1$ around $X$-axis varies, the inclination angle $\phi_1$ around $X$-axis has a sharp decrease from $\theta_1=-0.25^{\circ}$ to $0.25^{\circ}$, and $\phi_2$ around the $Y$-axis is zero.
Figure~\ref{fig:simu14}(c) shows the  disconnection densities profiles.
When the energy attains its minimum value at $\theta_1=0.0221^{\circ}$, the $(\bm{b}_1,h_1)$ disconnection density reaches a local maximum while the $(\bm{b}_2,h_2)$ disconnections density reaches the minimum (i.e., zero).
The $(\bm{b}_3,h_3)$ disconnection density is always zero.
Figures~\ref{fig:simu14}(e) and (h) show the results for $\theta_1=0$; there are two parallel arrays of disconnections, $(\bm{b}_1,h_1)$ and $(\bm{b}_2,h_2)$.
The $(\bm{b}_1,h_1)$ disconnections have positive step heights and tend to increase the inclination, while the$(\bm{b}_2,h_2)$ disconnections have negative step height and decrease the inclination.
The inclination increase from the $(\bm{b}_1,h_1)$ disconnections  cancel the inclination   decrease associated with the $(\bm{b}_2,h_2)$ disconnection.
Figures~\ref{fig:simu14}(f) and (i) show the oblique and plan views of the disconnection networks for $\theta_1=-0.1^{\circ}$.
The density of the $(\bm{b}_2,h_2)$ disconnections ($h_2=-3a^{\rm o}$) is larger than the density of the $(\bm{b}_1,h_1)$ disconnection ($h_1= a^{\rm o}$).
Hence, the interface has an effective positive inclination $\phi_1$ about the $X$-axis.

\begin{figure}[htbp]
\centering
\subfigure[]{  \includegraphics[width=0.32\linewidth]{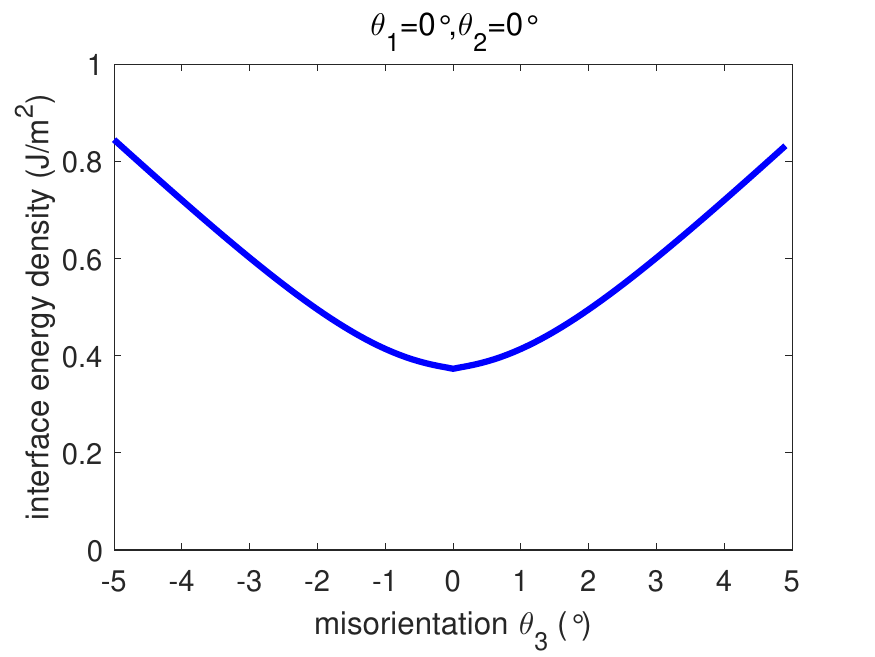}}
\subfigure[]{  \includegraphics[width=0.32\linewidth]{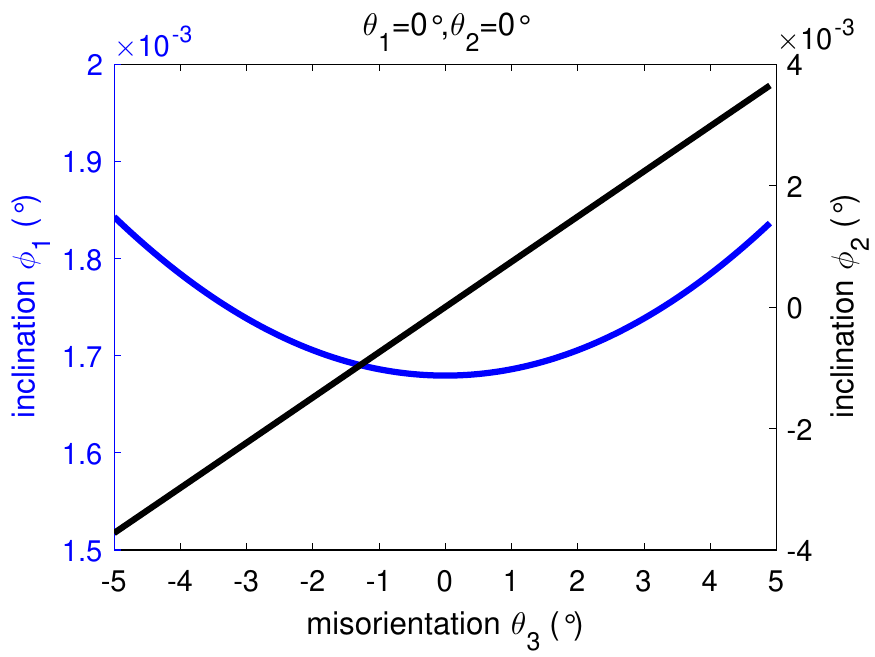}}
\subfigure[]{ \includegraphics[width=0.32\linewidth]{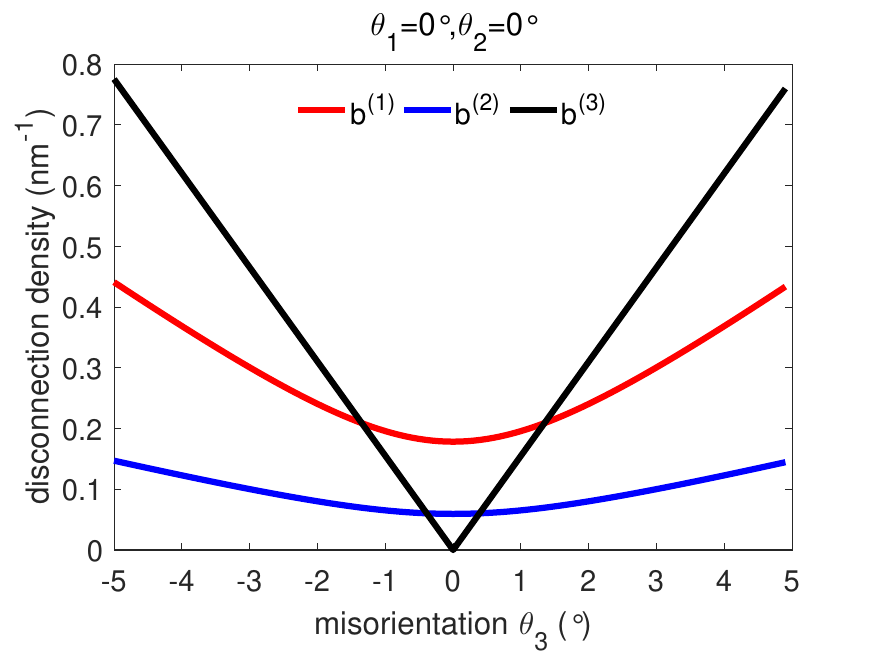}}
\subfigure[]{ \includegraphics[width=0.32\linewidth]{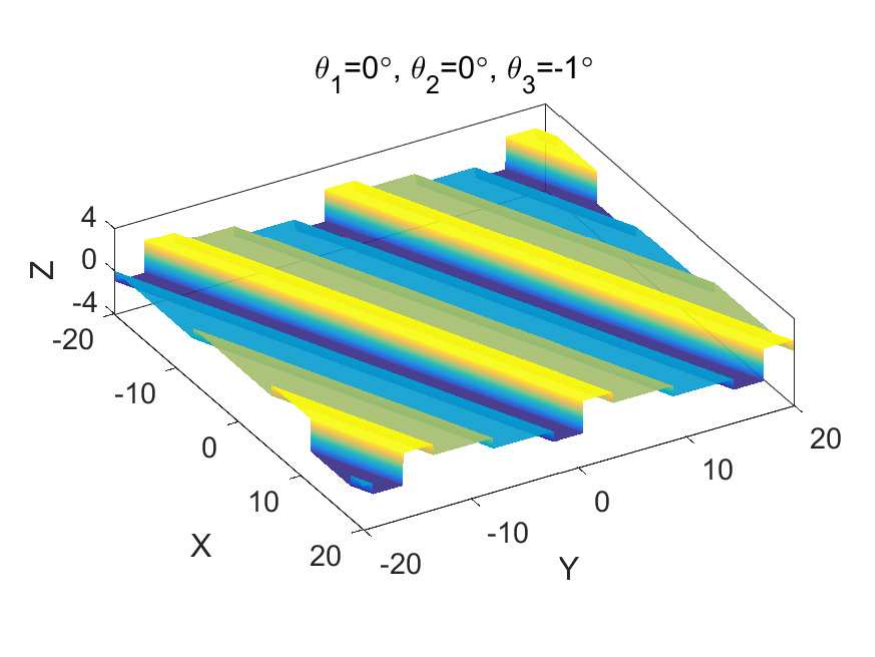}}
\subfigure[]{ \includegraphics[width=0.32\linewidth]{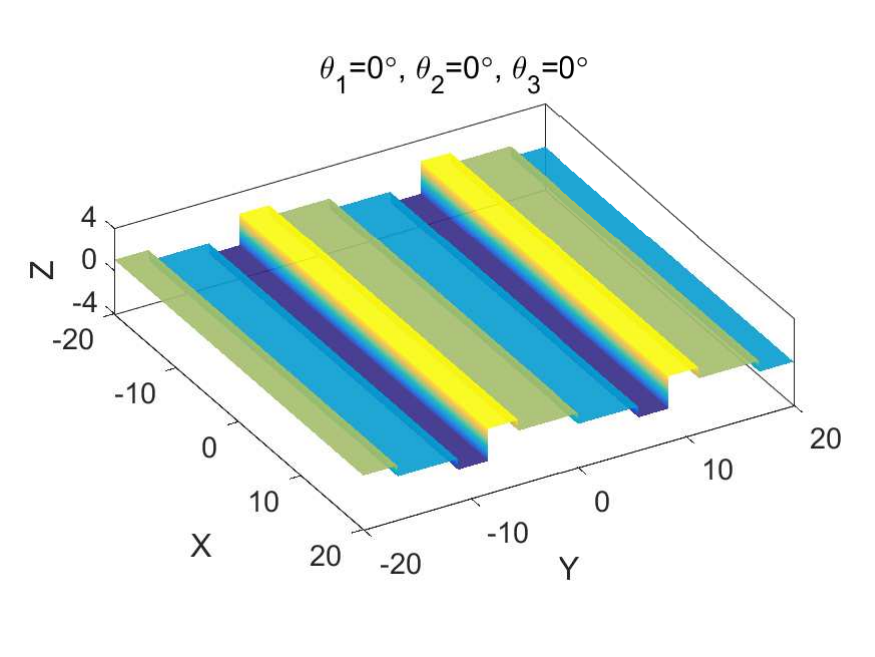}}
\subfigure[]{ \includegraphics[width=0.32\linewidth]{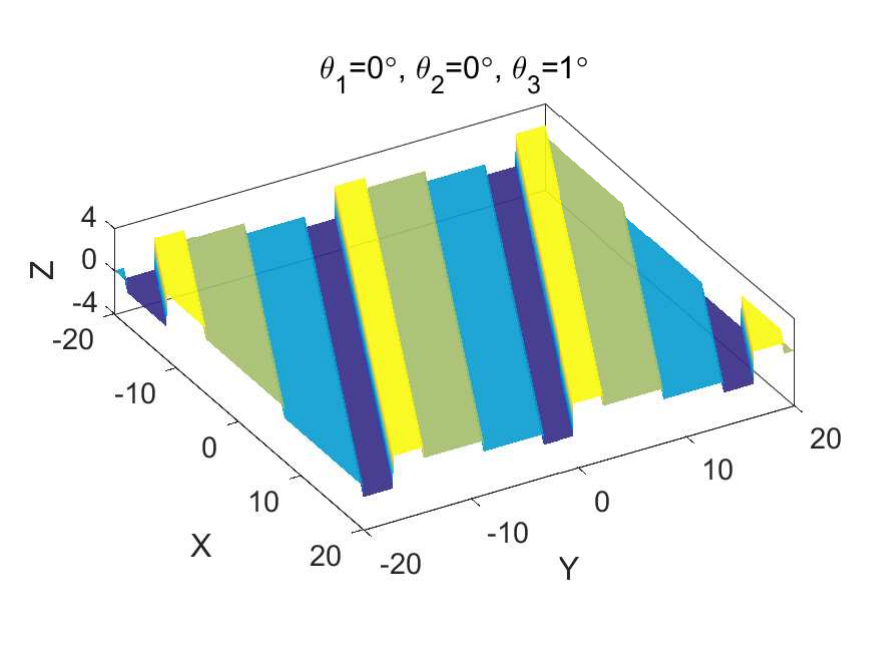}}
\subfigure[]{ \includegraphics[width=0.32\linewidth]{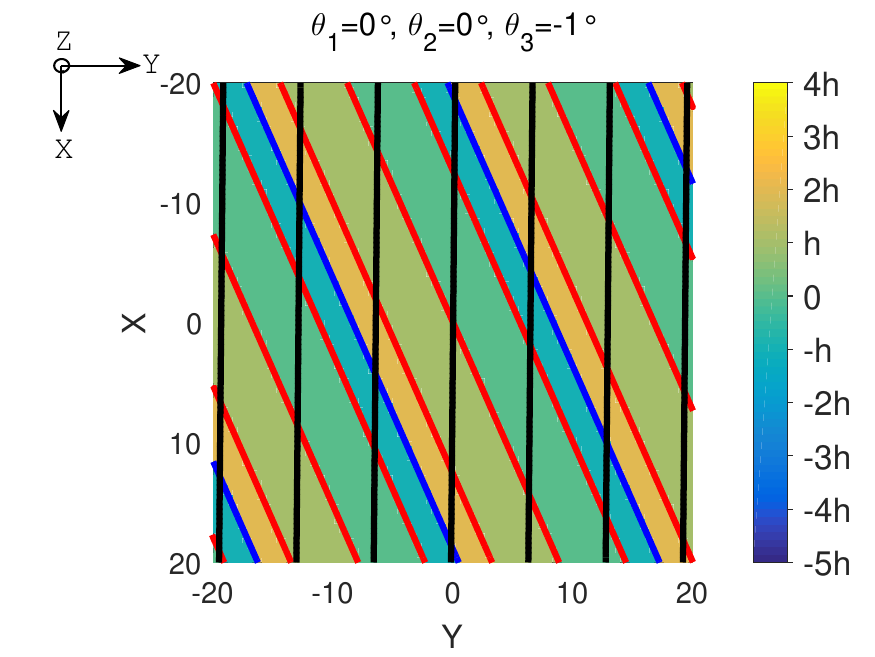}}
\subfigure[]{ \includegraphics[width=0.32\linewidth]{structure_misorientation_tau0_ZrO2_2d_v1.pdf}}
\subfigure[]{ \includegraphics[width=0.32\linewidth]{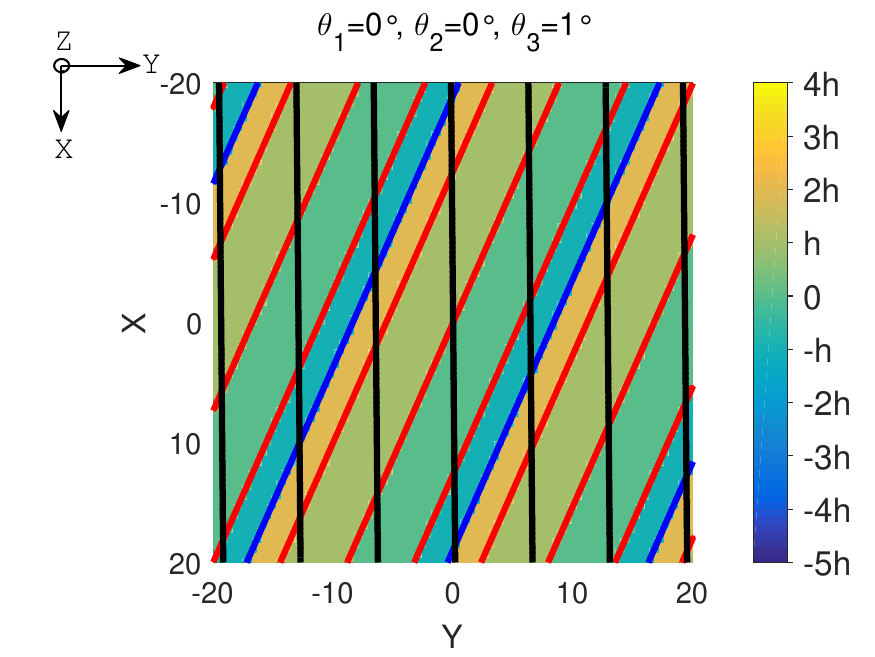}}
\caption{The interface structure of $(100)$ orthorhombic/monoclinic interface in zirconia, as a function of  twist misorientation angle $\theta_3$: (a)  interfacial energy (b)  inclinations, (c)  disconnection densities; (d-f) oblique view of the interface disconnection structure and height/step profile;
(g-i) plan views of the  interface structures along the $Z$-direction (normal to the reference interface, where the red, blue and black lines respectively denote disconnection types $(\bm{b}_1,h_1)$, $(\bm{b}_2,h_2)$, and $(\bm{b}_3,h_3)$, and the shading between the lines indicates the height profile),
for twist misorientation $\theta_3=-1^{\circ}$, $0^{\circ}$, and $1^{\circ}$, $\theta_1=0$, and $\theta_2=0$.
}
\label{fig:simu15}
\end{figure}

Figure~\ref{fig:simu15} shows the interface structures for several misorientation   angles  $\theta_3$ (the other two misorientations are fixed as zero).
The interfacial energy curve and disconnection densities are both minimum  at $\theta_3=0$, and are symmetric about $\theta_3=0$, as shown in Figs.~\ref{fig:simu15}(a) and (c).
Figure~\ref{fig:simu15}(b) demonstrates that both the $\phi_1$ and $\phi_2$ inclinations are always very small for all  $\theta_3$ misorientations; this  indicates that the twist misorientation around the  $Z$-axis has little effect on the inclinations around the $X$ or $Y$-axes,  $\phi_1$  or $\phi_2$.
Figures~\ref{fig:simu15}(d)-(i) show the oblique and plan views of the interfacial disconnection networks for three misorientation angles $\theta_3=-1^{\circ}$, $0^{\circ}$, and $1^{\circ}$.
In Figs.~\ref{fig:simu15}(e) and (h),  we  see that when $\theta_3=0$, there are two parallel arrays of disconnections,  $(\bm{b}_1,h_1)$ and  $(\bm{b}_2,h_2)$.
The  $(\bm{b}_1,h_1)$ disconnections, with a positive step height, increase the interface inclination, while the $(\bm{b}_2,h_2)$  disconnections, with negative step heights, decrease it.
The total inclination increase by disconnections $(\bm{b}_1,h_1)$ cancels that of the total inclination decrease by disconnections $(\bm{b}_2,h_2)$ resulting in  zero inclination angle.
Figures~\ref{fig:simu15}(d), (g), (f), and (i) show that when there is a twist misorientation, the interfacial disconnection networks are also rotated; this is the result of the screw disconnection  components (the source of the twist).

%

\subsection{Au/Pd twist-misfit interface}

We apply our continuum model to predict the interface structure on  Au/Pd $(001)$ twist-misfit interface.
We align the $(001)$ plane of an Au lattice with the $(001)$ plane of a Pd lattice and set $[310]_{\rm Au} \parallel [100]_{\rm Pd}$ as the $X$-axis with unit vector $\bm{e}_1$ and $[1\bar{3}0]_{\rm Au} \parallel [0\bar{1}0]_{\rm Pd}$ as the $Y$-axis with unit vector $\bm{e}_2$.
The reciprocal densities of the coincidence sites referenced to the Au and Pd lattices are, respectively, $9$ and $10$; so, this interface is represented as $\sigma_{\rm Au} : \sigma_{\rm Pd} = 9:10$ in the literature~\cite{Balluffi1982}.
Consider four disconnection modes characterized by the Burgers vectors and step heights:
\begin{align}
&\bm{b}_1 = \frac{\sqrt{10}}{20}a_{\rm Au} \bm{e}_1
\quad\text{and}\quad
h_1 = \frac{1}{2}a_{\rm Au};
\label{eqn:BV_twist1}\\
&\bm{b}_2 = \frac{\sqrt{10}}{20}a_{\rm Au} \bm{e}_1
\quad\text{and}\quad
h_2 = -\frac{1}{2}a_{\rm Au};
\label{eqn:BV_twist2}\\
&\bm{b}_3 = \frac{\sqrt{10}}{20}a_{\rm Au} \bm{e}_2
\quad\text{and}\quad
h_3 = \frac{1}{2}a_{\rm Au};
\label{eqn:BV_twist3}\\
&\bm{b}_4 = \frac{\sqrt{10}}{20}a_{\rm Au} \bm{e}_2
\quad\text{and}\quad
h_4 = -\frac{1}{2}a_{\rm Au}.
\label{eqn:BV_twist4}
\end{align}
{(Note that, although the sign convention for the step height is arbitrary, once a convention is adopted, $(\bm{b}, h)$ and $(\bm{b}, -h)$ represent two distinct modes.)}
The lattice constants are: $a_{\rm Au}=0.40785$~nm and $a_{\rm Pd}=0.38902$~nm~\cite{Balluffi1982}.
{In the experiment of Hwang {\it et al.}~\cite{Hwang1980}, the Au/Pd thin-film bilayer was fabricated by depositing an Au film onto a tungsten grid that had previously been coated with a Pd film. Consequently, the Pd film was rigidly constrained by the substrate, whereas the Au film was constrained primarily through its contact with the Pd film. It is therefore natural to adopt the Pd lattice as the reference lattice in the present analysis.}
By choosing Pd as the reference lattice, the lattice correspondence parameters are $\Omega_1 = \frac{\sqrt{10}a_{\rm pd}}{3a_{\rm Au}}$, $\Omega_2 = \frac{\sqrt{10}a_{\rm pd}}{3a_{\rm Au}}$, and $\Omega_3 = \frac{a_{\rm pd}}{a_{\rm Au}}$ when the Au lattice is rotated $\theta_3^0=18.43^{\circ}$  around $[001]$ relative to the Pd lattice.
Therefore, the transformation matrix $\mathbf U_{\alpha}$ is
\begin{flalign}
\mathbf U_{\alpha}=
\mathbf{R}_{-\theta_3^0}
\left(
\begin{array}{lll}
\Omega_1 & 0 & 0\\
0 & \Omega_2 & 0 \\
0 & 0 & \Omega_3
\end{array}
\right)
\end{flalign}
where $\mathbf{R}_{-\theta_3^0}= \left(
\begin{array}{lll}
\cos\theta_3^0 & \sin\theta_3^0  & 0\\
-\sin\theta_3^0  & \cos\theta_3^0  & 0\\
0 & 0 & 1
\end{array}
\right)$ is the rotation matrix.

\subsubsection{Equilibrium interface structure}

\begin{figure}[H]
\centering
\subfigure[]{ \includegraphics[width=0.4\linewidth]{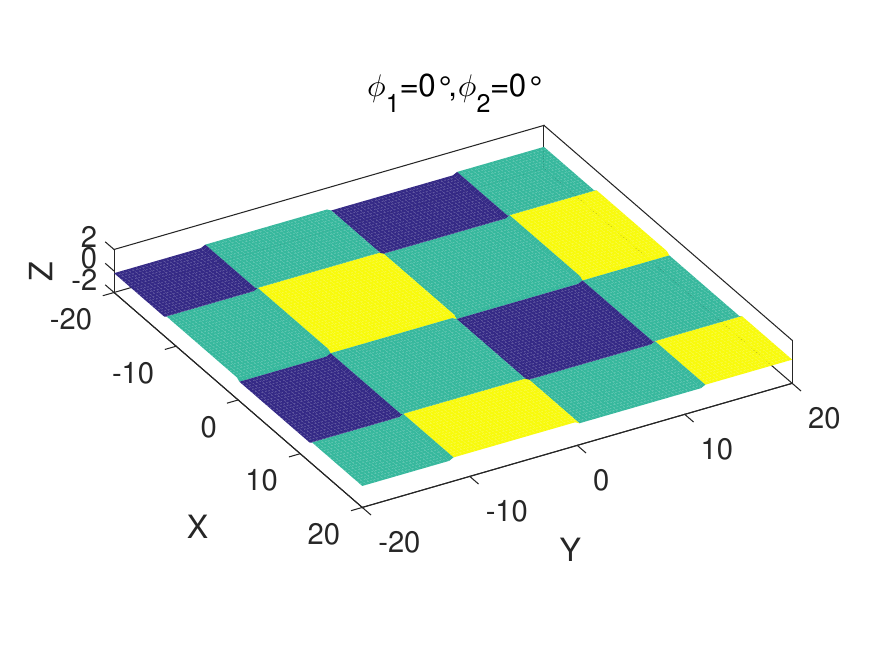}}
\subfigure[]{ \includegraphics[width=0.4\linewidth]{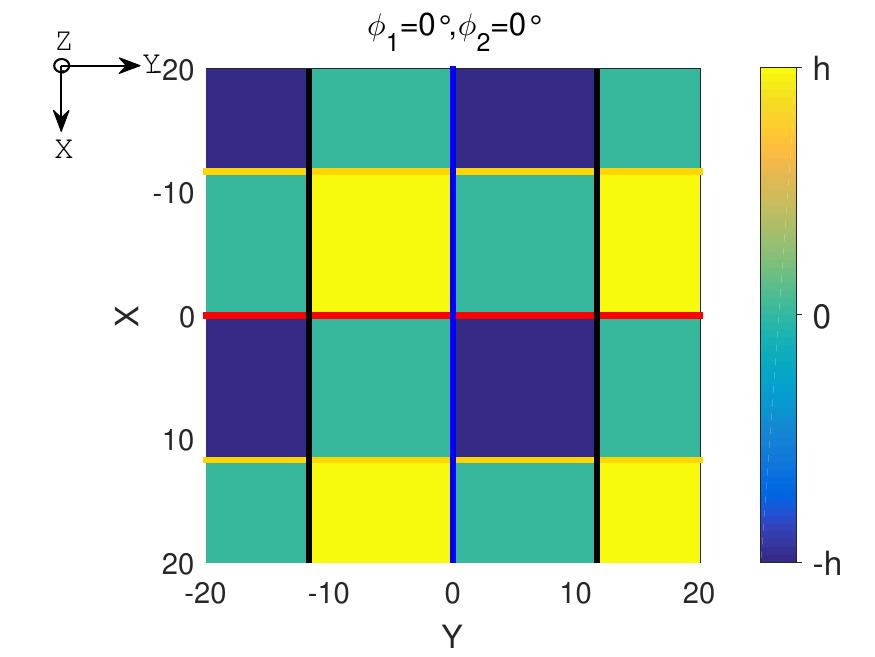}}
\caption{The equilibrium structure of Au/Pd (001) twist-misfit interface of the CSL $\sigma_{\rm Au} : \sigma_{\rm Pd} = 9 : 10$. (a) Oblique view and (b) plan view of the interface structure along the $Z$-direction (normal to the reference interface), where the red, yellow, blue, and black lines, respectively, denote disconnection type $(\bm{b}_1,h_1)$, $(\bm{b}_2,h_2)$, $(\bm{b}_3,h_3)$, and $(\bm{b}_4,h_4)$, and the shading between the lines indicates the height profile.
}
\label{fig:simu30}
\end{figure}

The interfacial energy attains minimum value at the equilibrium interfacial structure with inclination $\bm{\phi} = \bm{0}$, misorientation angles $\theta_1=0$, $\theta_2=0$, and $\theta_3=\theta_3^0=18.43^{\circ}$,
and the disconnection network shown in Fig.~\ref{fig:simu30}.
The disconnection network structure consists of all four modes of dislocations, where the two arrays of disconnections, $(\bm{b}_1,h_1)$ and $(\bm{b}_2,h_2)$, are parallel to each other, respectively, represented by the reciprocal vectors $\bm{N}_1= \bm{N}_2 = (0.0428, 0)$~nm$^{-1}$;
and two arrays of disconnections, $(\bm{b}_3,h_3)$ and $(\bm{b}_4,h_4)$, are parallel, represented by the reciprocal vectors $\bm{N}_3 = \bm{N}_4 = (0, 0.0428)$~nm$^{-1}$.
The two arrays of disconnections, $(\bm{b}_1,h_1)$ and $(\bm{b}_2,h_2)$, are perpendicular to the two arrays of disconnections, $(\bm{b}_3,h_3)$ and $(\bm{b}_4,h_4)$.
In this disconnection network, the dislocation characters of the disconnections are all screw dislocations in both the $X$ and $Y$ directions with equal inter-dislocation distances  $D_1=D_2=D_3=D_4=23.4$~nm.

The obtained disconnection structure agrees with the theoretical prediction and experimental
observations \cite{Balluffi1982,Hwang1980}.
There are two pairs of disconnection arrays that perpendicular to each other. Each pair of disconnection arrays has the same Burgers vectors and opposite step heights.
It is consistent with the experimental observations and the topography of the CSL/DCS lattice model \cite{Balluffi1982,Hwang1980}.
The equilibrium inclination angles, $\phi_1=0$ and $\phi_2=0$, and the misorientation twist angle $\theta_3=18.43^{\circ}$ also agree with the theoretical CSL/DCS lattice model and the experimental observations \cite{Balluffi1982,Hwang1980}.

\subsubsection{Interfaces with fixed inclinations}

\begin{figure}[htbp]
\centering
\subfigure[]{  \includegraphics[width=0.33\linewidth, trim=3cm 9.5cm 3.5cm 8cm, clip]{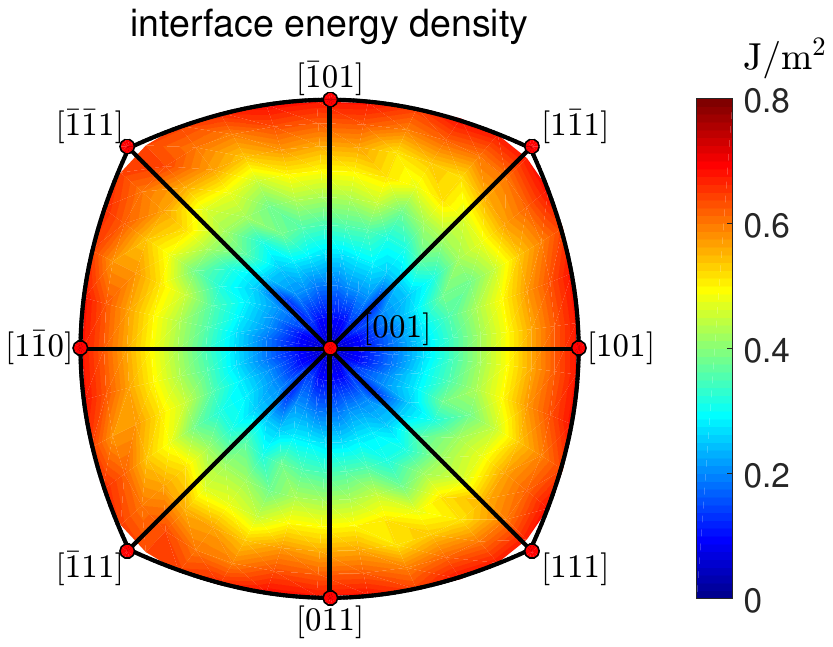}}
\subfigure[]{  \includegraphics[width=0.33\linewidth, trim=3cm 9.5cm 3.5cm 8cm, clip]{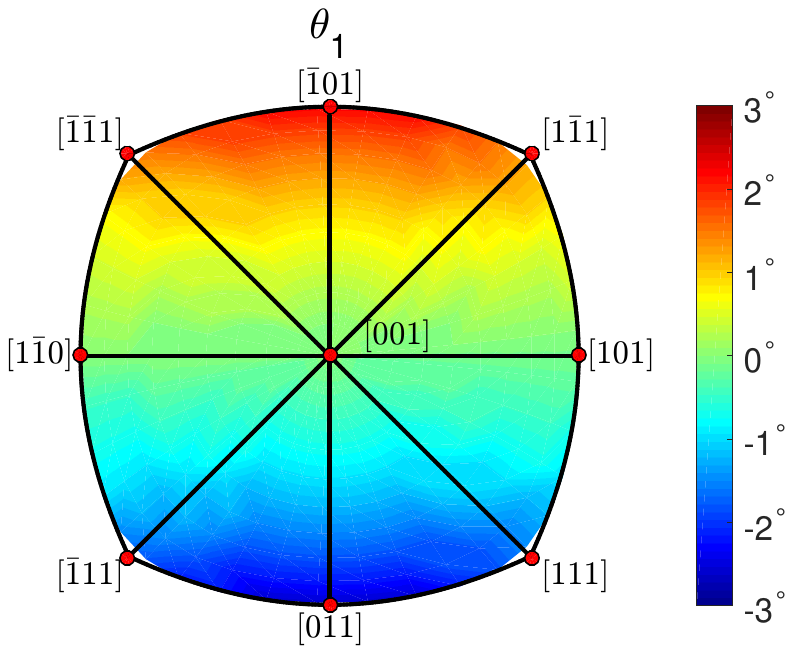}}
\subfigure[]{  \includegraphics[width=0.33\linewidth, trim=3cm 9.5cm 3.5cm 8cm, clip]{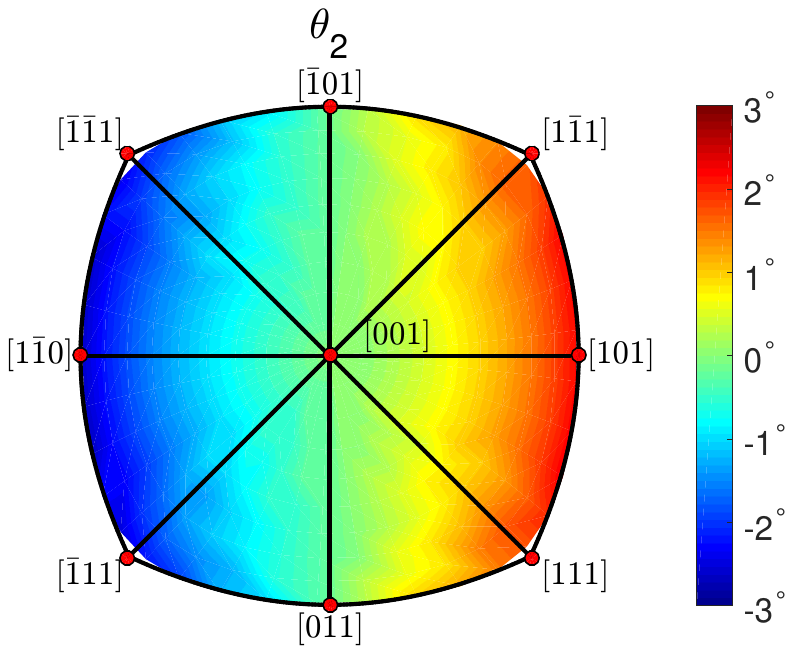}}
\subfigure[]{  \includegraphics[width=0.33\linewidth, trim=3cm 9.5cm 3.5cm 8cm, clip]{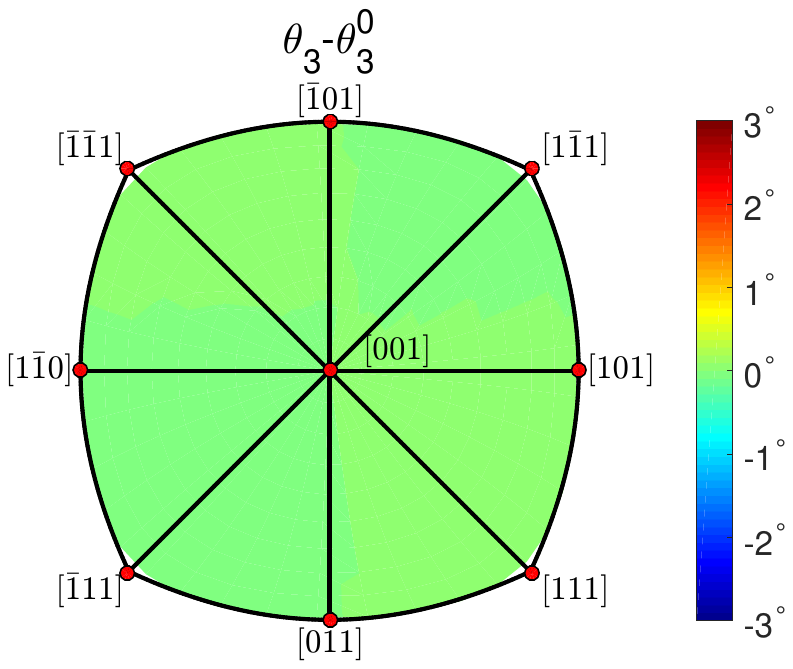}}
\caption{(a) Interfacial energy and (b-d) misorientation angles of Au/Pd (001) twist-misfit interface of the CSL $\sigma_{\rm Au} : \sigma_{\rm Pd} = 9:10$, as functions of inclination angles $\phi_1$ and $\phi_2$.
 $\phi_1$ and $\phi_2$ are the inclination angles relative to the $X$ and $Y$ axes of the tangent plane of the stereographic projection sphere. $\phi_1$ is increasing from bottom to top, and $\phi_2$ is increasing from left to right.
}
\label{fig:simu31}
\end{figure}

We study the interface disconnection network structures and misorientations at fixed interface inclination $\bm{\phi}$; {This is exactly the case studied in the experiment of Hwang {\it et al.}~\cite{Hwang1980}.}
These interfacial structure properties are obtained by the energy gradient flows in Eqs.~\eqref{eqn:evolution01} and \eqref{eqn:evolution02}.
The interfacial energy and misorientation angles are demonstrated in Fig.~\ref{fig:simu31}, as functions of the inclination angles ($\phi_1$ and $\phi_2$).
It can be observed that the global minimum value of the interfacial energy is attained at the inclination $\bm{\phi} = \bm{0}$, at which the interface stays at the equilibrium state.
Figures~\ref{fig:simu31}(b-d) demonstrate how the misorientation angles ($\theta_1$, $\theta_2$,  and $\theta_3$) vary with the interface inclination ($\phi_1$ and $\phi_2$).
The variation of interface inclination mainly leads to the change of misorientation angles $\theta_1$ and $\theta_2$ (tilt angle), and little variation of misorientation angle $\theta_3$ (twist angle).
As inclination angle $\phi_1$ increases, the misorientation tilt angle $\theta_1$ increases linearly with little change in tilt angle $\theta_2$ and twist angle $\theta_3$.
As inclination angle $\phi_2$ increases, the misorientation tilt angle $\theta_2$ increases linearly with little change in tilt angle $\theta_1$ and twist angle $\theta_3$.
The misorientation twist angle $\theta_3$ maintains $\theta_3^0$ while $\phi_1$ or $\phi_2$ changes.
Interface inclination variations lead to changes in misorientations around the same axis, which again reflects the relation between the interface inclination and misorientation angles.

\begin{figure}[htbp]
\centering
\subfigure[]{ \includegraphics[width=0.32\linewidth]{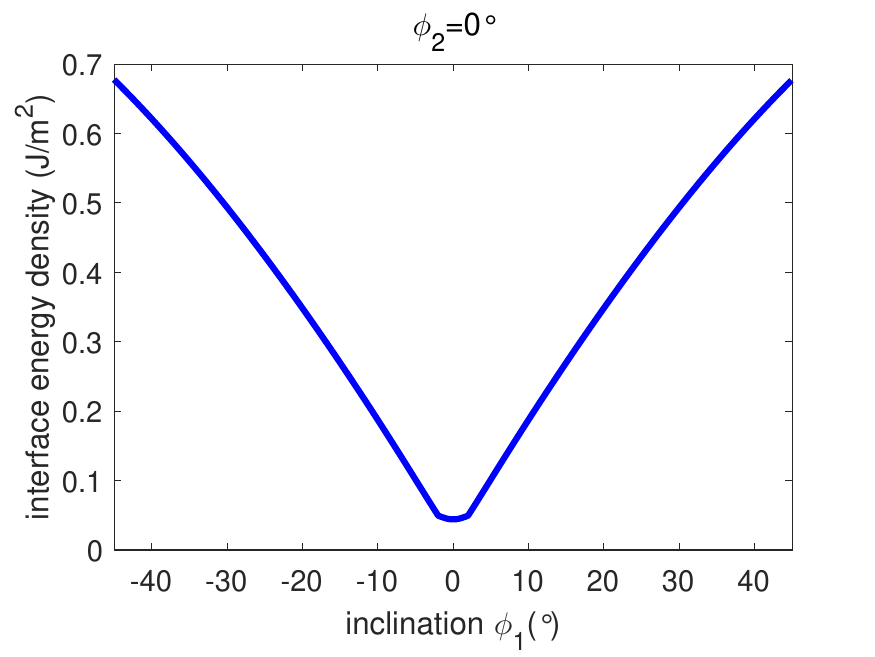}}
\subfigure[]{ \includegraphics[width=0.32\linewidth]{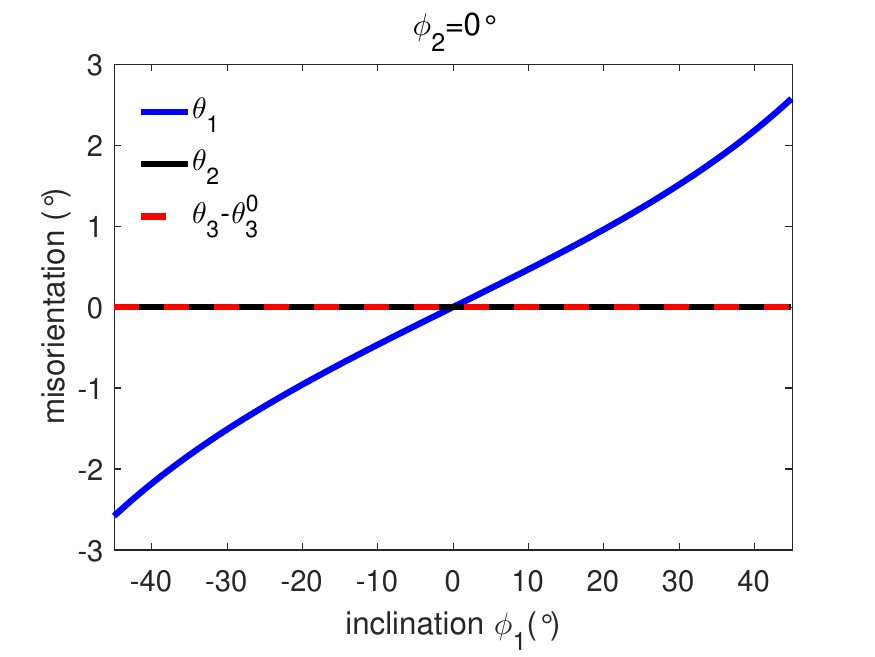}}
\subfigure[]{ \includegraphics[width=0.32\linewidth]{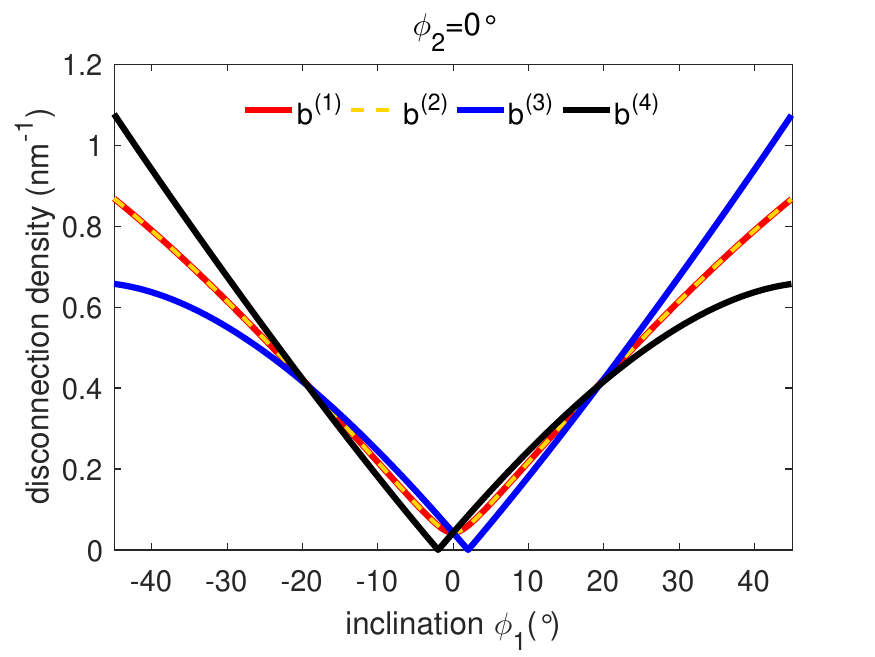}}
\subfigure[]{ \includegraphics[width=0.32\linewidth]{structure_inclination00_3d.pdf}}
\subfigure[]{ \includegraphics[width=0.32\linewidth]{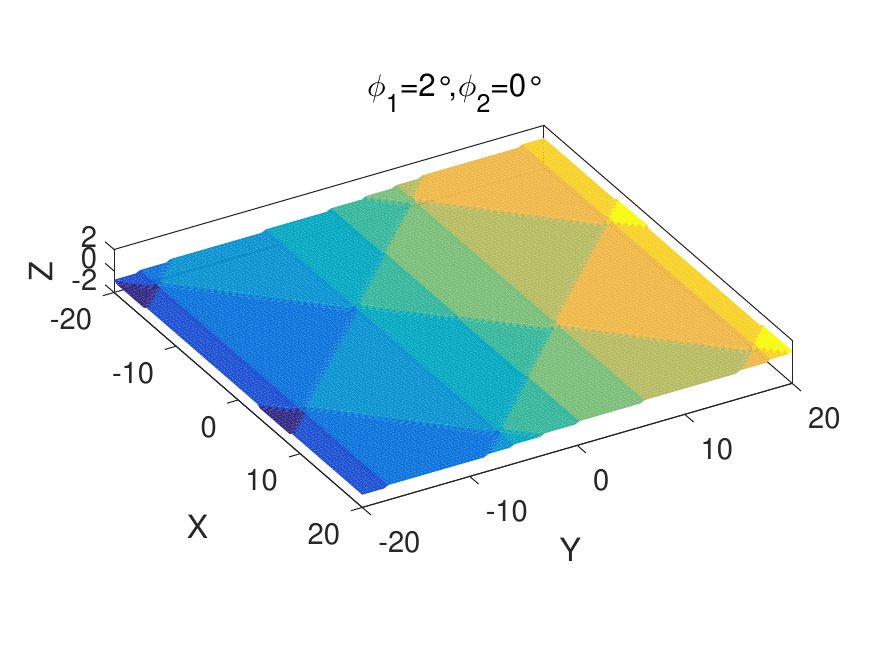}}
\subfigure[]{ \includegraphics[width=0.32\linewidth]{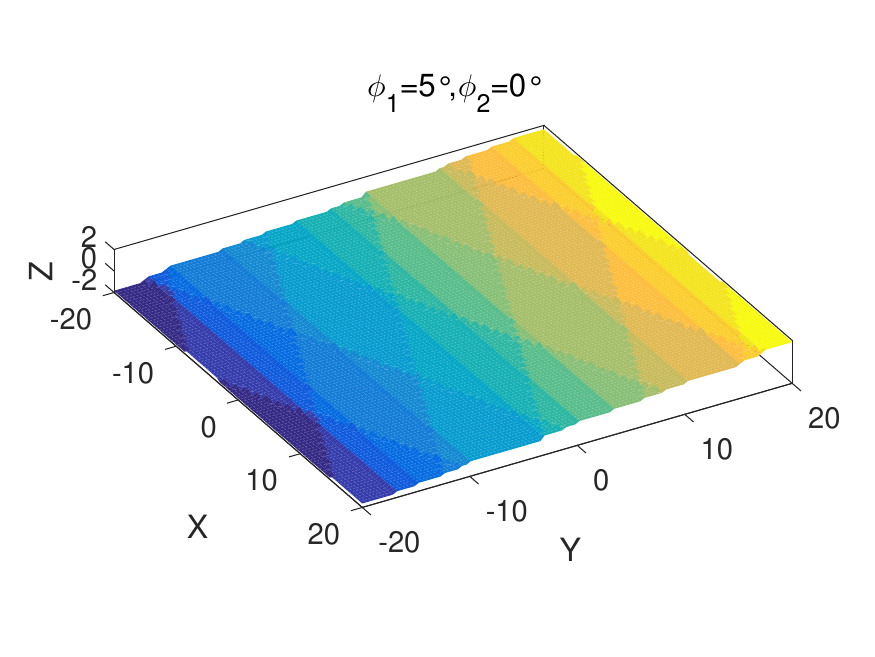}}
\subfigure[]{ \includegraphics[width=0.32\linewidth]{structure_inclination00_2d_nomesh.pdf}}
\subfigure[]{ \includegraphics[width=0.32\linewidth]{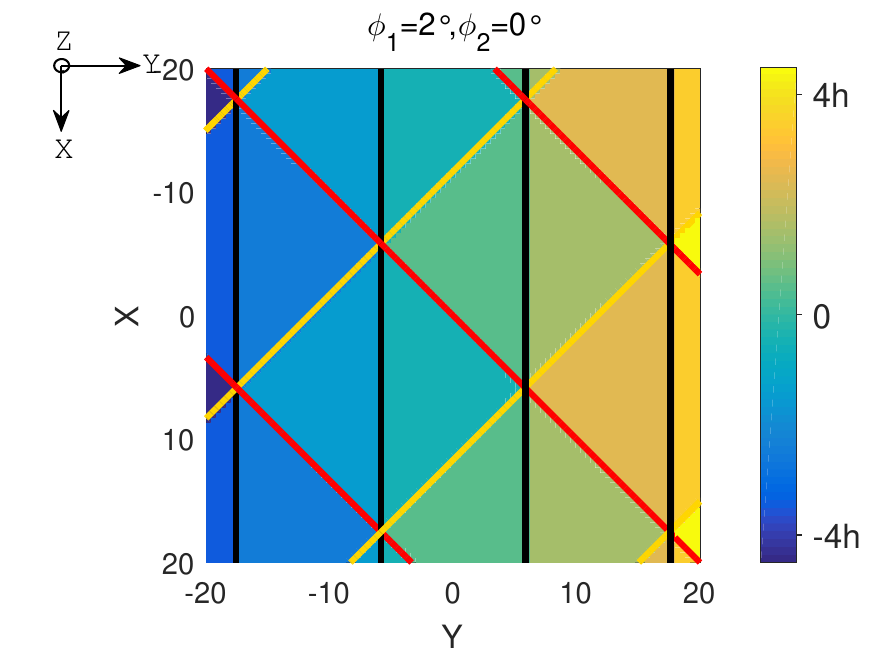}}
\subfigure[]{ \includegraphics[width=0.32\linewidth]{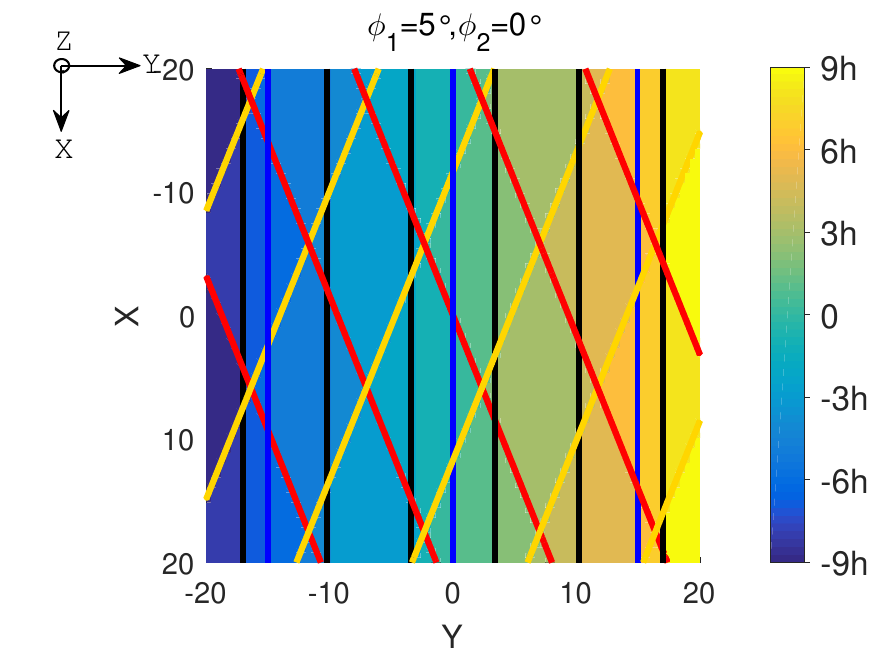}}
\caption{The interface structure of the Au/Pd (001) twist-misfit interface of the CSL $\sigma_{\rm Au} : \sigma_{\rm Pd} = 9:10$ for $\phi_2=0$, as a function of inclination $\phi_1$: (a) interfacial energy, (b) misorientation angles, (c) disconnection densities; (d-f) oblique view of the interface disconnection structure and height/step profile;
 (g-i) plan views of the interface structures along the $Z$-direction (normal to the reference interface), where the red, yellow, blue, and black lines, respectively, denote disconnection type $(\bm{b}_1,h_1)$, $(\bm{b}_2,h_2)$, $(\bm{b}_3,h_3)$, and $(\bm{b}_4,h_4)$, and the shading between the lines indicates the height profile,
at inclinations $\phi_1=0$, $2^{\circ}$, and $5^{\circ}$, and $\phi_2=0$.}
\label{fig:simu32}
\end{figure}

We also study the interface structures at specific inclinations.
Figure~\ref{fig:simu32} shows how the interface structures vary with the inclination angle $\phi_1$ with zero $\phi_2$.
As $\phi_1$ increases, the interfacial energy firstly decreases almost linearly to the global minimum at $\phi_1=0$, and then increases symmetrically about $\phi_1=0$, as shown in Fig.~\ref{fig:simu32}(a).
The misorientation tilt angle $\theta_1$ increases linearly as $\phi_1$ increases, while $\theta_2$ and $\theta_3$ maintain unchanged, as shown in Fig.~\ref{fig:simu32}(b), which indicates that the change of inclination $\phi_1$ around the $X$-axis only leads to the change of misorientation around the same $X$-axis.
Figure~\ref{fig:simu32}(c) shows the changes of the disconnection densities as inclination $\phi_1$ varies.
The curves for the $(\bm{b}_1,h_1)$ and $(\bm{b}_2,h_2)$ disconnection densities are exactly overlapped, which firstly decrease to a minimum at $\phi_1=0$ and then increase.
The curves for the $(\bm{b}_3,h_3)$ and $(\bm{b}_4,h_4)$ disconnection densities are symmetric with each other about $\phi_1=0$, and both firstly decrease to zero at, respectively, $\phi_1=2^{\circ}$ and  $\phi_1=-2^{\circ}$, and then increase.
It suggests the trend of the interfacial energy change is an overall result of the disconnection network.

\begin{figure}[htbp]
\centering
\subfigure[]{  \includegraphics[width=0.32\linewidth]{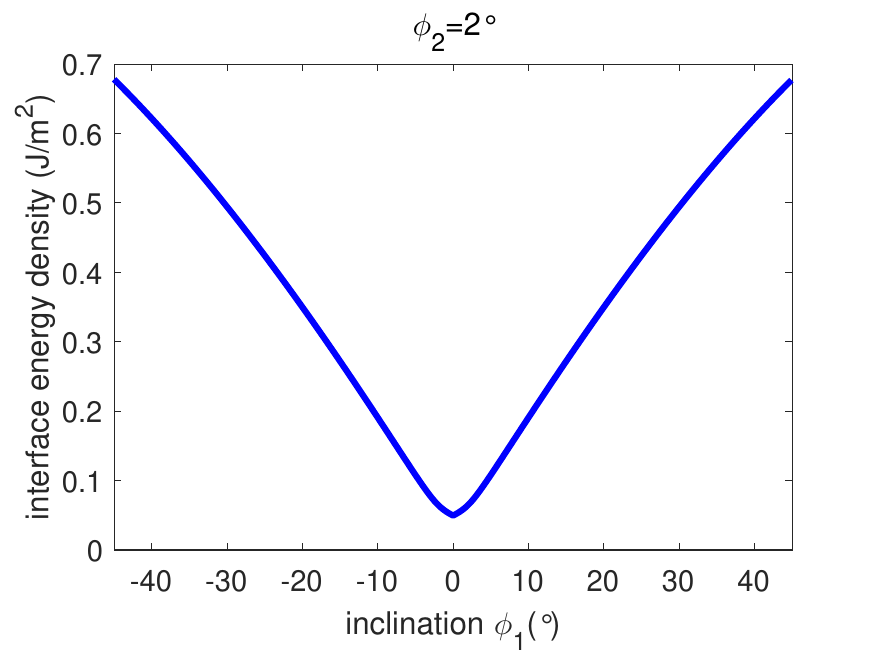}}
\subfigure[]{  \includegraphics[width=0.32\linewidth]{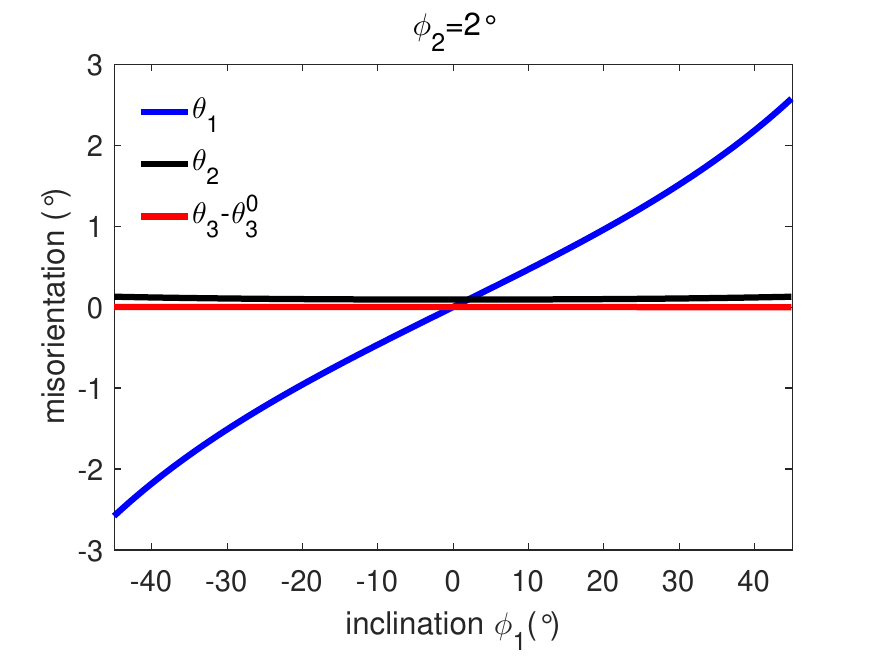}}
\subfigure[]{  \includegraphics[width=0.32\linewidth]{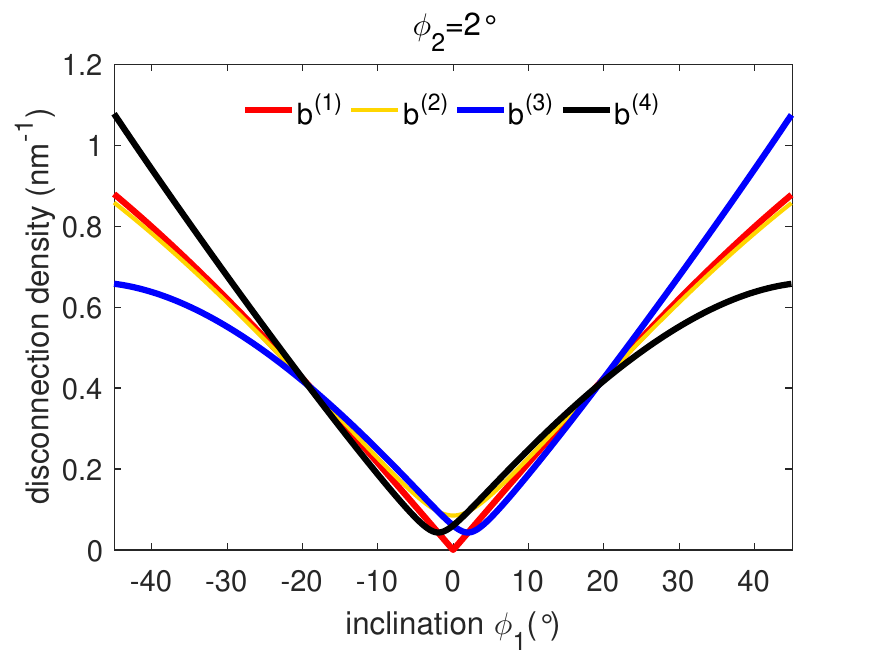}}
\subfigure[]{ \includegraphics[width=0.32\linewidth]{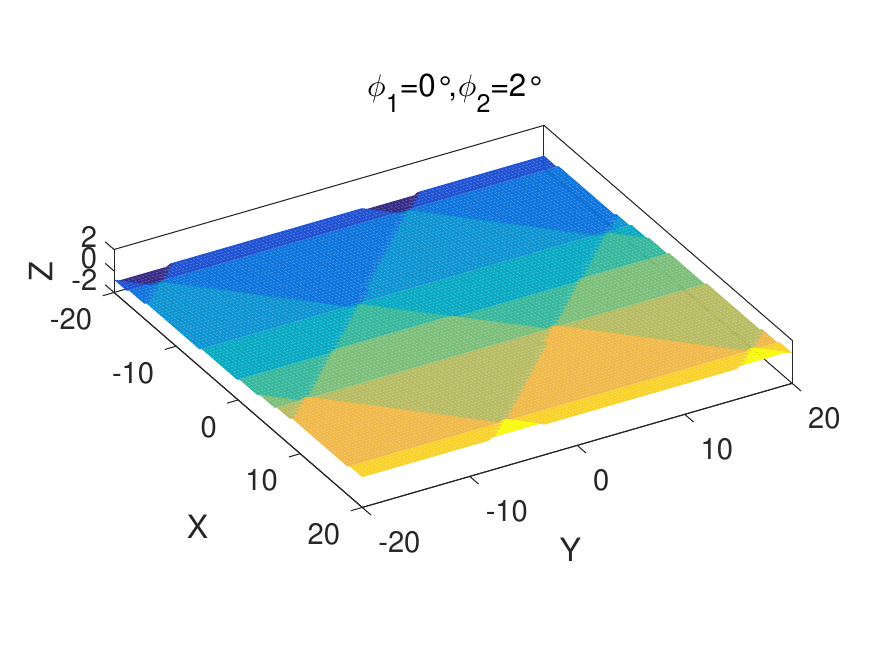}}
\subfigure[]{ \includegraphics[width=0.32\linewidth]{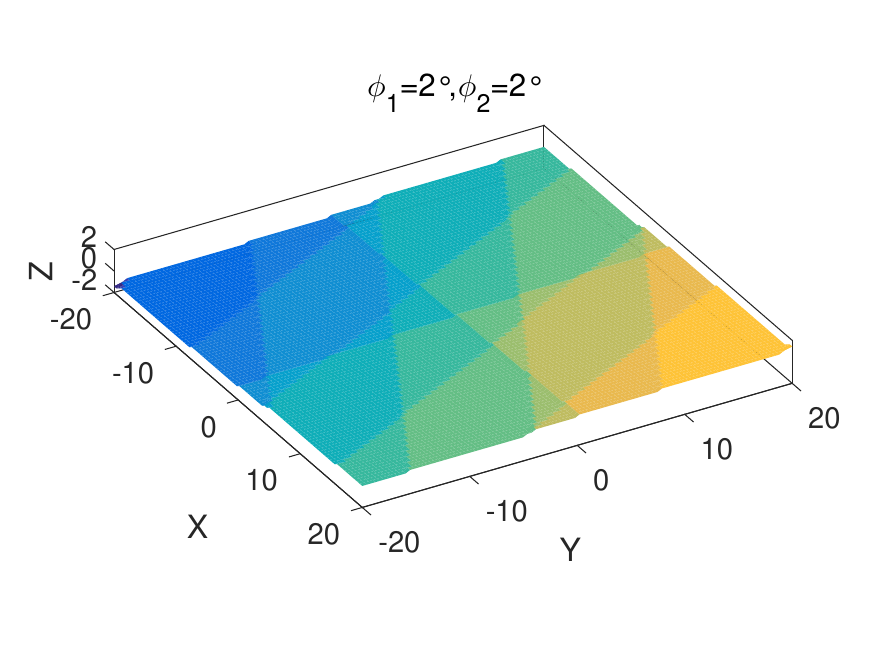}}
\subfigure[]{ \includegraphics[width=0.32\linewidth]{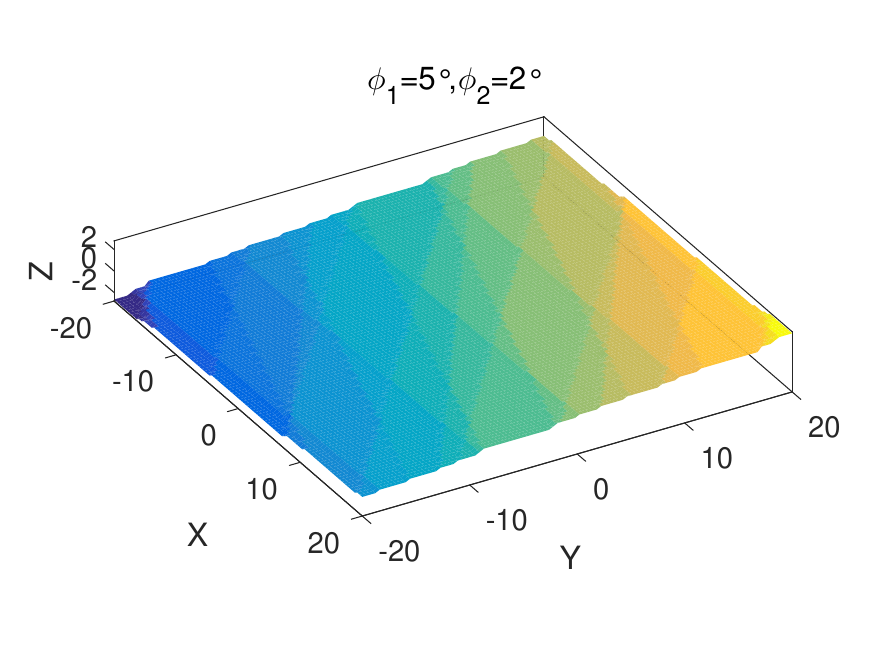}}
\subfigure[]{ \includegraphics[width=0.32\linewidth]{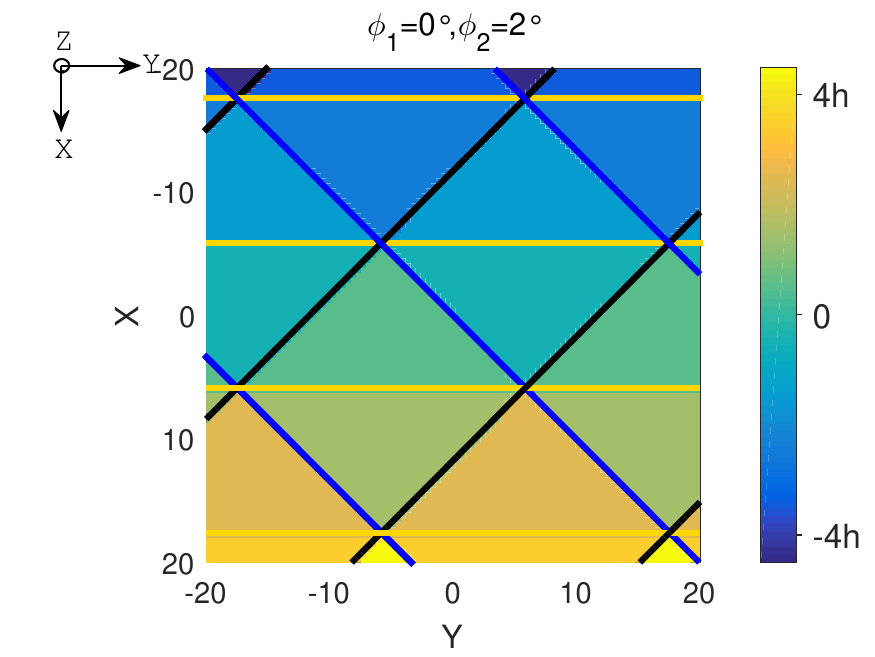}}
\subfigure[]{ \includegraphics[width=0.32\linewidth]{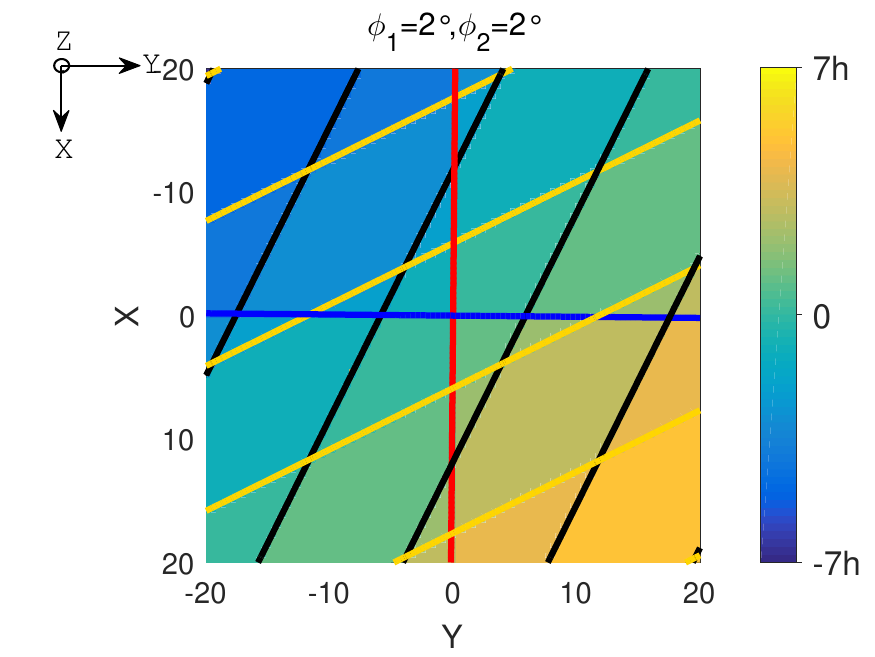}}
\subfigure[]{ \includegraphics[width=0.32\linewidth]{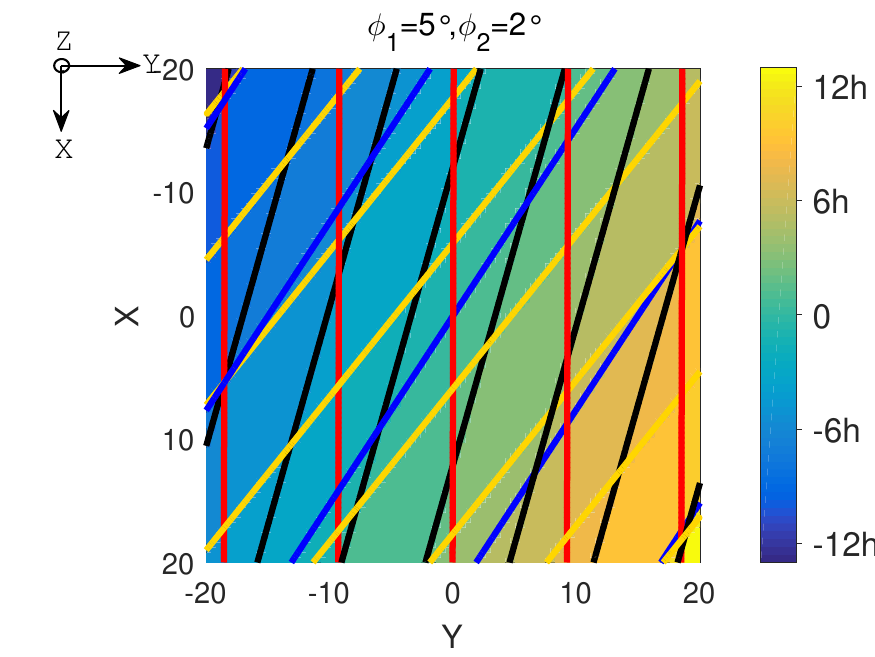}}
\caption{The interface structure of the Au/Pd (001) twist-misfit interface of the CSL $\sigma_{\rm Au} : \sigma_{\rm Pd} = 9:10$for $\phi_2=2^{\circ}$, as a function of inclination $\phi_1$: (a) interfacial energy, (b) misorientation angles, (c) disconnection densities; (d-f) oblique view of the interface disconnection structure and height/step profile;
 (g-i) plan views of the interface structures along the $Z$-direction (normal to the reference interface), where where the red, yellow, blue, and black lines, respectively, denote disconnection type $(\bm{b}_1,h_1)$, $(\bm{b}_2,h_2)$, $(\bm{b}_3,h_3)$, and $(\bm{b}_4,h_4)$, and the shading between the lines indicates the height profile,
at inclinations $\phi_1=0$, $2^{\circ}$, and $5^{\circ}$, and $\phi_2=2^{\circ}$.}
\label{fig:simu33}
\end{figure}

Figures~\ref{fig:simu32}(d)-(f) and (g)-(i) are, respectively, the oblique and plan views of interface disconnection networks at inclination angles $\phi_1=0$, $2^{\circ}$, and $5^{\circ}$, and $\phi_2=0$.
When $\phi_1=\phi_2=0$, there are four modes of disconnections: the $(\bm{b}_1,h_1)$ and $(\bm{b}_2,h_2)$  disconnections  parallel to the $Y$-axis, and the $(\bm{b}_3,h_3)$ and $(\bm{b}_4,h_4)$ disconnections parallel to the $X$-axis.
The dislocation characters of all the four disconnections are edge dislocations.
Each of these parallel disconnections have equal densities and opposite step height, and therefore the net step height of $(\bm{b}_1,h_1)$ and $(\bm{b}_2,h_2)$ disconnections is zero in the $Y$-direction, and the net step height of $(\bm{b}_3,h_3)$ and $(\bm{b}_4,h_4)$ disconnections is zero in the $X$-direction, which constitutes a flat interface with zero inclination.
While the interface is inclined with $\phi_1=2^{\circ}$ or $\phi_1=5^{\circ}$, the $(\bm{b}_1,h_1)$ and $(\bm{b}_2,h_2)$ disconnection lines are rotated to have the mixed dislocation characters with both edge and screw components, as shown in Figs.~\ref{fig:simu32}(e), (f), (h), and (i).
The screw components of $(\bm{b}_1,h_1)$ and $(\bm{b}_2,h_2)$ disconnections, which inclines the interface around the $Y$-axis, cancel out each other, remaining their edge components are effective for inclining the interface around the $X$-axis.
Meanwhile, the $(\bm{b}_3,h_3)$ and $(\bm{b}_4,h_4)$ disconnection lines are always parallel to the $X$-axis, and also incline the interface around the $X$-axis.
On the other hand, as the inclination angle $\phi_1$ increases from $\phi_1=0\rightarrow2^{\circ}\rightarrow5^{\circ}$, the densities of disconnections $(\bm{b}_1,h_1)$, $(\bm{b}_2,h_2)$, and $(\bm{b}_4,h_4)$ increases, and the $(\bm{b}_3,h_3)$ density firstly decreases as $\phi_1=0\rightarrow2^{\circ}$, and then increases as $\phi_1=2^{\circ}\rightarrow5^{\circ}$, consistent to the disconnection density variations in  Fig.~\ref{fig:simu32}(c).

Figure~\ref{fig:simu33} shows the interface structures as functions of the inclination angle $\phi_1$ at fixed $\phi_2=2^{\circ}$.
Comparing to the cases of fixed  $\phi_2=0$ in Fig.~\ref{fig:simu32}, the interfacial energy profile has a shaper corner at the energy minimum attained $\phi_1=0$; and the misorientations $\theta_2$ keeps a non-zero value as the inclination $\phi_1$ varies, which corresponds to the non-zero inclination $\phi_2=2^{\circ}$.
The densities of  $(\bm{b}_1,h_1)$ and $(\bm{b}_2,h_2)$ disconnections are not identical any more, but both attain the  minimum densities at $\phi_1=0$, where $(\bm{b}_1,h_1)$ disconnection density has a zero minimum and $(\bm{b}_2,h_2)$ disconnection density minimum is non-zero.
Densities of $(\bm{b}_3,h_3)$ and $(\bm{b}_4,h_4)$ disconnections do not attain zero as the minimum values, as compared to the $\phi_2=0$ in Fig.~\ref{fig:simu32}(c).
Figures~\ref{fig:simu33}(d)-(f) and (g)-(i) are, respectively, the oblique and plan views interface disconnection networks with inclination angles $\phi_1=0$, $2^{\circ}$, and $5^{\circ}$.
When the inclinations are $\phi_1=0$ and $\phi_2=2^{\circ}$, there are three modes of disconnections in the disconnection network due to the zero density of $(\bm{b}_1,h_1)$ disconnection;
the $(\bm{b}_3,h_3)$ and $(\bm{b}_4,h_4)$ disconnections are rotated to the mixed dislocation as shown in Figs.~\ref{fig:simu33}(d) and (g), in which the screw components of $(\bm{b}_3,h_3)$ and $(\bm{b}_4,h_4)$ disconnections cancel out each other, remaining the edge components to incline the interface around the $Y$-axis.
Meanwhile, $(\bm{b}_2,h_2)$ disconnections are parallel to the $Y$-axis, which also incline the interface around the $Y$-axis.
It can be observed that the disconnection network in Figs.~\ref{fig:simu33}(d), (g) and Figs.~\ref{fig:simu32}(e), (h) are the same if the $X$ and $Y$ axes are exchanged, because one corresponds to the inclination around $X$-axis and the other corresponds to the inclination around $Y$-axis.
As the inclination $\phi_1$ increases from $\phi_1=0^{\circ}\rightarrow2^{\circ}\rightarrow5^{\circ}$, the densities of $(\bm{b}_1,h_1)$, $(\bm{b}_2,h_2)$, $(\bm{b}_4,h_4)$  disconnections are increased, and the density of $(\bm{b}_3,h_3)$ disconnection first decreases and then increases, which are consistent to the disconnection density variations in  Fig.~\ref{fig:simu33}(c).
The disconnection lines of $(\bm{b}_2,h_2)$, $(\bm{b}_3,h_3)$, and $(\bm{b}_4,h_4)$ have counterclockwise rotation and $(\bm{b}_1,h_1)$ disconnection lines keep in parallel to $X$-axis, both contributing to the increasing inclination around the $X$-axis.
In this process, for each mode of disconnections, the overall inclined height, which can be calculated by the reciprocal vector multiplying the step height, is increased.

\subsubsection{Interfaces with fixed misorientations}

\begin{figure}[htbp]
\centering
\subfigure[]{  \includegraphics[width=0.32\linewidth]{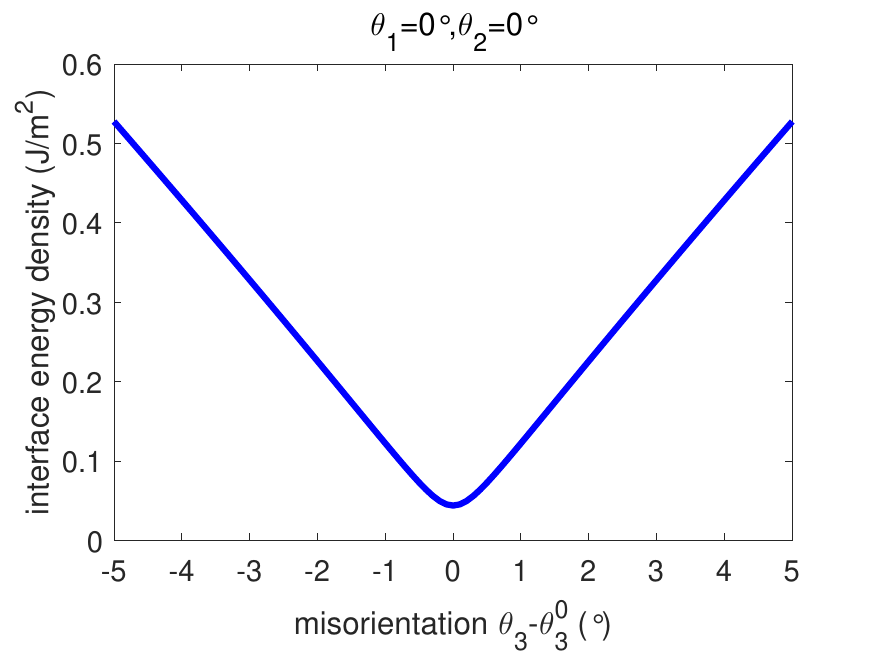}}
\subfigure[]{  \includegraphics[width=0.32\linewidth]{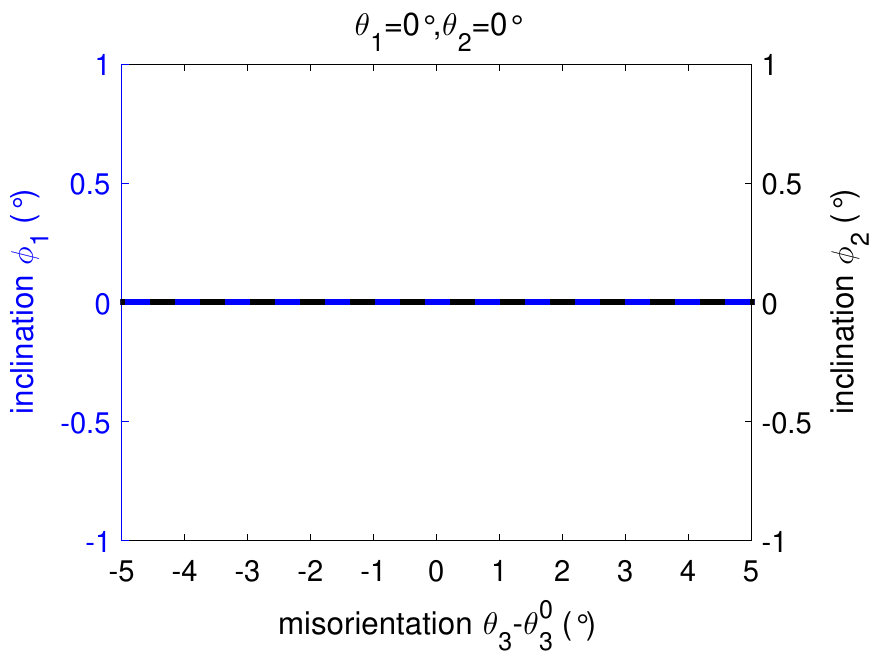}}
\subfigure[]{  \includegraphics[width=0.32\linewidth]{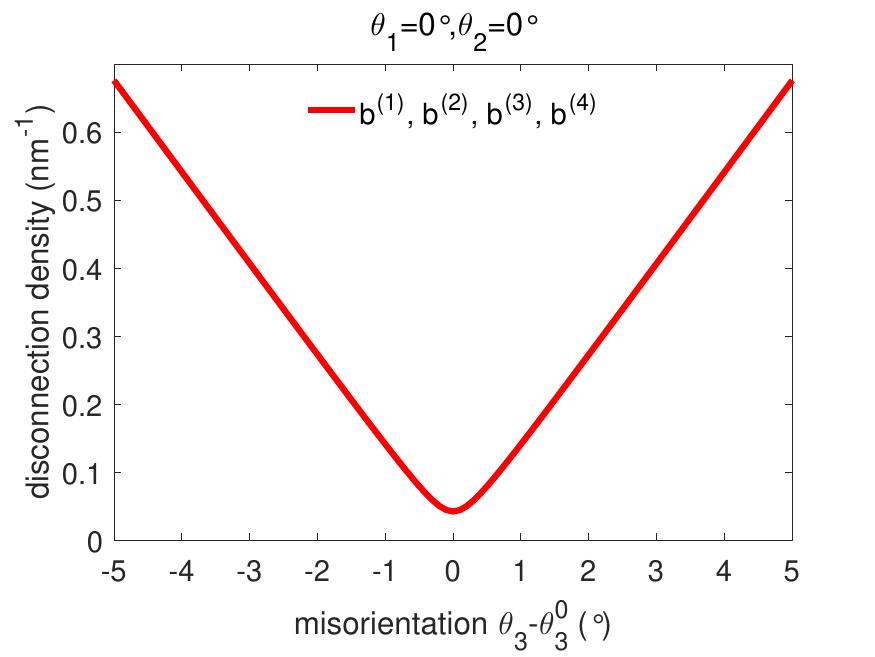}}
\subfigure[]{ \includegraphics[width=0.32\linewidth]{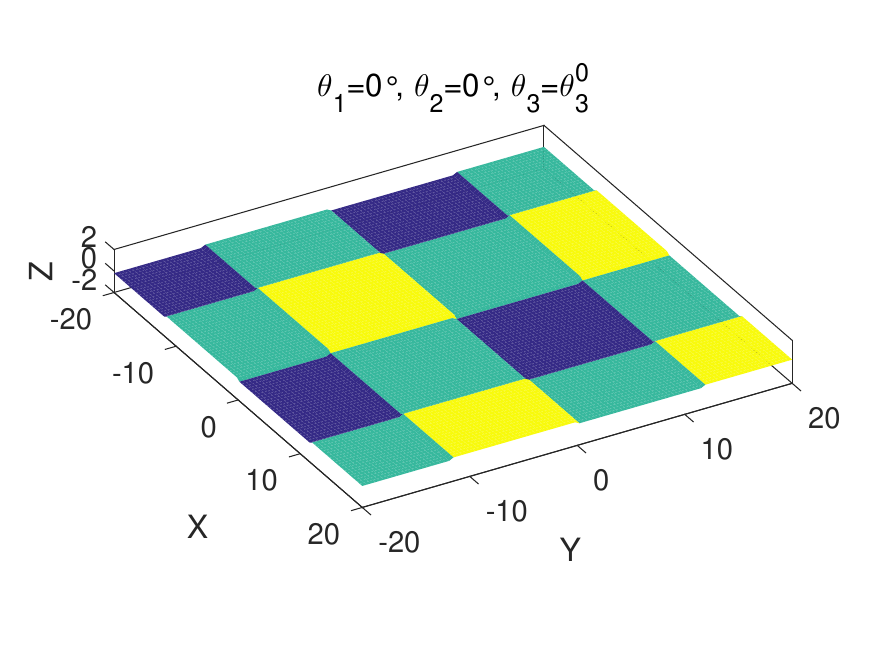}}
\subfigure[]{ \includegraphics[width=0.32\linewidth]{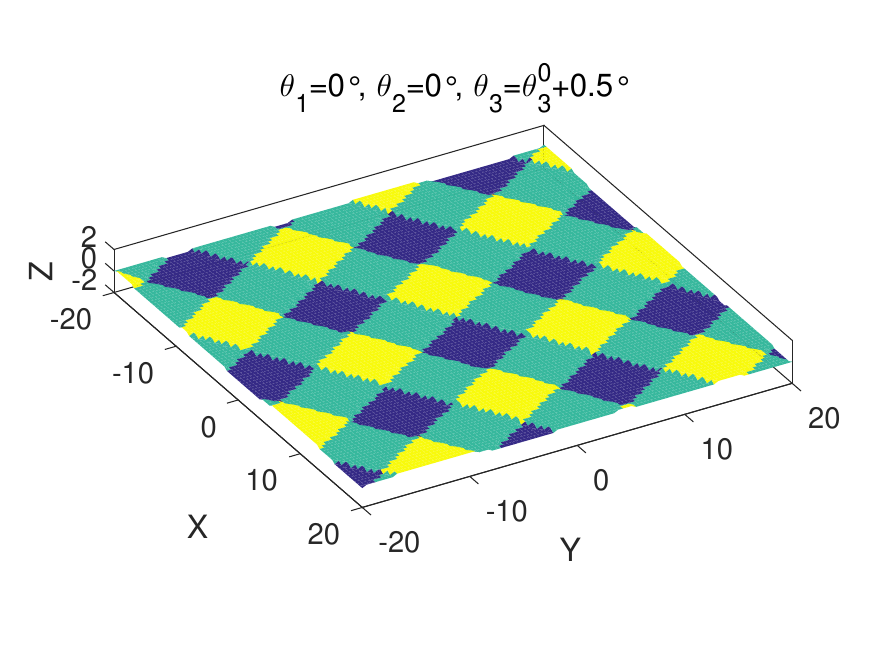}}
\subfigure[]{ \includegraphics[width=0.32\linewidth]{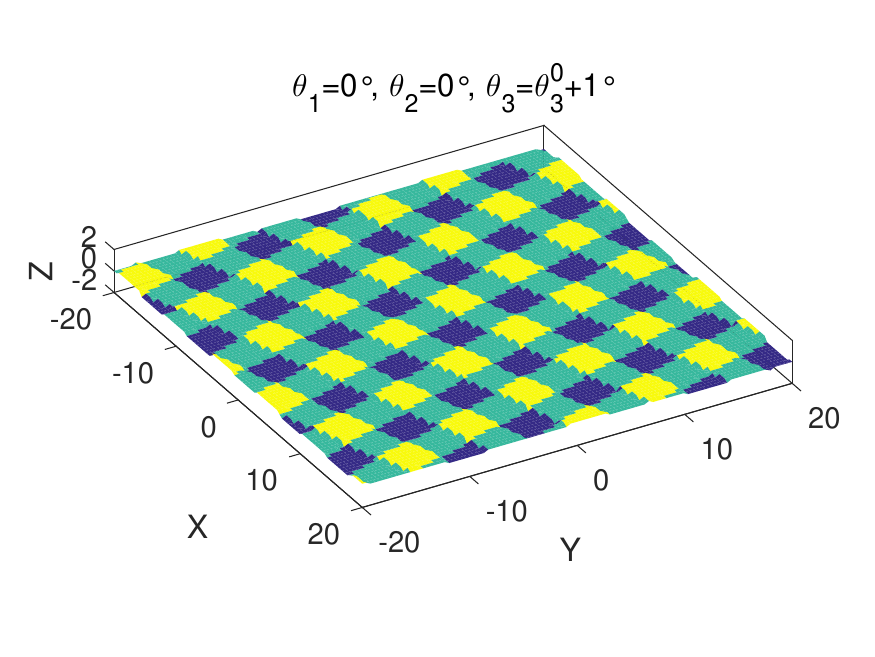}}
\subfigure[]{ \includegraphics[width=0.32\linewidth]{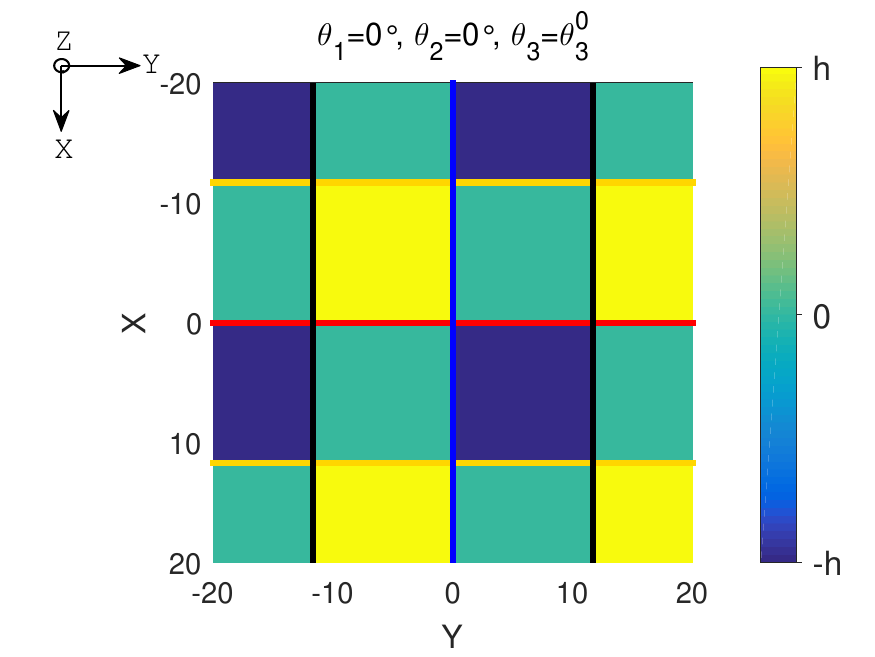}}
\subfigure[]{ \includegraphics[width=0.32\linewidth]{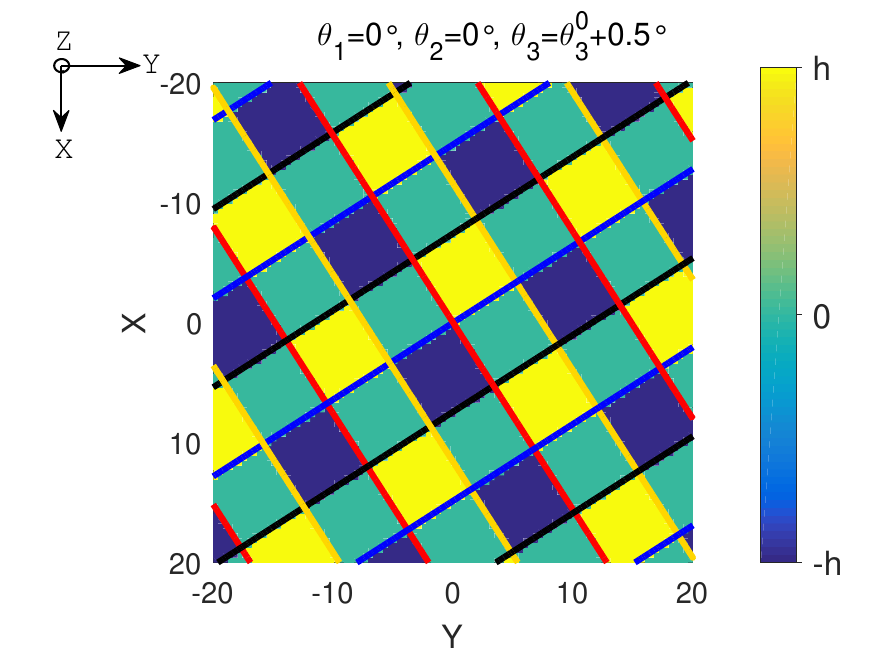}}
\subfigure[]{ \includegraphics[width=0.32\linewidth]{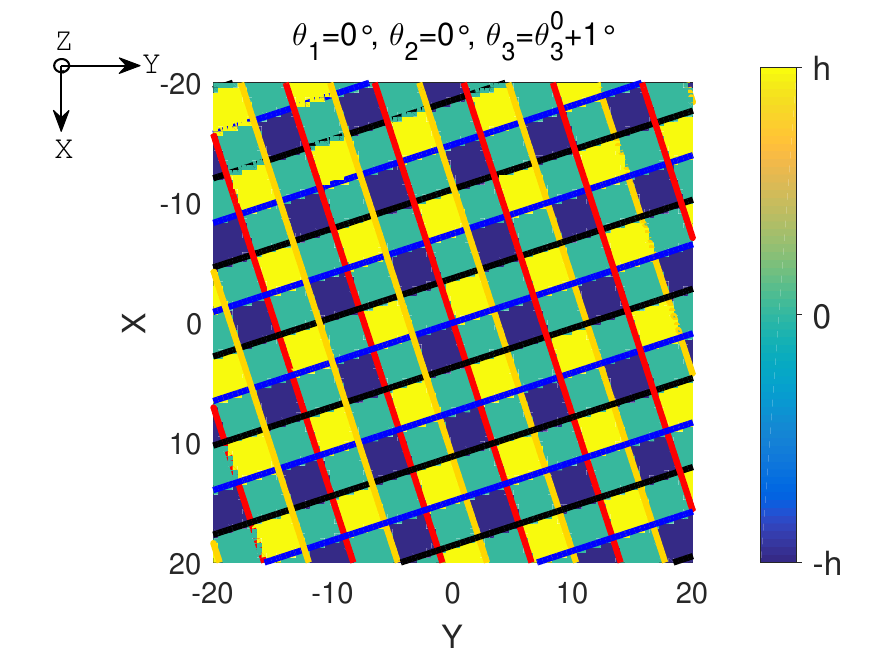}}
\caption{The interface structure of the Au/Pd (001) twist-misfit interface of the CSL $\sigma_{\rm Au} : \sigma_{\rm Pd} = 9 : 10$, as a function of the twist misorientation $\theta_3$: (a) interfacial energy, (b) inclination angles, (c) disconnection densities; (d-f) oblique view of the interface disconnection structure and height/step profile;
 (g-i) plan views of the interface structures along the $Z$-direction (normal to the reference interface), where the red, yellow, blue and black lines respectively denote disconnection type $(\bm{b}_1,h_1)$, $(\bm{b}_2,h_2)$, $(\bm{b}_3,h_3)$, and $(\bm{b}_4,h_4)$, and the shading between the lines indicates the height profile,
for misorientation $\theta_3=\theta_3^0$, $\theta_3^0+0.5^{\circ}$, and $\theta_3^0+1^{\circ}$, $\theta_1=0$, and $\theta_2=0$.}
\label{fig:simu36}
\end{figure}

We now predict the interfacial energy, inclination, and disconnection densities as the misorientation varies.
As the tilt misorientation $\theta_1$ varies, these interface structure properties are similar to the cases where the inclination angle $\phi_1$ varies, as shown in Fig.~\ref{fig:simu32}, which is due to the observed linear relation between the misorientation angle $\theta_1$ and inclination angle $\phi_1$ in Fig.~\ref{fig:simu32}(b).
Similarly, the cases where the tilt misorientation $\theta_2$ varies have the similar trend to the change of inclination angle $\phi_2$.

As misorientation twist angle $\theta_3$ varies, Figure~\ref{fig:simu36} shows how the interface structures vary, with $\theta_1$ and $\theta_2$ fixed as zeros.
The interfacial energy attains the minimum values at $\theta_3=\theta_3^0$, and its profile is symmetric about $\theta_3-\theta_3^0=0$, as shown in Fig.~\ref{fig:simu36}(a).
Figure~\ref{fig:simu36}(b) shows that the inclinations $\phi_1$ and $\phi_2$ do not change as misorientation $\theta_3$ varies, which indicates that the twist misorientation around $Z$-axis affects neither the inclination $\phi_1$ around $X$-axis nor $\phi_2$ around $Y$-axis.
All the densities of disconnections are identical as shown in Fig.~\ref{fig:simu36}(c), and attain the minimum value at $\theta_3=\theta_3^0$.
Figures~\ref{fig:simu36}(d)-(f) and (g)-(i), respectively, show the oblique and plan views of the interfacial disconnection networks with three misorientation angles $\theta_3=\theta_3^0$, $\theta_3^0+0.5^{\circ}$, and $\theta_3^0+1^{\circ}$.
As the twist angle $\theta_3-\theta_3^0$ is increased from $\theta_3-\theta_3^0=0\rightarrow 0.5^{\circ}\rightarrow 1^{\circ}$, the interfacial disconnection network keeps the same pattern with a clockwise rotation around $Z$-axis and increasing disconnection densities of all four modes.

\section{Conclusions and discussion}\label{sec:conclusion}

In this paper, we have developed a continuum model to predict the hetero-interface structures in terms of disconnection network, the interface inclination, and misorientation angles. The model is based on the interfacial energy minimization with respect to all possible disconnection modes, subject to the constraint of the generalized Frank-Bilby equation~\cite{Tao2023}.
The generalized Frank-Bilby equation describes the requirement that the Burgers vector and step height contents, characterizing disconnection distribution, must accommodate the interface geometry and morphology. Numerical method for solving this constrained minimization problem is proposed by using the penalty method and gradient minimization. This continuum model represents a generalization of the methods of predicting dislocation structures of low-angle grain boundaries~\cite{Zhang2017,Qin2022} and semicoherent interfaces~\cite{Zhang2021110277}.

{ The interfacial energy due to local energy of dislocation networks used in this paper is based on isotropic elasticity \cite{Zhu2014175,Zhang2017}, and has been shown being able to predict the dislocation network structures of low angle grain boundaries accurately \cite{Zhang2017}. More accurate interfacial energy due to equilibrium dislocation networks can be calculated using anisotropic elasticity following the methods in \cite{Dai2013} for low angle grain boundaries and for dislocation networks of hetero-interfaces \cite{Demkowicz2013}. Since the anisotropic elasticity problem has to be solved for each fixed solution of the generalized Frank-Bilby equation, direct incorporation of these anisotropic elasticity approaches in our continuum will be quite computationally expensive.
Recalling that this paper focuses on the disconnection structure of hetero-interfaces, we do not believe there will be significant changes in the disconnection structure with anisotropic elasticity based on the available results on low angle grain boundaries. These can be explored in the future work.

The classical geometrical energy parameters $P$ \cite{Bollmann1970}, $Q$, and $R$ \cite{Ecob1980} were used in Sangghaleh and Demkowicz \cite{quantizedFB2018} and  parameter R was used in Gordon and Sills \cite{Gordon-Sills2024} for the approximations of the hetero-interface energy. By their definitions, "these parameters were assumed to vary monotonically with the energy in the range of interest, and do not necessarily have to be proportional to the energy" \cite{Bollmann1970,Ecob1980}, as also remarked in \cite{quantizedFB2018,Gordon-Sills2024}. One limitation of these energy parameters is that they cannot give  the $\theta\log\theta$ form of the energy of low-angle grain boundaries based on the Read-Shockley dislocation theory \cite{Sutton1995,Hirth2001,ReadShockley1950}, where $\theta$ is the misorientation angle across the interface.  While the interfacial energy $\gamma$ in our continuum model was derived based discrete dislocation model \cite{Zhu2014175,Zhang2017} and can recover the Read-Shockley low-angle grain boundary energy.
Also note that dislocation reaction in the dislocation network on a hetero-interface can be further incorporated in our continuum model following the approach in \cite{Zhang2021110277}.
}

Finally, using our continuum model, we perform numerical simulations of the equilibrium configurations of three different hetero-interfaces: the orthorhombic/monoclinic interface in zirconia, Au/Pd (001) twist-misfit interfaces, and $\alpha$/$\beta$ interface in Ti alloys (given in Appendix).
Comparisons with experiments, atomistic simulations, and theoretical predictions show that our continuum model provides excellent predictions of the hetero-interface structures including disconnection network, inclination angles, and misorientation angles.
We systematically study and analyze the hetero-interface structure properties as functions of the inclination and misorientation angles.
{The energy minimization framework enables the determination of energetically favorable interface structures under conditions of fixed misorientation, fixed inclination, or full relaxation. These cases represent different levels of configurational relaxation permitted by the boundary conditions on the minority phase. For an embedded phase, the interface morphology (inclination) may evolve, whereas the orientation of the phase remains largely constrained. When the phase exists as a particle supported on a substrate, its orientation can adjust freely, while the substrate surface fixes the interface inclination.}
As a result, we find that the disconnection networks are arranged to satisfy the inclination angles $\phi_1$ around the $X$-axis and $\phi_2$ around the $Y$-axis through using the disconnection step character.
The inclination angle increase causes the misorientation angle increase around the same rotation axis
($\phi_1\leftrightarrow \theta_1$ around $X$-axis, $\phi_2\leftrightarrow \theta_2$ around $Y$-axis; see Fig.~\ref{fig:simu12},  Fig.~\ref{fig:simu32}, and Fig.~3 in Supplementary Material.)
The change of misorientation $\theta_3$ causes the rotation of disconnection network around $Z$-axis, as shown in Fig.~\ref{fig:simu15}, Fig.~\ref{fig:simu36}, and Fig.~5 in Supplementary Material.

\begin{figure}[htbp]
\centering
\subfigure[]{  \includegraphics[width=0.32\linewidth]{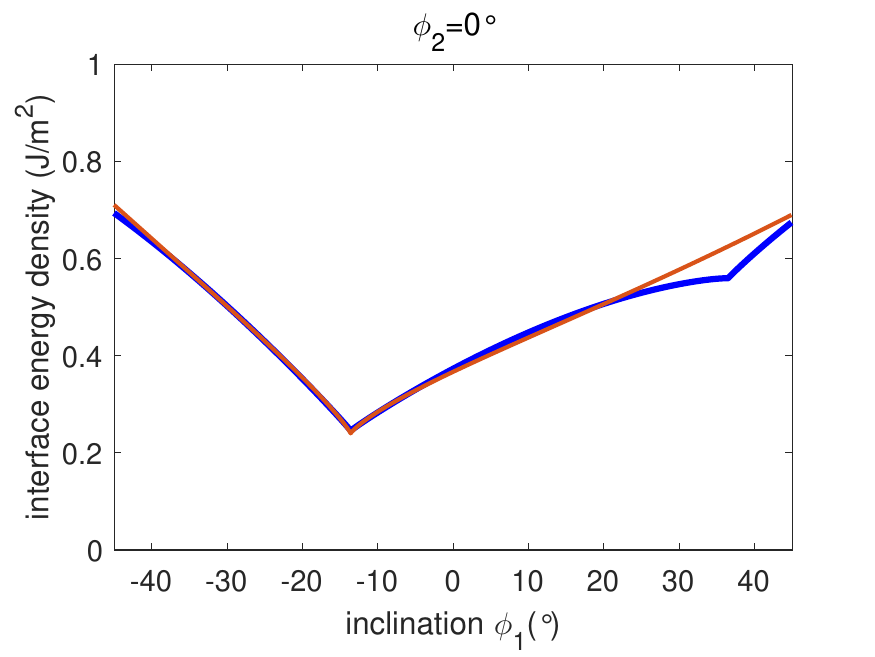}}
\subfigure[]{  \includegraphics[width=0.32\linewidth]{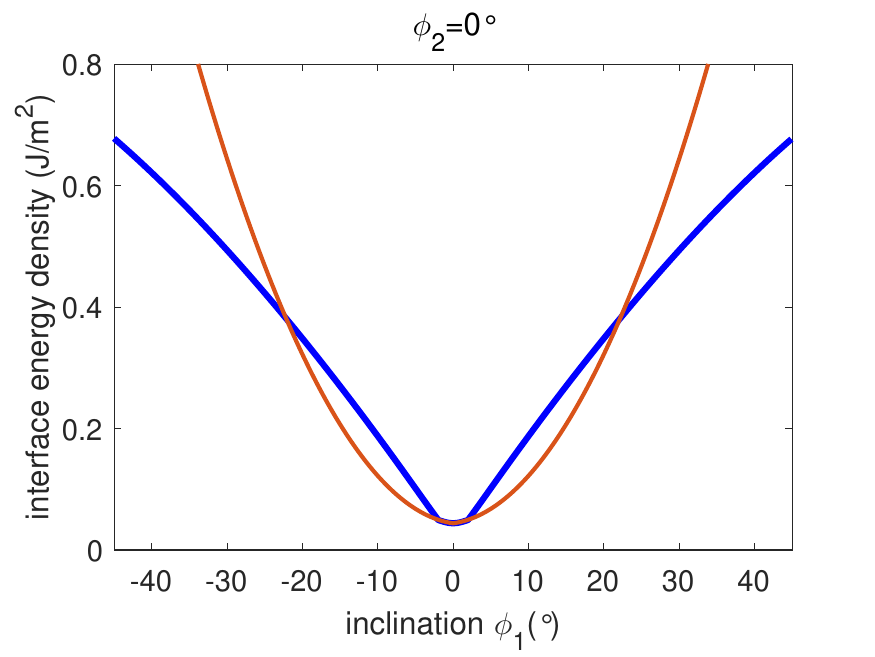}}
\subfigure[]{  \includegraphics[width=0.32\linewidth]{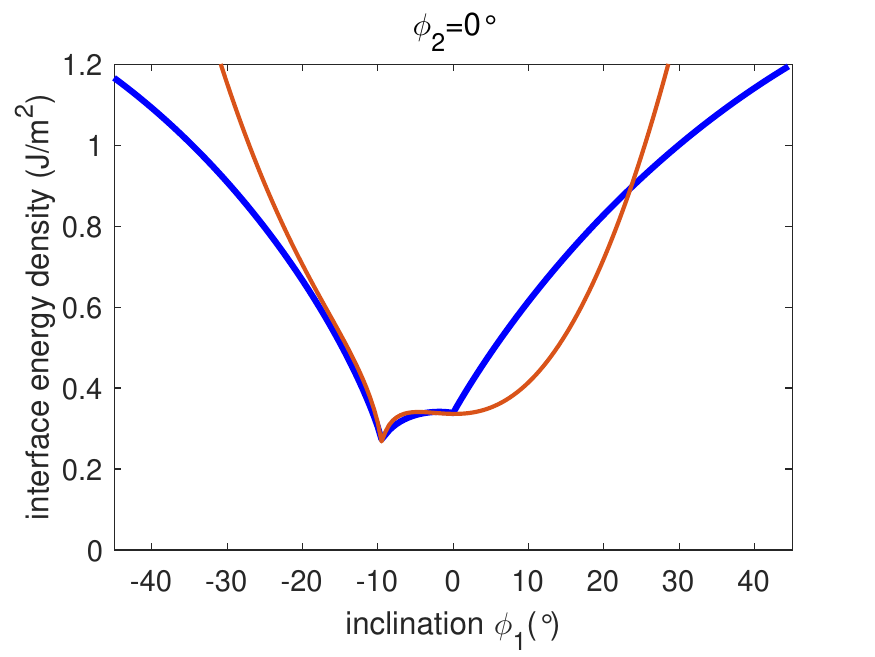}}
\caption{Comparisons of the exact energy (blue curve) and the simplified energy (red curve) of the three examples in this paper.}
\label{fig:fit}
\end{figure}

The interfacial energy can be designed as a simplified function of the overall inclination angle $\phi$ which is angle between the interface and the reference plane as a function of $\phi_1$ and $\phi_2$,
\begin{equation}
\gamma
=
\gamma_0+c_1(\phi-\phi_0)+c_2(\phi-\phi_0)^2-c_3\mu|\phi-\phi_0|\log|\phi-\phi_0|+c_4\frac{\gamma_s}{\beta}|\phi-\phi_0|,
\end{equation}
where $\gamma_0$ is the minimum energy, $\phi_0$ is the inclination angle at energy minimum, $\beta$ is the overall shear-coupling factor, $c_i$ ($i=1,2,3,4$) are the coefficients.
Applying this expression to the three examples (the third example is given in Appendix) in this paper where the inclination $\phi=\phi_1$, we have their corresponding simplified expression as
\begin{align}
&\gamma_1 = 0.24-0.2(\phi_1-\phi_0)+0.16(\phi_1-\phi_0)^2-0.15|\phi_1-\phi_0|\log(|\phi_1-\phi_0|)+0.48|\phi_1-\phi_0|,
\\
&\gamma_2 = 0.044+2(\phi_1-\phi_0)^2+0.1|\phi_1-\phi_0|,
\\
&\gamma_3 = 0.26-(\phi_1-\phi_0)+6(\phi_1-\phi_0)^2-1.46|\phi_1-\phi_0|\log(|\phi_1-\phi_0|)+2.16|\phi_1-\phi_0|.
\end{align}
We examine these simplified energy functions to their corresponding exact energy profiles of the three examples above, as shown in Fig.~\ref{fig:fit}.  These simplified energy functions (red curves) approximate the exact energy (blue curves) well in the region near the energy minimum.

\section*{Acknowledgments}
This work was supported by the Hong Kong Research Grants Council Collaborative Research Fund C1005-19G (Y. Xiang and D.J. Srolovitz) and General Research Fund 16309825 (Y. Xiang), National Natural Science Foundation of China 12201423, 12571464, 12426311, Shenzhen Science and Technology Innovation Program RCYX20231211090222026, JCYJ20241202124209011, Research Team Cultivation Program of Shenzhen University 2023QNT011 and Huayuan Computing Technology Co., Ltd. (L.C. Zhang), and National Natural Science Foundation of China 12401569 and Shanghai Sailing Program, China 24YF2712700 (X.X. Qin).

\section*{Author Contributions}
L.C. Zhang, X.X. Qin, D. J. Srolovtiz and Y. Xiang developed the continuum model.
L.C. Zhang, X.X. Qin and Y. Xiang derived analytical formulations and designed the numerical scheme.
L.C. Zhang, J. Han, D.S. Tao. and D. J. Srolovtiz designed simulation examples and analysed the implications.
L.C. Zhang performed the numerical simulations and prepared all the figures.
L.C. Zhang, J. Han, D. J. Srolovtiz and Y. Xiang wrote the manuscript.

\section*{Competing Interests}
The authors declare no competing interests.

\bibliographystyle{elsarticle-num}
\bibliography{ref}

\end{document}


\maketitle

\section{Supplementary example application: $\alpha$/$\beta$ interface in Ti alloys}
\label{sec:simu_preticipate}

We apply our continuum model to predict the interface structure of the $\alpha$/$\beta$ (HCP/BCC) $(\bar{1}\bar{2}1)_{\beta} \parallel (0\bar{1}10)_{\alpha}$, $[\bar{1}11]_{\beta} \parallel [\bar{2}110]_{\alpha}$ interface that is common in martensitic transformations of Ti alloys.
We set $[101]_{\beta} \parallel [0001]_{\alpha}$ as the $X$-axis with  unit vector $\bm{e}_1$ and $[\bar{1}11]_{\beta} \parallel [\bar{2}110]_{\alpha}$ as the $Y$-axis with unit vector $\bm{e}_2$.
Consider three sets of disconnections on the interface with Burgers vectors and step heights
\begin{align}
&\bm{b}_1 = \frac{\sqrt{3}}{12}a_{\rm Ti}^{\beta} \bm{e}_2 + \left(\frac{\sqrt{6}}{3} a_{\rm Ti}^{\beta} - \frac{\sqrt{3}}{2} a_{\rm Ti}^{\alpha}\right) \bm{e}_3
\quad\text{and}\quad
h_1 = \sqrt{\frac{2}{3}}a_{\rm Ti}^{\beta};
\label{eqn:BV_step7}\\
&\bm{b}_2 = a_{\rm Ti}^{\alpha} \bm{e}_2
\quad\text{and}\quad
h_2 = 0;
\\
&\bm{b}_3 = \frac{1}{2}c_{\rm Ti}^{\alpha} \bm{e}_1
\quad\text{and}\quad
h_3 = 0.
\label{eqn:BV_step9}
\end{align}
The lattice constants of the HCP phase are $a_{\rm Ti}^{\alpha}=0.2943$~nm, $c_{\rm Ti}^{\alpha}=0.468$~nm, and for BCC Ti $a_{\rm Ti}^{\beta}=0.3196$~nm~\cite{Shi20124172,Mills1998295}.
By choosing the $\beta$ phase as the reference lattice, the lattice correspondence parameters are $\Omega_1 = c_{\rm Ti}^{\alpha}/\sqrt{2}a_{\rm Ti}^{\beta}$, $\Omega_2 = a_{\rm Ti}^{\alpha}/a_{\rm Ti}^{\beta}$, and $\Omega_3 = \sqrt{3}a_{\rm Ti}^{\alpha}/\sqrt{2}a_{\rm Ti}^{\beta}$.

\subsection{Equilibrium interface structure}

The minimum value of the interfacial energy occurs at interface inclination angles $\phi_1=-9.45^{\circ}$ and $\phi_2=0$ and misorientation angles $\theta_1=-5.27^{\circ}$, $\theta_2=0$,  and $\theta_3=0$.
The disconnection structure  in three dimensional side-view and two dimensional top-view obtained by using our continuum model  are shown in Supplementary Figs.~\ref{fig:simu20}(a) and (b).
The disconnection structure is a rectangular network that consists of one array of disconnections $(\bm{b}_1,h_1)$, represented by the reciprocal vectors $\bm{N}_1=(0, 0.6292)$~nm$^{-1}$ and the other is with Burgers vector $\bm{b}_3$, represented by the reciprocal vectors $\bm{N}_3=(0.1463, 0)$~nm$^{-1}$.
The $(\bm{b}_1,h_1)$ disconnections  are of edge character lying in the $X$-direction with separations  $D_1=1/{N_1} = 1.5893 $~nm and the $(\bm{b}_3,h_3)$ disconnections are also of edge character lying in the $Y$-direction with inter-dislocation distances  $D_3=1/{N_3} = 6.8353$~nm.
$(\bm{b}_2,h_2)$ disconnections are absent.

The predicted disconnection structure agrees with previous theoretical predictions and experimental observations \cite{Dahmen198263, Aaronson19912857, Aaronson19912873, Mills1998295, Zheng201825}.
The $\bm{b}_3$ disconnections are c-type dislocations in these references.
There is no dislocation with Burgers vector $\bm{b}_2$, i.e., there is no a-type dislocation on this interface.
The misorientation angle $\theta_1=-5.27^{\circ}$ agrees with the theoretical value of $5.26^{\circ}$ of the well-known Pitsch-Schrader Burgers orientation relationship
\cite{Dahmen198263}.
The  inter-disconnection separation, 6.84 nm, agrees with the experimental observation
of $\sim 7$~nm \cite{Mills1998295}.
The inter-disconnection separation of $\sim1.59$ nm in the other directions, corresponding to an iinclination angle $\phi_1=9.45^{\circ}$ is very close to
the experimental observation of 1.4 nm  with a corresponding inclination  $10.32^{\circ}$) in a different alloy with slightly different crystallographic parameters \cite{Zheng201825}.
The inclination angle $\phi_1=-9.45^{\circ}$ is smaller than the theoretical prediction and some experimental observation of $14.85^{\circ}$ \cite{Aaronson19912857, Aaronson19912873}.
Note that experimental observations show that this habit plane is not completely planar
(i.e., with non-uniform disconnection spacing), and the experimental observations are not completely consistent with each other \cite{Zheng201825}.
This may be a result of incompletely equilibrated dislocation structures.

\begin{figure}[htbp]
\centering
\subfigure[]{ \includegraphics[width=0.4\linewidth]{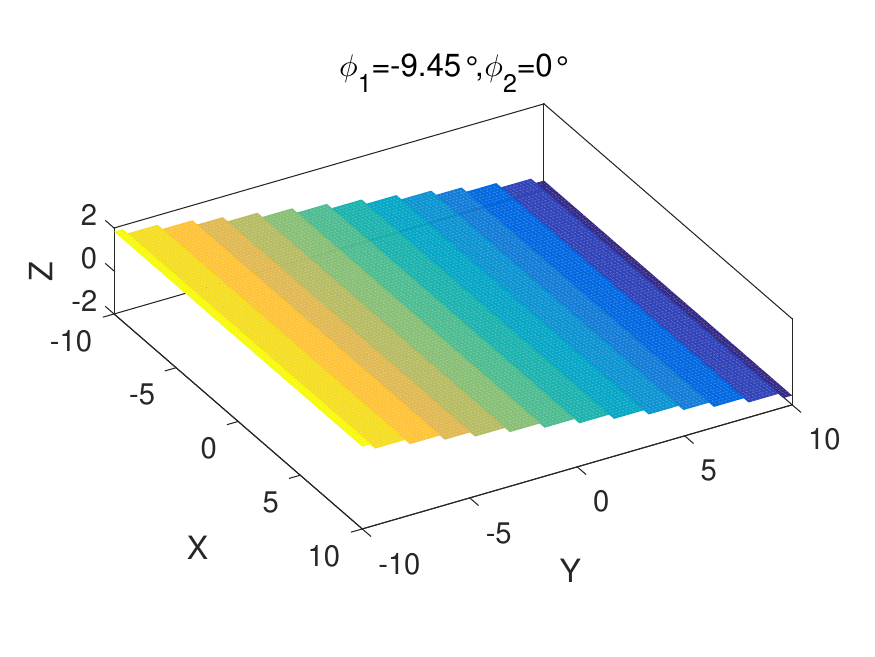}}
\subfigure[]{ \includegraphics[width=0.4\linewidth]{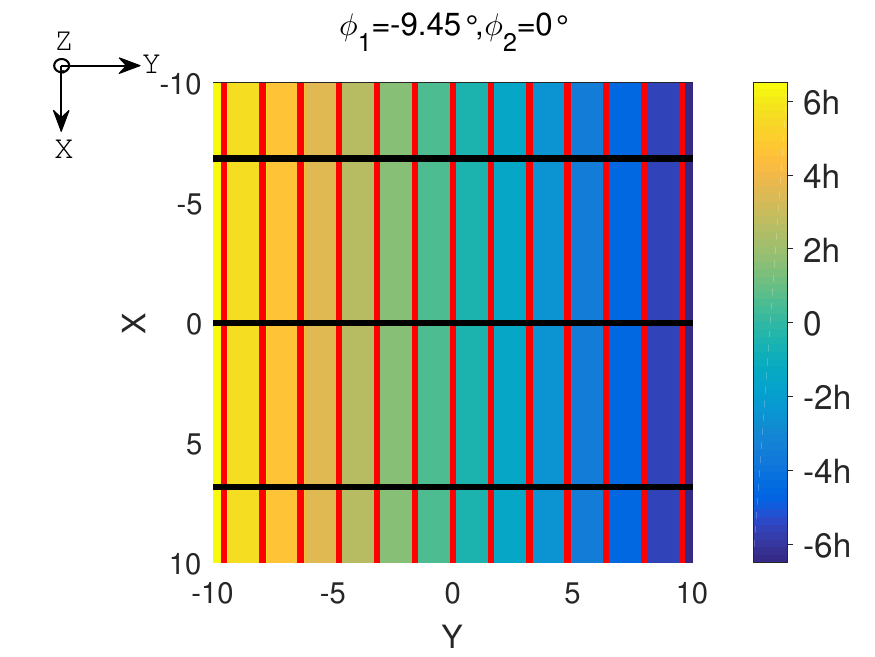}}
\caption{
Equilibrium interface in Ti alloys referenced to the Burgers orientation relationship: $(\bar{1}\bar{2}1)_{\beta} \parallel (0\bar{1}10)_{\alpha}$, $[\bar{1}11]_{\beta} \parallel [\bar{2}11]_{\alpha}$. (a) Oblique view and
(b) plan view of the interface structure along the $Z$-direction (normal to the reference interface, where the red and black lines respectively denote disconnection type $(\bm{b}_1,h_1)$ and $(\bm{b}_3,h_3)$ and the shading between the lines indicates the height profile.
}
    \label{fig:simu20}
\end{figure}

\subsection{Interfaces with fixed inclination angles}\label{sec:simu2_2}

We now investigate interface disconnection structures and misorientations at fixed interface inclination $\phi_1,\phi_2$. The energy gradient flows for determining the disconnection network and misorientation angles are obtained by minimizing the interfacial energy through Eqs. (10)(12) in the main article.

As shown in  Supplementary Fig.~\ref{fig:simu21}, the interfacial energy, misorientation angles and disconnection densities are functions of the interface inclination angles $\phi_1,\phi_2$.
The  global minimum in  the interfacial energy, corresponding to the equilibrium state, is attained at an inclination angle $\phi_1=-9.45^{\circ},\phi_2=0$, as seen in  Supplementary Fig.~\ref{fig:simu21}(a).
Supplementary Figs.~\ref{fig:simu21}(b-d) show how the crystal misorientations across the interface vary with interface inclination ($\phi_1,\phi_2$).
Varying  the inclination angles $\phi_1,\phi_2$ result in only very small changes in the misorientation around$(\theta_1=\theta_1^0,\theta_2,\theta_3)=(-5.27^{\circ},0,0)$.
As  inclination angle $\phi_1$ increases, the misorientation tilt angle $\theta_1$ increases linearly with little change in $\theta_2$ and the misorientation twist angle $\theta_3$ increases when $\phi_2<0$  and decreases when $\phi_2>0$.
As  inclination angle $\phi_2$ increases, the misorientation tilt angles $\theta_2$ increases linearly with little change in $\theta_1$; while the misorientation twist angle $\theta_3$ decreases for $\phi_1<0$ and increases for $\phi_1>0$.
This again reflects the relation between the interface inclination and misorientation angles; i.e.,  that changes in interface inclinations lead to changes in misorientations around the same axis.
The interface inclination $\phi_1$ around the $X$-axis and $\phi_2$ around the $Y$-axis both modify the twist about the $Z$-axis.

\begin{figure}[htbp]
\centering
\subfigure[]{  \includegraphics[width=0.33\linewidth, trim=3cm 9.5cm 3.5cm 8cm, clip]{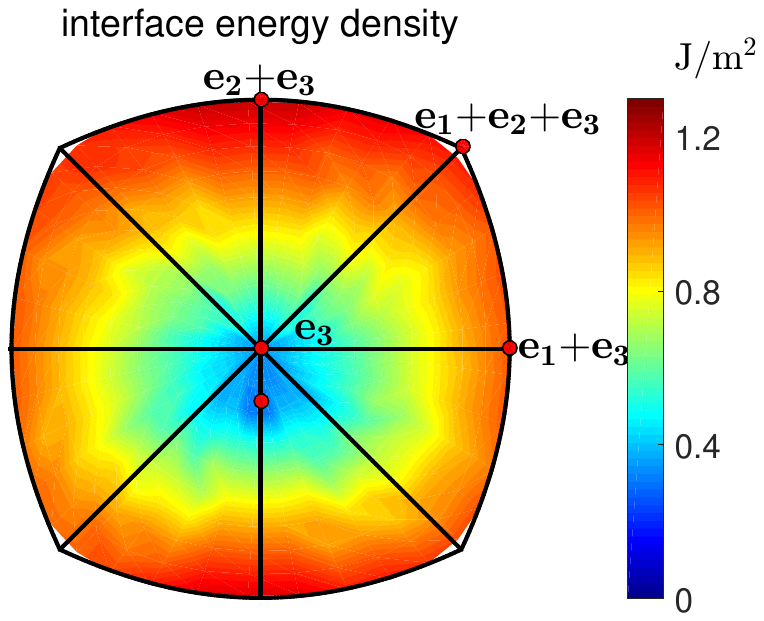}}
\subfigure[]{  \includegraphics[width=0.33\linewidth, trim=3cm 9.5cm 3.5cm 8cm, clip]{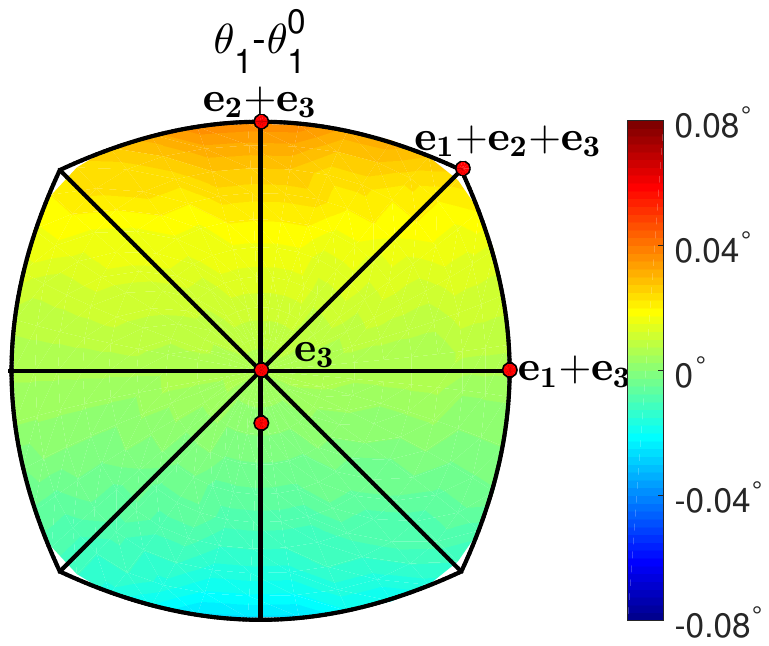}}
\subfigure[]{  \includegraphics[width=0.33\linewidth, trim=3cm 9.5cm 3.5cm 8cm, clip]{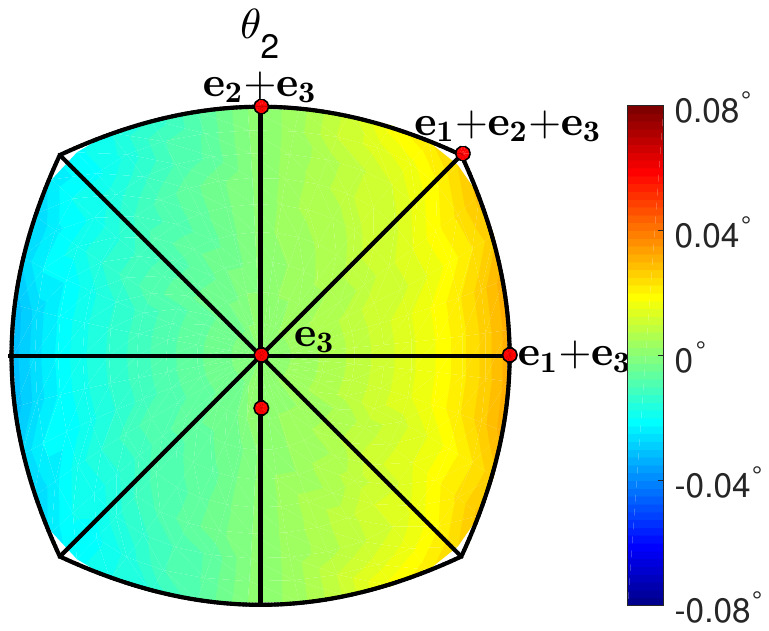}}
\subfigure[]{  \includegraphics[width=0.33\linewidth, trim=3cm 9.5cm 3.5cm 8cm, clip]{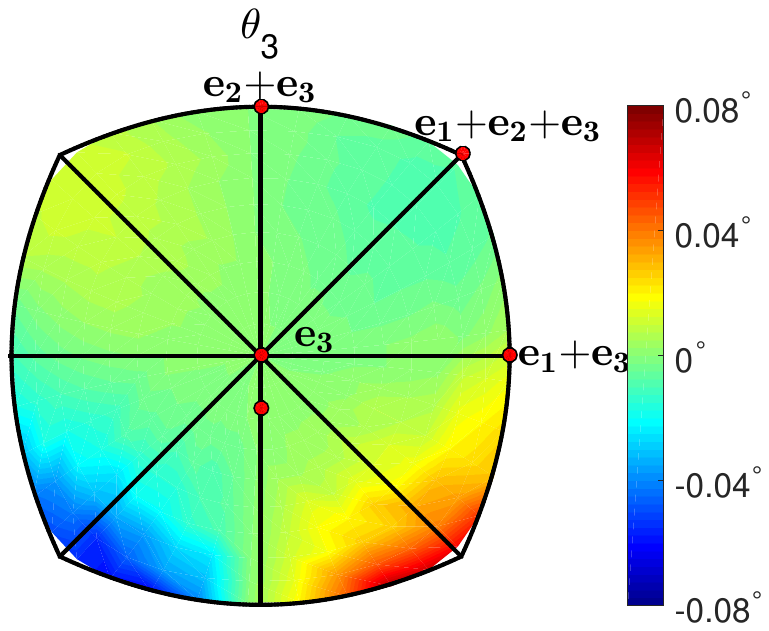}}
\caption{The (a) interfacial energy and (b-d) misorientation angles of $\alpha$-$\beta$ interface in titanium with Burgers OR $(\bar{1}\bar{2}1)_{\beta}||(0\bar{1}10)_{\alpha}$, $[\bar{1}11]_{\beta}||[\bar{2}110]_{\alpha}$, as functions of inclination angles $\phi_1,\phi_2$.
$\phi_1,\phi_2$ are the inclination angles relative to the $X$, $Y$ axes of the tangent plane of the stereographic projection sphere. $\phi_1$ is increasing from bottom to top, and $\phi_2$ is increasing from left to right.
}
\label{fig:simu21}
\end{figure}

\begin{figure}[htbp]
\centering
\subfigure[]{ \includegraphics[width=0.32\linewidth]{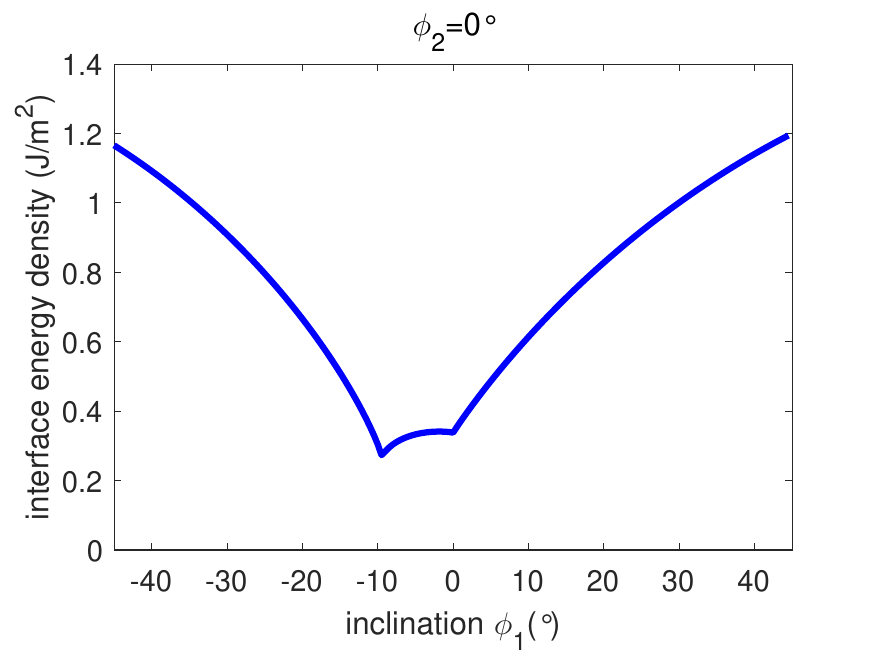}}
\subfigure[]{ \includegraphics[width=0.32\linewidth]{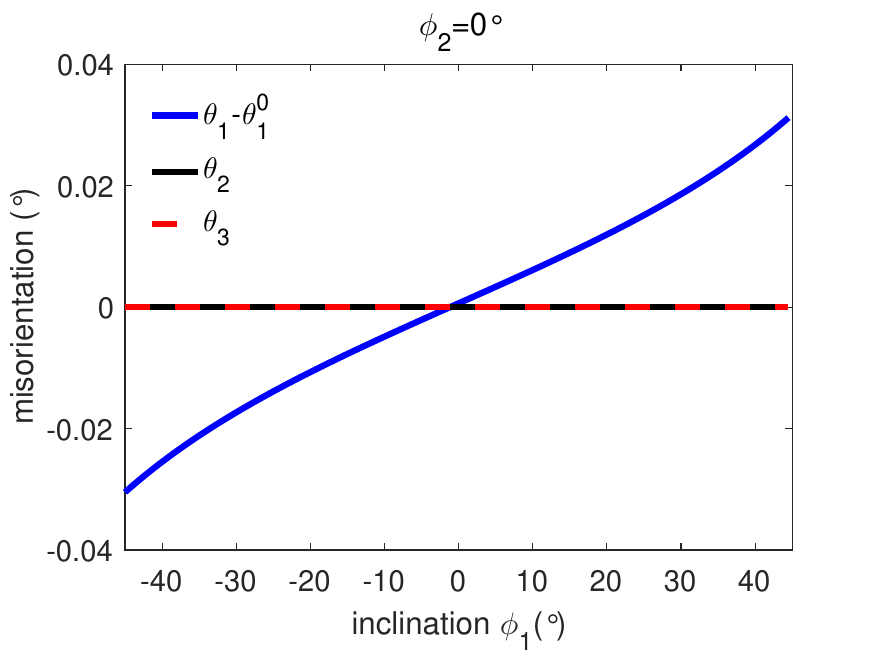}}
\subfigure[]{ \includegraphics[width=0.32\linewidth]{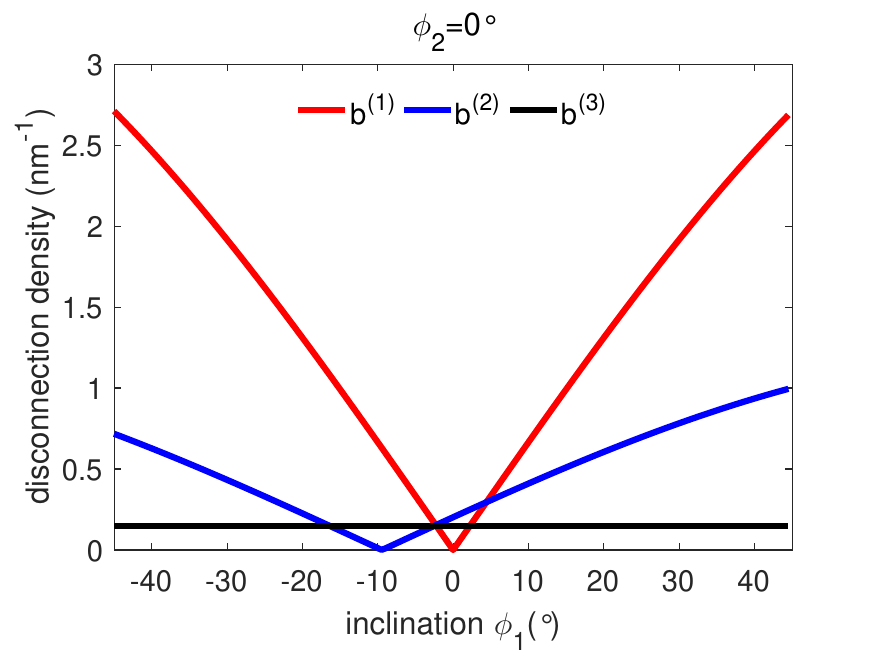}}
\subfigure[]{ \includegraphics[width=0.32\linewidth]{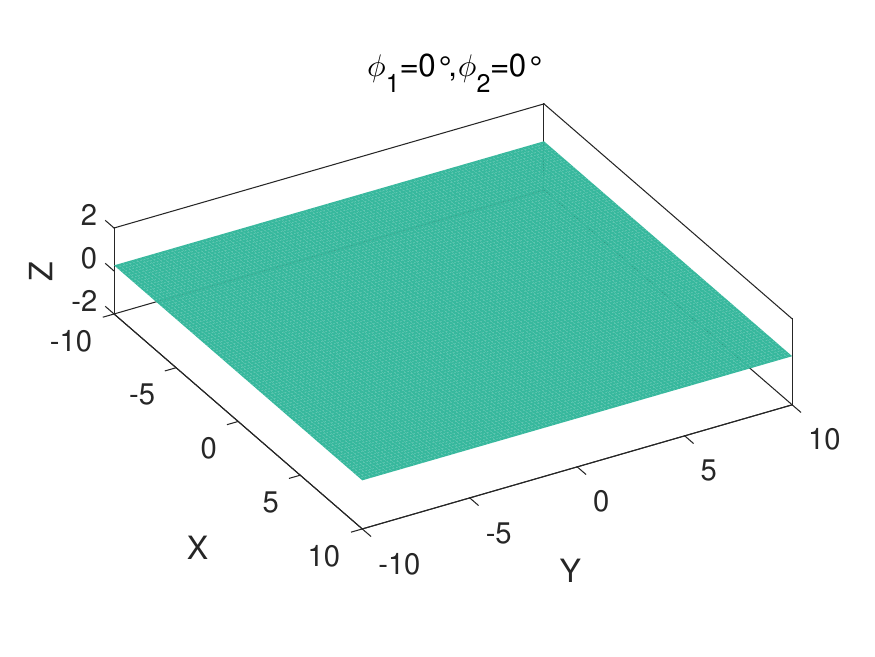}}
\subfigure[]{ \includegraphics[width=0.32\linewidth]{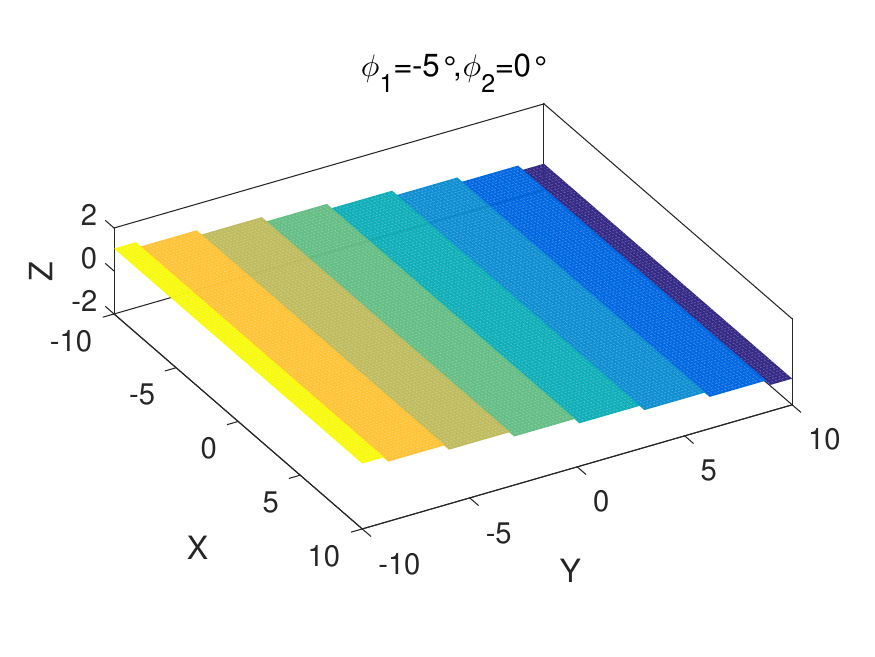}}
\subfigure[]{ \includegraphics[width=0.32\linewidth]{structure_inclination_phi00_ti3d.pdf}}
\subfigure[]{ \includegraphics[width=0.32\linewidth]{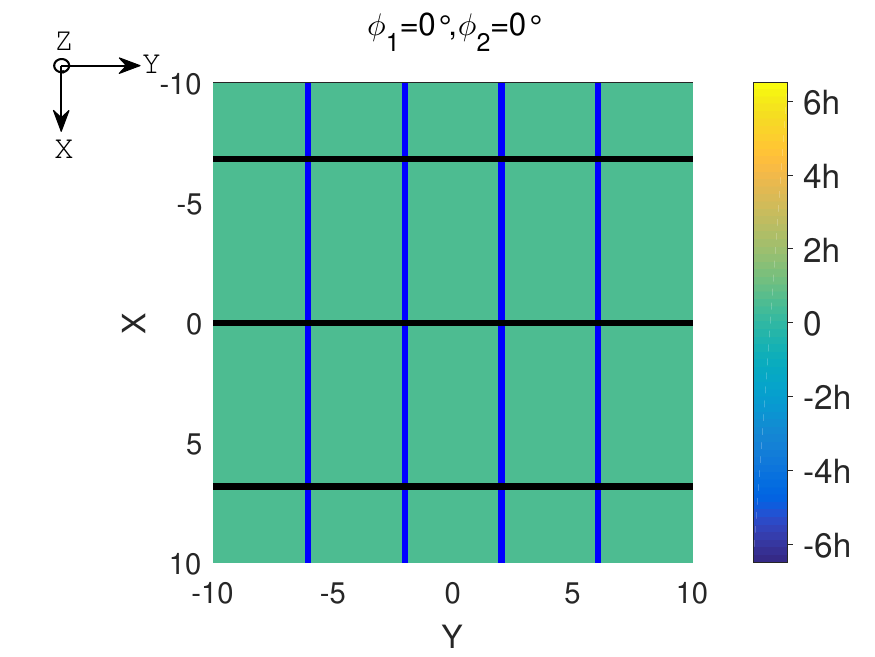}}
\subfigure[]{ \includegraphics[width=0.32\linewidth]{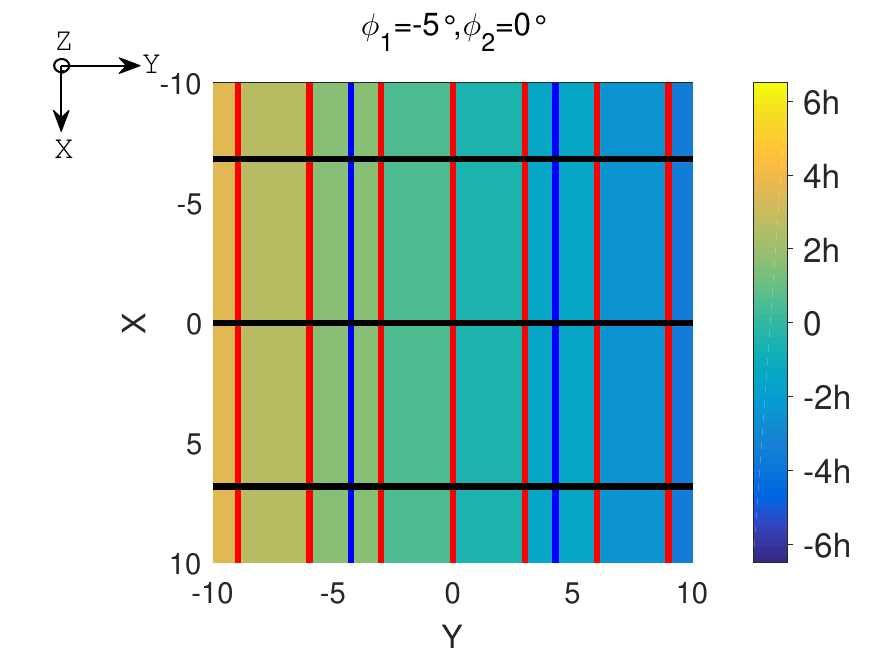}}
\subfigure[]{ \includegraphics[width=0.32\linewidth]{structure_inclination_phi00_ti2d.pdf}}
\caption{$\alpha$/$\beta$ interface in Ti for $\phi_2=0$, as a function of inclination $\phi_1$: (a) interfacial energy, (b) misorientation angles, (c) disconnection densities; (d-f) oblique view of the interface disconnection structure and height/step profile;
(g-i) plan views of the interface structures along the $Z$-direction (normal to the reference interface, where the red, blue and black lines respectively denote disconnection types $(\bm{b}_1,h_1)$, $(\bm{b}_2,h_2)$ and $(\bm{b}_3,h_3)$, and the shading between the lines indicates the height profile),
at interface inclinations $\phi_1=0$, $-5^{\circ}$, and $-9.45^{\circ}$, and $\phi_2=0$.
}
\label{fig:simu22}
\end{figure}

We also examine the interface structures at specific inclinations.
Supplementary Fig.~\ref{fig:simu22} demonstrates how the interface structures vary with interface inclination angle $\phi_1$ at $\phi_2=0$.
As $\phi_1$ increases, the interfacial energy  decreases to a cusped global minimum ($\phi_1=-9.45^{\circ}$) and rises to another cusp at $\phi_1=0$.
As $\phi_1$ varies, the misorientation angle $\theta_1$ slightly increase near $\theta_1^0=-5.27^{\circ}$ while  $(\theta_2,\theta_3)=(0,0)$.
Supplementary Fig.~\ref{fig:simu22}(c) suggests that the energy change comes from the disconnection network.
We can see that when the energy is  minimum  (at $\phi_1=-9.45^{\circ}$), the $(\bm{b}_2,h_2)$ disconnection density is minimum (i.e., zero).
As  inclination angle $\phi_1$ increases, the $(\bm{b}_1,h_1)$ disconnection density  first decreases to a minimum (zero) at $\phi_1=0$, then increases monotonically;
while the $(\bm{b}_2,h_2)$ disconnection density decreases to a minimum (i.e., zero) at $\phi_1=-9.45^{\circ}$ then increases monotonically.
The $(\bm{b}_3,h_3)$ disconnection density remains constant at $0.1463$ nm$^{-1}$ for all inclinations, $\phi_1$.

Supplementary Figs.~\ref{fig:simu22}(d)-(f) and (g)-(i) show oblique and plan views  of the disconnection networks at the interface, respectively, for inclination angles $\phi_1=0$, $-5^{\circ}$, and $-9.45^{\circ}$ -- all with $\phi_2=0$.
We  see that when $(\phi_1,\phi_2)=(0,0)$, there are only two disconnection  with Burgers vectors $\bm{b}_2,\bm{b}_3$ and zero step heights;  these two sets of disconnection lines are perpendicular to each other.
When the interface is inclined  with $\phi_1=-5^{\circ}$ or $\phi_1=-9.45^{\circ}$, another array of disconnections $(\bm{b}_1,h_1)$ appears, as shown in Supplementary Figs.~\ref{fig:simu22}(d)(e) -- these new disconnections are parallel to the  $(\bm{b}_2,h_2)$ disconnections.
As the inclination $\phi_1$ decreases from $\phi_1=0\rightarrow-5^{\circ}$ and $-9.45^{\circ}$, the density of $(\bm{b}_1,h_1)$ disconnections  increases, the density of disconnections with Burgers vector $\bm{b}_2$  decreases, and the density of $\bm{b}_3$ disconnections remains fixed.
These specific cases reflect the essential mechanism of interface inclination; i.e., that  the disconnection density and its step height distribution must be consistent with the prescribed inclination and cancel all long range elastic fields.

\begin{figure}[!htbp]
\centering
\subfigure[]{ \includegraphics[width=0.32\linewidth]{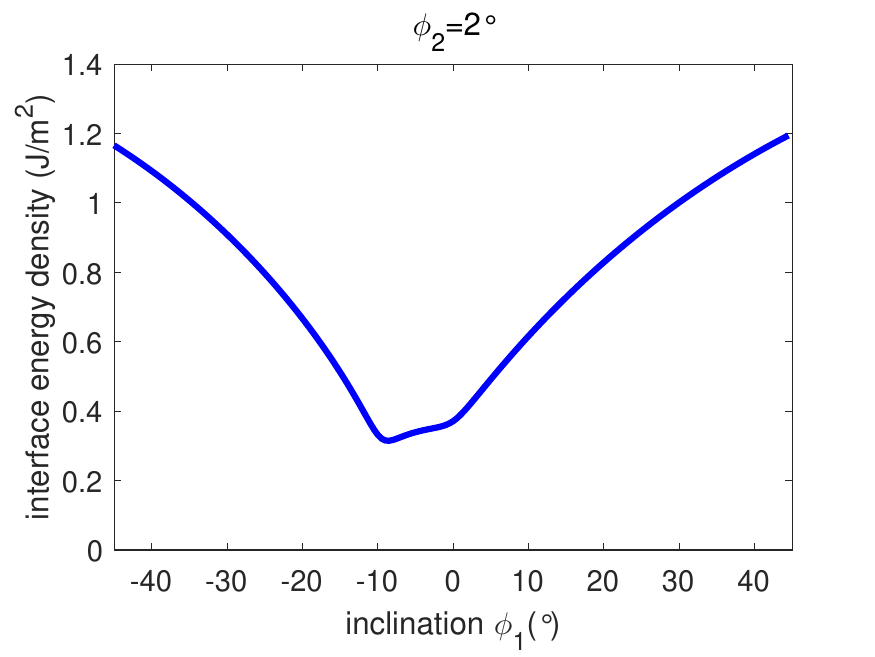}}
\subfigure[]{ \includegraphics[width=0.32\linewidth]{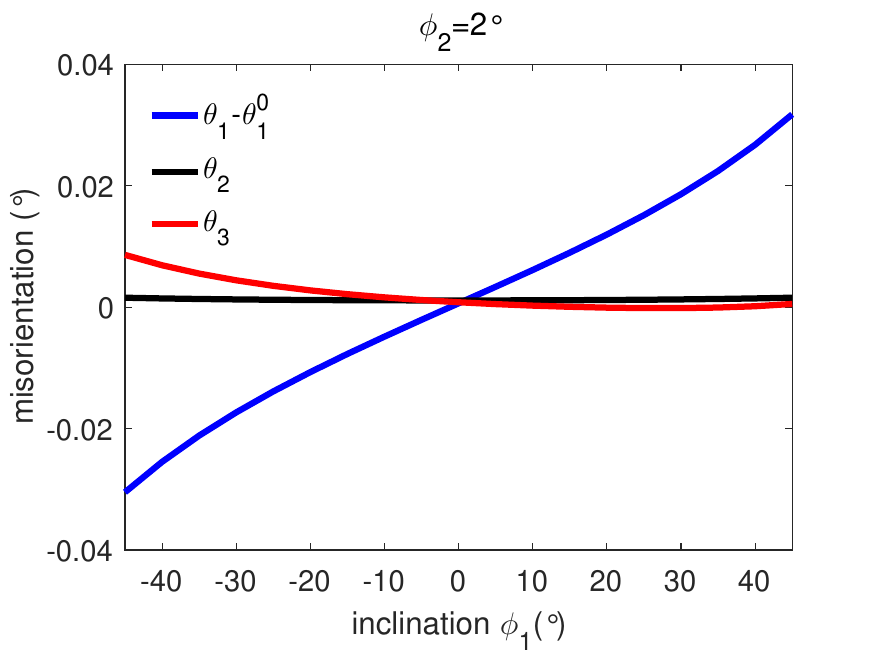}}
\subfigure[]{ \includegraphics[width=0.32\linewidth]{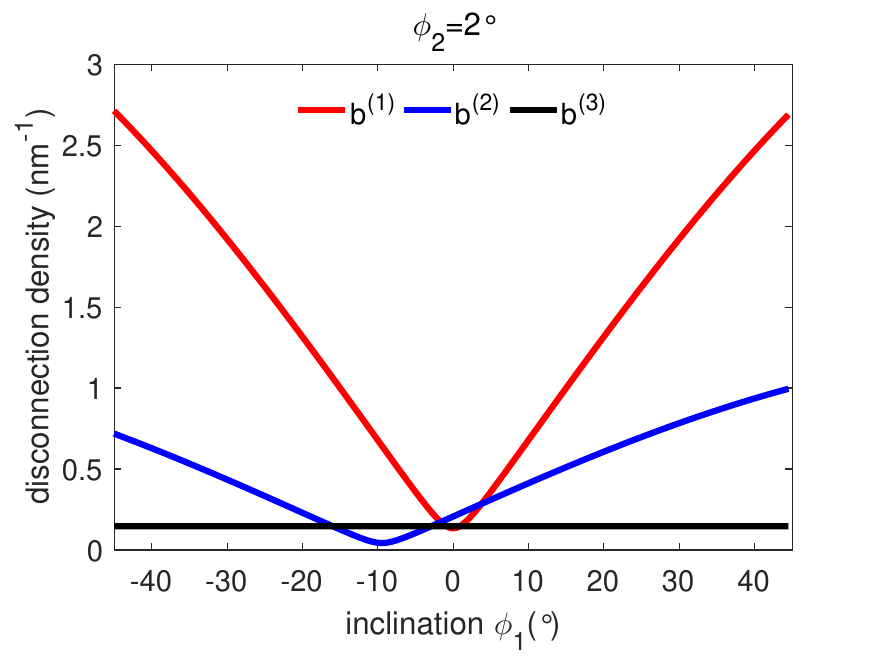}}
\subfigure[]{ \includegraphics[width=0.32\linewidth]{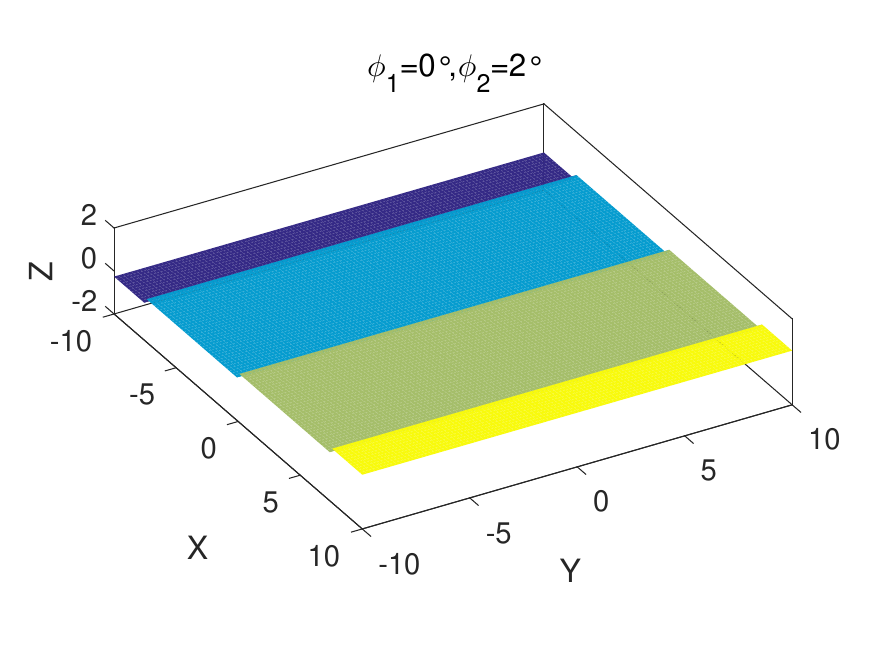}}
\subfigure[]{ \includegraphics[width=0.32\linewidth]{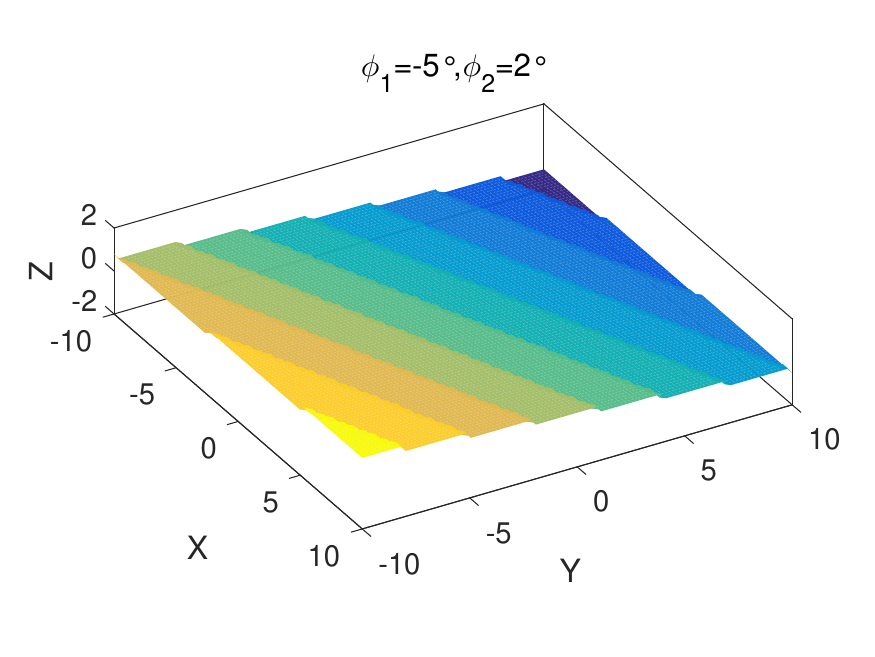}}
\subfigure[]{ \includegraphics[width=0.32\linewidth]{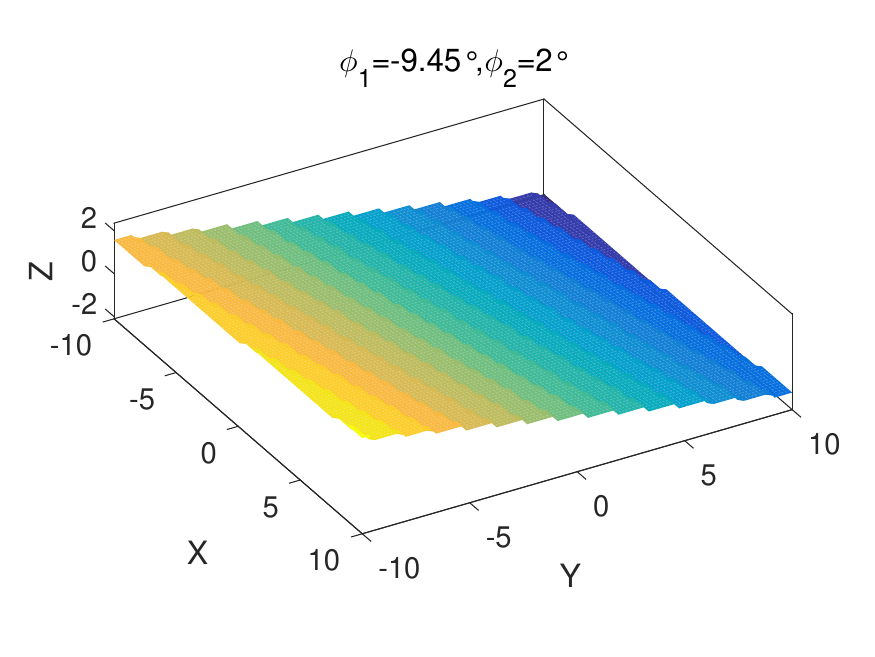}}
\subfigure[]{ \includegraphics[width=0.32\linewidth]{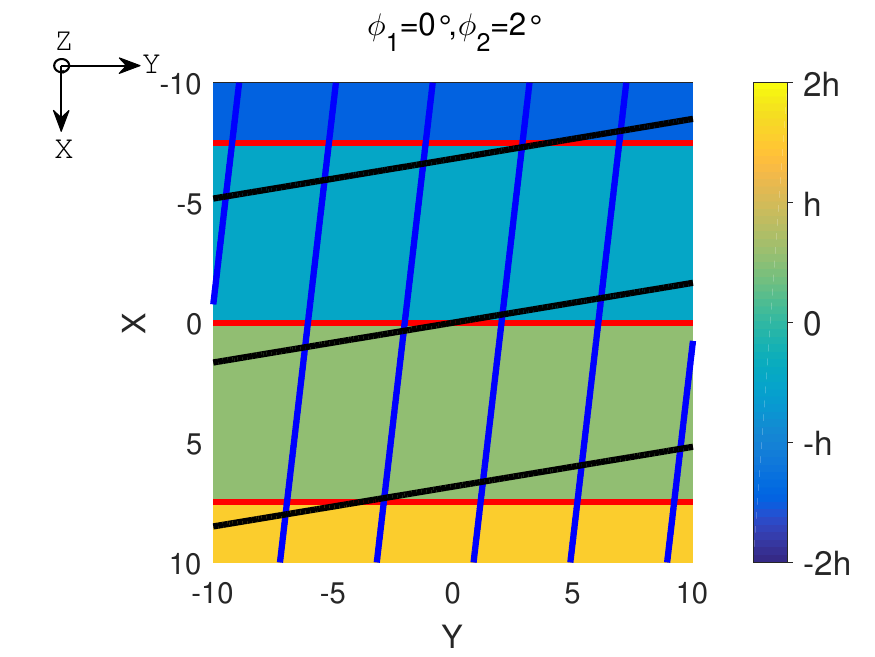}}
\subfigure[]{ \includegraphics[width=0.32\linewidth]{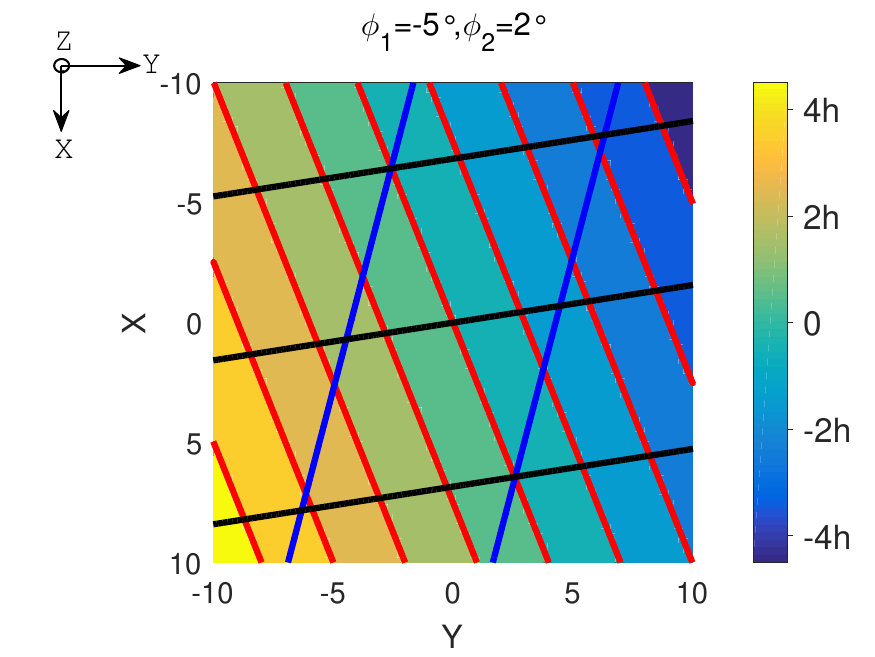}}
\subfigure[]{ \includegraphics[width=0.32\linewidth]{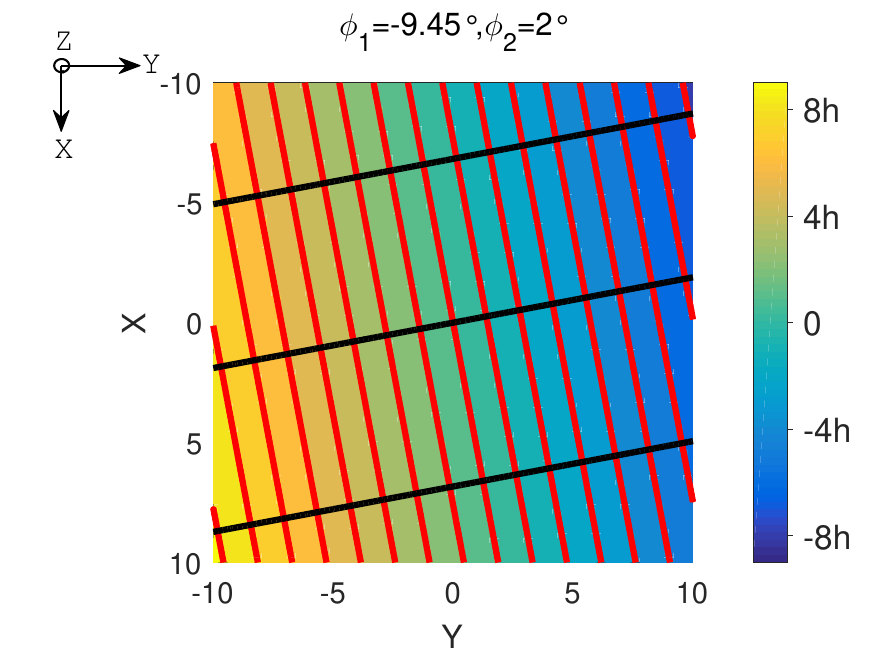}}
\caption{$\alpha$/$\beta$ interface in Ti for $\phi_2=2^{\circ}$, as a function of  inclination $\phi_1$: (a) interfacial energy, (b) misorientation angles, (c) disconnection densities; (d-f) oblique view of the interface disconnection structure and height/step profile;
(g-i) plan views of the interface structures along the $Z$-direction (normal to the reference interface, where the red, blue and black lines respectively denote disconnection types $(\bm{b}_1,h_1)$, $(\bm{b}_2,h_2)$ and $(\bm{b}_3,h_3)$, and the shading between the lines indicates the height profile),
at inclinations $\phi_1=0$, $-5^{\circ}$, and $-9.45^{\circ}$, and $\phi_2=2^{\circ}$. }
\label{fig:simu23}
\end{figure}

Supplementary Fig.~\ref{fig:simu23} shows the interface disconnection structures as a function of inclination angle $\phi_1$ at fixed $\phi_2=2^{\circ}$.
The interfacial energy exhibits  a global minimum at $\phi_1=-9.45^{\circ}$ (see Supplementary Fig.~\ref{fig:simu23}(a)), which is similar to the $\phi_2=0$ case.
In this case,  the misorientation angles $\theta_2,\theta_3$ are non-zero (see Supplementary Fig.~\ref{fig:simu23}(b)) as  compared with the $\phi_2=0$ in Supplementary Fig.~\ref{fig:simu22}, and the misorientation $\theta_1$ is similar to the case of $\phi_2=0$ that has slightly linear increase.
At the minimum energy inclination $\phi_1=-9.45^{\circ}$, the  $(\bm{b}_2,h_2)$ disconnections density is also minimum (albeit non-zero, unlike in the $\phi_2=0$ case).
The $(\bm{b}_1,h_1)$ disconnection density  is a (non-zero) minimum at $\phi_1=0$; and the  $(\bm{b}_3,h_3)$  disconnection density constant at $0.1463$~nm$^{-1}$.
The interfacial energy and  densities of all three disconnections modes are similar to the case of  $\phi_2=0$ as shown  in Supplementary Figs.~\ref{fig:simu23}(a) and (c).
Supplementary Fig.~\ref{fig:simu23}(d)-(f) and (g)-(i) are oblique and plan views of the interface disconnection networks for inclination angles $\phi_1=0$, $-5^{\circ}$, and $-9.45^{\circ}$.
All three sets of disconnections are present in the interface and the sets of  disconnection lines are not always parallel or perpendicular to each other as for $\phi_2=0$.
Compared to the $\phi_2=0$ cases, where the dislocation character of the disconnections are pure edge, here the $(\bm{b}_1,h_1)$ disconnections have non-zero screw character (see Supplementary Figs.~\ref{fig:simu23}(d) and (g)).
As the inclination $\phi_1$ decreases from $\phi_1=0\rightarrow-5^{\circ}$ and $-9.45^{\circ}$, the $(\bm{b}_1,h_1)$ and $(\bm{b}_2,h_2)$ disconnection lines are rotated with both edge and screw components, and the density of $(\bm{b}_1,h_1)$ disconnections increases, while the density of $(\bm{b}_2,h_2)$ disconnections decreases.
The $(\bm{b}_3,h_3)$ disconnection lines remain unchanged.
Here the disconnection step character of the edge component contributes to the inclination $\phi_1$, while the screw component contributes to the inclination $\phi_2$.

\subsection{Interfaces with fixed misorientation angles}

We further examine the interfacial energy,  inclination angle and  disconnection densities for different misorientations.
As the misorientation $\theta_1$ varies, these interface structure properties are similar to the cases of the inclination angle $\phi_1$ varying, as shown in Supplementary Fig.~\ref{fig:simu22}, which is due to the observed linear relation between the misorientation angle $\theta_1-\theta_1^0$ and inclination angle $\phi_1$ in Supplementary Fig.~\ref{fig:simu22}(b).
Similarly, the cases of the misorientation $\theta_2$ varying have the similar trend to the change of inclination angle $\phi_2$.

\begin{figure}[htbp]
\centering
\subfigure[]{  \includegraphics[width=0.32\linewidth]{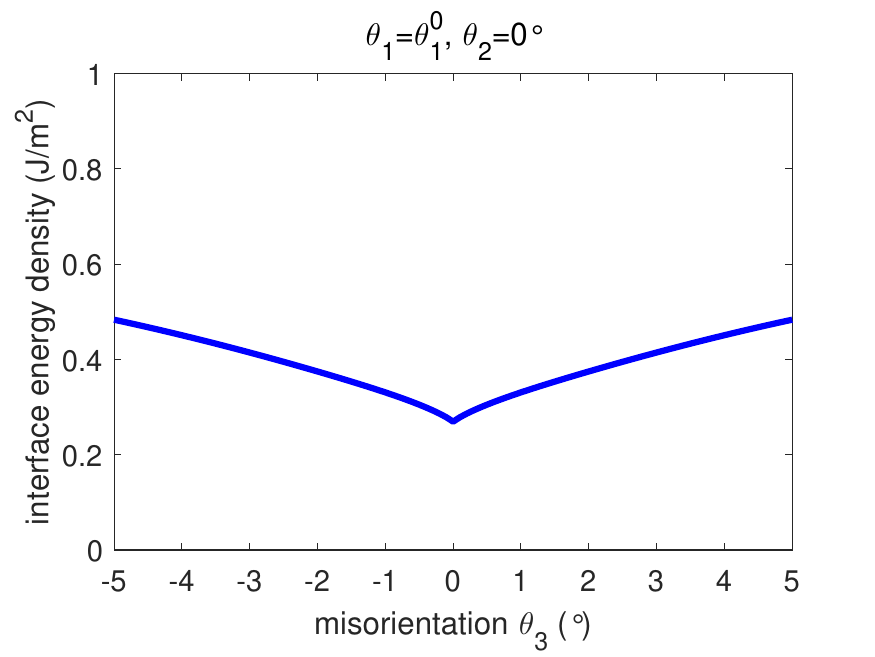}}
\subfigure[]{  \includegraphics[width=0.32\linewidth]{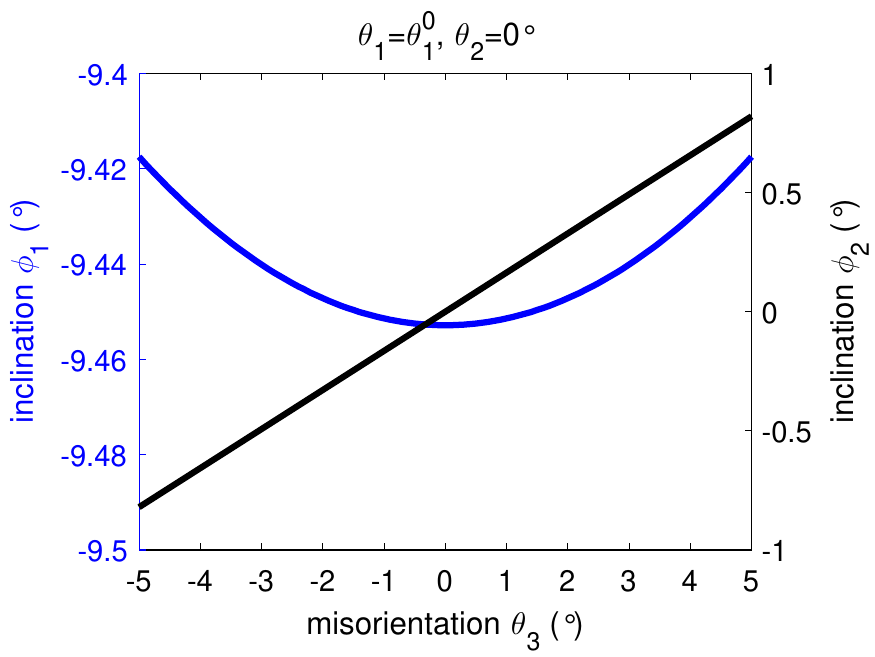}}
\subfigure[]{ \includegraphics[width=0.32\linewidth]{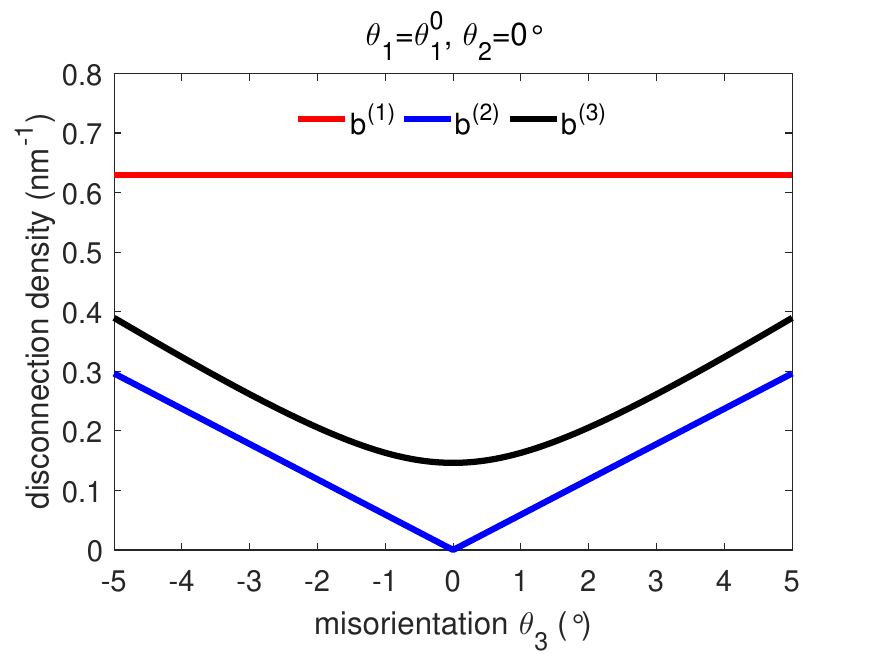}}
\subfigure[]{ \includegraphics[width=0.32\linewidth]{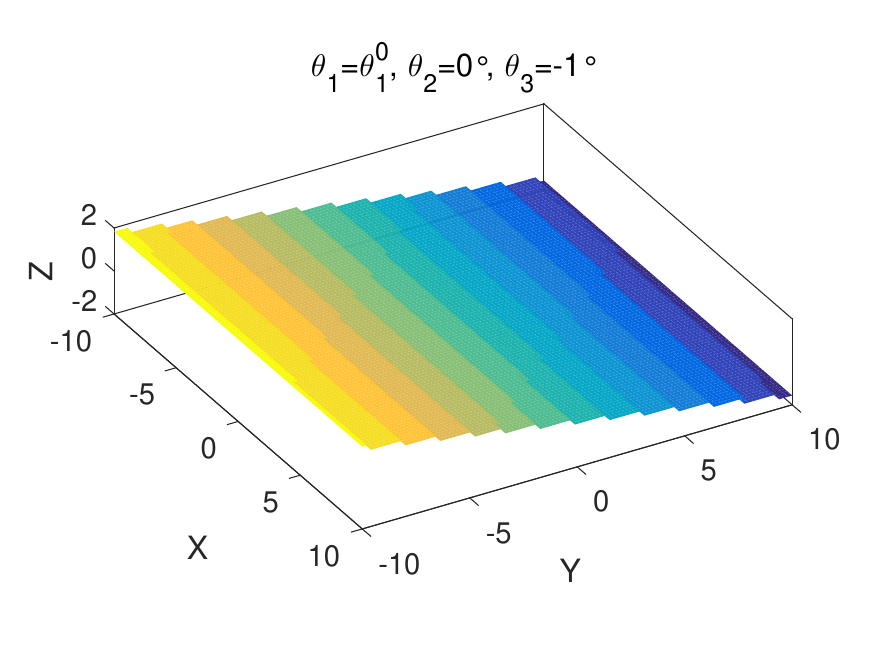}}
\subfigure[]{ \includegraphics[width=0.32\linewidth]{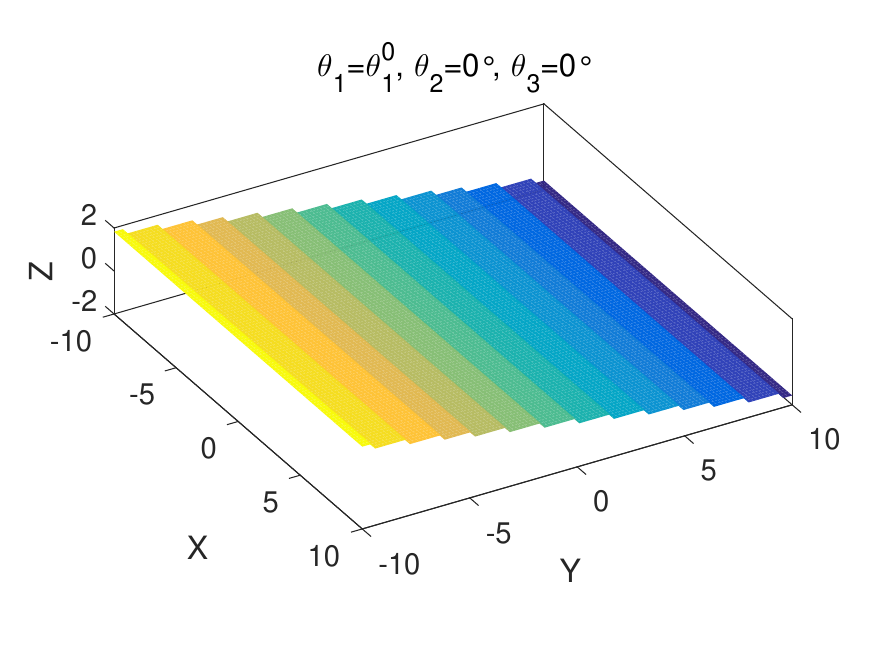}}
\subfigure[]{ \includegraphics[width=0.32\linewidth]{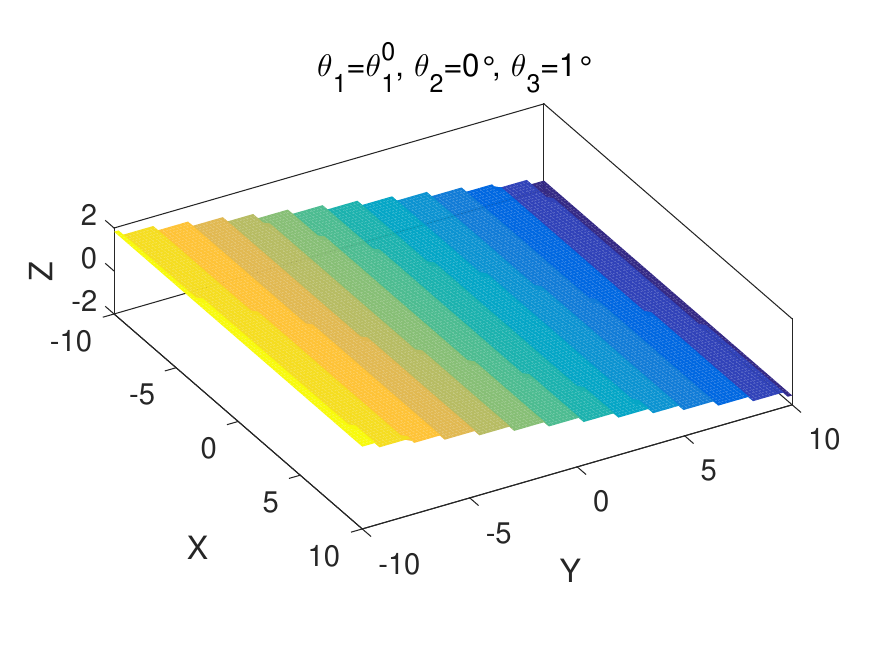}}
\subfigure[]{ \includegraphics[width=0.32\linewidth]{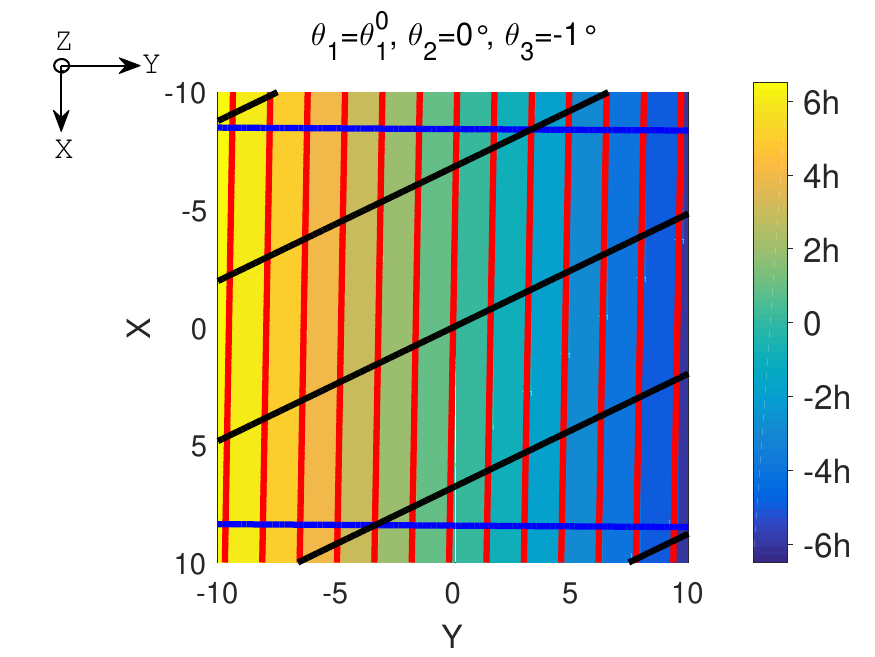}}
\subfigure[]{ \includegraphics[width=0.32\linewidth]{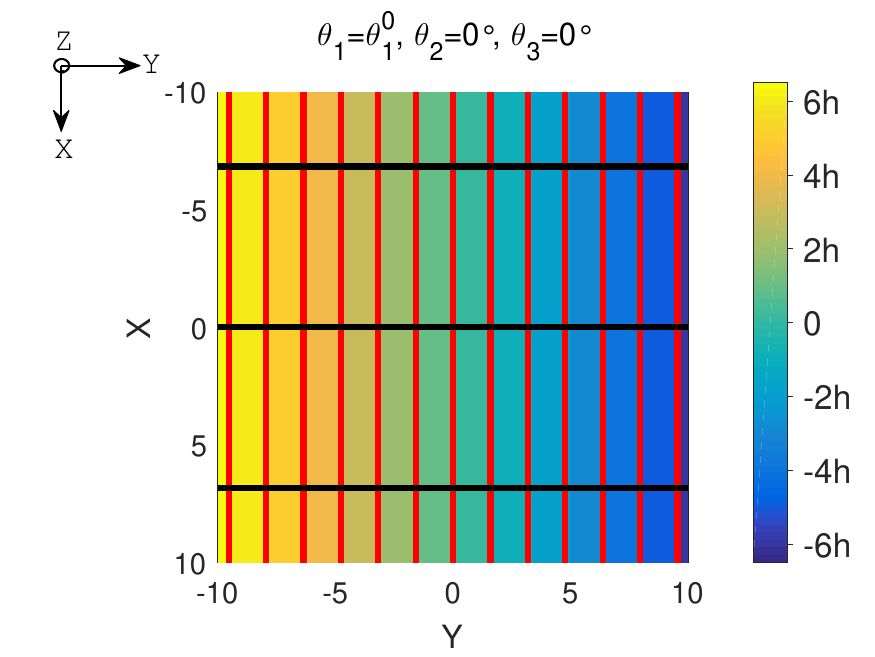}}
\subfigure[]{ \includegraphics[width=0.32\linewidth]{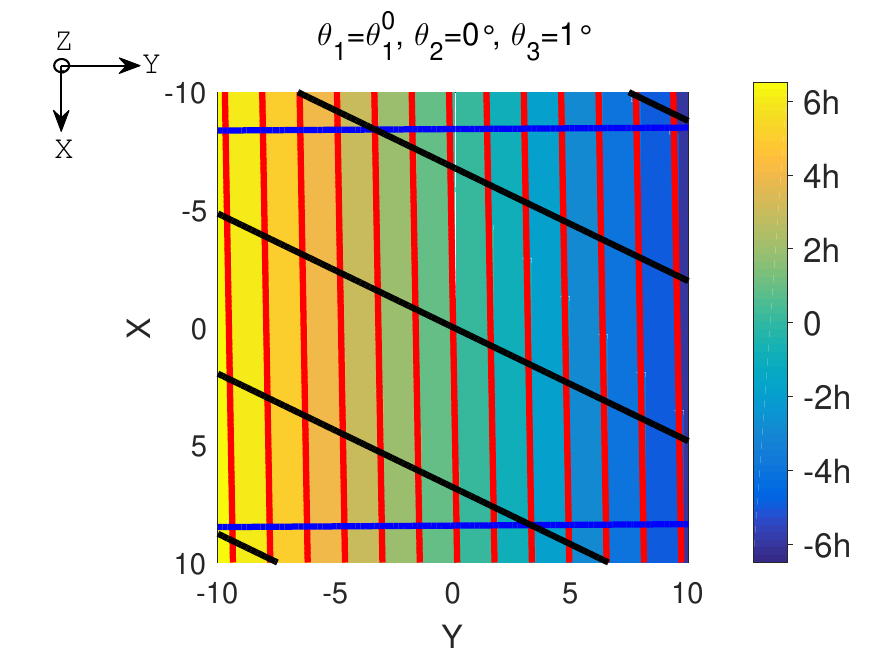}}
\caption{$\alpha$-$\beta$ interface in Ti, as a function of twist misorientation angle $\theta_3$: (a) interfacial energy, (b) misorientation angles, (c) disconnection densities; (d-f) oblique view of the interface disconnection structure and height/step profile;
(g-i) plan views of the interface structures along the $Z$-direction (normal to the reference interface, where the red, blue and black lines respectively denote disconnection types $(\bm{b}_1,h_1)$, $(\bm{b}_2,h_2)$ and $(\bm{b}_3,h_3)$, and the shading between the lines indicates the height profile),
at misorientation $\theta_3=-1^{\circ}$, $0^{\circ}$, and $1^{\circ}$, $\theta_1=\theta_1^0=-5.27^{\circ}$, and $\theta_2=0$. }
\label{fig:simu25}
\end{figure}

For the misorientation twist angle $\theta_3$'s variation, Supplementary Fig.~\ref{fig:simu25} demonstrates the corresponding change of the interface structure properties, with the other two misorientations are fixed as  $\theta_1=\theta_1^0=-5.27^{\circ}$, $\theta_2=0$.
The interfacial energy curve and disconnection density curves attain the minimum values at $\theta_3=0$, and are symmetric about $\theta_3=0$, as shown in Supplementary Figs.~\ref{fig:simu25}(a) and (c). The misorientation twist angle $\theta_3$ does not affect the density of disconnection $(\bm{b}_1,h_1)$, which is a constant value in Supplementary Fig.~\ref{fig:simu25}(c).
We can see in Supplementary Fig.~\ref{fig:simu25}(b) that both the inclinations $\phi_1$, $\phi_2$ vary very small near $\phi_1=-9.45^{\circ}$, $\phi_2=0$, where the interfacial energy attains the global minimum.
It indicates that the twist misorientation around $Z$-axis almost does not affect either the inclination $\phi_1$ around $X$-axis or $\phi_2$ around $Y$-axis.
Supplementary Figs.~\ref{fig:simu25}(d)-(i) show the oblique and plan views of the interfacial disconnection networks with three misorientation angles $\theta_3=-1^{\circ}$, $0^{\circ}$, and $1^{\circ}$.
In Supplementary Figs.~\ref{fig:simu25}(e) and (h), we can see that when $\theta_3=0$,  there are one array of disconnections $(\bm{b}_1,h_1)$ and one array of dislocations with Burgers vectors  $\bm{b}_3$ which are  perpendicular to each other.
Supplementary Figs.~\ref{fig:simu25}(d), (g) and Supplementary Figs.~\ref{fig:simu25}(f), (i) show that when there is a misorientation twist angle, the dislocation lines with Burgers vector $\bm{b}_2$ appear and are perpendicular to the arrays of disconnection $(\bm{b}_1,h_1)$; meanwhile, the lines of dislocation arrays with Burgers vector  $\bm{b}_3$ in the interfacial disconnection networks are rotated;
it is the resultant disconnection screw components that yield the twist.

\bibliographystyle{elsarticle-num}
\bibliography{ref}